\documentclass{IEEEtran}

\usepackage[cmex10]{amsmath}
\usepackage{amssymb}

\usepackage{graphicx}

\newtheorem{defi}{Definition}
\newtheorem{theorem}{Theorem}
\newtheorem{lemma}{Lemma}
\newtheorem{remark}{Remark}
\newtheorem{corollary}{Corollary}

\newcommand{\Uc}{\mathcal{U}}
\newcommand{\Xc}{\mathcal{X}}
\newcommand{\Yc}{\mathcal{Y}}

\newcommand{\Cc}{\mathcal{C}}
\newcommand{\Mc}{\mathcal{M}}
\newcommand{\Tc}{\mathcal{T}}
\newcommand{\Vc}{\mathcal{V}}
\newcommand{\Qc}{\mathcal{Q}}
\newcommand{\Rc}{\mathcal{R}}
\newcommand{\Fc}{\mathcal{F}}
\newcommand{\Nc}{\mathcal{N}}

\begin{document}
%
\title{On the Capacity of Interference Channel with Causal and Non-causal Generalized Feedback at the Cognitive Transmitter}
%
%

\author
{
Mahtab Mirmohseni, Student Member, IEEE, Bahareh Akhbari, and Mohammad Reza Aref\\
\thanks{This work was partially supported by Iran National Science Foundation (INSF) under contract No. 88114/46-2010 and by Iran Telecom Research Center (ITRC) under contract No. T500/17865.

The authors are with the Information Systems and Security Lab (ISSL), Department of Electrical Engineering, Sharif University of Technology, Tehran,
Iran (e-mail: mirmohseni@ee.sharif.edu, b\_akhbari@ee.sharif.edu, and aref@sharif.edu).}}

%

\maketitle
\begin{abstract}
In this paper, taking into account the effect of link delays, we investigate the capacity region of the Cognitive Interference Channel (C-IFC), where cognition can be obtained from either causal or non-causal generalized feedback. For this purpose, we introduce the Causal Cognitive Interference Channel With Delay (CC-IFC-WD) in which the cognitive user's transmission can depend on $L$ future received symbols as well as the past ones. We show that the CC-IFC-WD model is equivalent to a classical Causal C-IFC (CC-IFC) with link delays.
Moreover, CC-IFC-WD extends both genie-aided and causal cognitive radio channels and bridges the gap between them. First, we derive an outer bound on the capacity region for the arbitrary value of $L$ and specialize this general outer bound to the strong interference case. Then, under strong interference conditions, we tighten the outer bound. To derive the achievable rate regions, we concentrate on three special cases: 1)~Classical CC-IFC ($L=0$), 2)~CC-IFC without delay ($L=1$), and 3)~CC-IFC with unlimited look-ahead in which the cognitive user non-causally knows its entire received sequence. In each case, we obtain a new inner bound on the capacity region.
The derived achievable rate regions under special conditions reduce to several previously known results. Moreover, we show that the coding strategy which we use to derive an achievable rate region for the classical CC-IFC achieves the capacity for the classes of degraded and semi-deterministic classical CC-IFC under strong interference conditions. Furthermore, we extend our achievable rate regions to the Gaussian case. Providing some
numerical examples for Gaussian CC-IFC-WD, we compare the performances of the different strategies and investigate the rate gain of the cognitive link for different delay values. We show that one can achieve larger rate regions in the ``without delay'' and ``unlimited look-ahead'' cases than in the classical CC-IFC; this improvement is likely due to the fact that, in the former two cases, the cognitive user can cooperate more effectively with the primary user by knowing the current and future received symbols.
\end{abstract}



%
\IEEEpeerreviewmaketitle


\section{Introduction}
Interference management is one of the key issues in wireless networks wherein multiple source-destination pairs share same medium and interfere with each other. Interference Channel (IFC) \cite{Carl78} is the simplest model for this scenario, with two independent transmitters sending messages to their intended receivers. However, users with cognitive radio technology may sense the medium and use the obtained data to adapt their transmissions to cooperate with other users and improve their own rates as well as the rates of others. Cognitive Interference Channel (C-IFC) refers to a two-user IFC in which the cognitive user (secondary user) has the ability to obtain the message being transmitted by the other user (primary user), in either a non-causal or causal manner. C-IFC was first introduced in \cite{DevrMitTar06}, where for the non-causal C-IFC an achievable rate region was derived by combining Gel'fand-Pinsker (GP) binning \cite{GelfPin80} and a well-known simultaneous superposition coding scheme (rate splitting) applied to the IFC \cite{HanKob81}, which allows the receivers to decode part of the non-intended message.

Non-causal C-IFC, also termed genie-aided C-IFC, in which the cognitive user has non-causal full or partial knowledge of the other user's transmitted message, has been widely investigated in \cite{JoviVis09}-\cite{CaoChen09} and the studies represented in the references therein. Yet, capacity results are known only in special cases. For an overview on the capacity results of the non-causal C-IFC, see \cite{RiniIT11}, which contains the strongest results for the non-causal channel model. In the Causal C-IFC (CC-IFC), the cognitive user can exploit knowledge of the primary user's message from the causally received signals (information overheard by the feedback link from the channel and not that sent back from the receivers). Due to the complex nature of the problem, although CC-IFC is a more realistic and appropriate model for practical applications than the non-causal C-IFC, CC-IFC has been far less investigated in comparison to the latter \cite{CaoChen09}. In \cite{DevrMitTar06}, achievable rate regions for the CC-IFC that consist of non-cooperative causal transmission protocols have been characterized. An improved rate region for CC-IFC employing a cooperative coding strategy based on the block Markov superposition coding (full Decode-and-Forward (DF) \cite{CoveElg79}) and GP coding has been derived in \cite{SeyXinLian09}. Also, inner and outer bounds on the capacity region of CC-IFC have been derived in \cite{CaoChen08}. However, the problem of finding the capacity region of CC-IFC remains open. A more general model in which both transmitters are causally cognitive has been proposed in \cite{Tuni07}, called Interference Channel with Generalized Feedback (IFC-GF). The \emph{generalized feedback}, in contrast to the output feedback, refers to the information overheard by the transmitter(s) over the channel and not to the information sent back by the receiver(s). Different achievable rate regions for IFC-GF have been obtained in \cite{Tuni07}-\cite{YangTuni08}, combining the methods of rate splitting, block Markov superposition coding and GP binning. Moreover, outer bounds on the capacity region of the Gaussian cognitive Z-interference channel were derived in the causal case \cite{CaoChenGlob09}. It is noteworthy that in IFC-GF, cooperation between transmitters is performed using the links which share the same band as the links in IFC. Another scenario for transmitters cooperation is the case in which the cooperative links are orthogonal to each other as well as the links in IFC, termed conferencing. Multiple Access Channel (MAC) with conferencing was first studied by Willems \cite{Wil83}, and in \cite{MariYatKra07} is extended to the compound MACs with conferencing encoders.
\begin{figure}[tb]
  \centering
  \includegraphics[width=9cm]{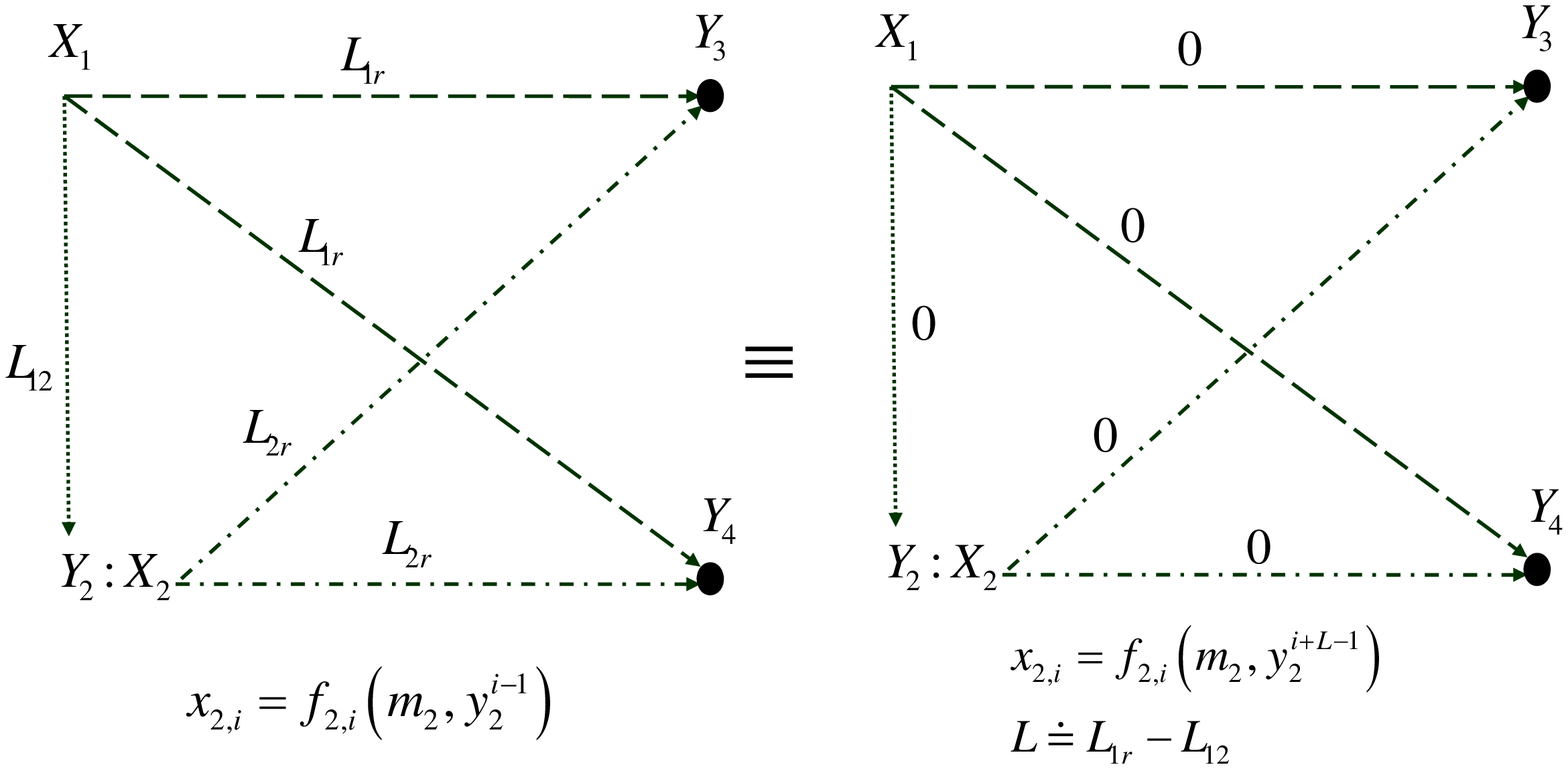}
  \caption{(Left) Graphic representation for CC-IFC with link delays, (Right) Graphic representation for CC-IFC-WD. Two channel models are equivalent.}
  \label{fig:channel_graph}
\end{figure}

\begin{figure}[tb]
  \centering
  \includegraphics[width=7cm]{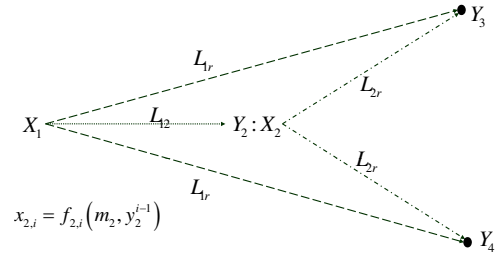}
  \caption{An example for the configuration of Fig.~\ref{fig:channel_graph}}
  \label{fig:channel_example}
\end{figure}

CC-IFC reveals the characteristics of the broadcast, multiple-access and relay channels. Since an arbitrarily long delay is required to achieve the capacity, link delays have no effect on the capacity of broadcast and multiple-access channels. However, relaying structure may be changed by introducing link delays, and this can change the capacity of channels with relays \cite{ElgaHasMam07}. Consider the classical CC-IFC in Fig.~\ref{fig:channel_graph}~(Left) and suppose that there are delays of $L_{1r}$ units on the links between the primary user and the
receivers, of $L_{2r}$ units on the links between the cognitive user and the receivers, and of $L_{12}$ units on the link between transmitters. We refer to this channel as \emph{CC-IFC with link delays}. We assume that all link delays are positive integers and the cognitive user hears the primary user's transmitted signal earlier than do the receivers, i.e., $L_{12}\leq L_{1r}$. A simple example which satisfies this
assumption is shown in Fig.~\ref{fig:channel_example}. We use this channel to obtain an information theoretical model which extends genie-aided and causal cognitive radio channels.

In order to obtain the information theoretical limits of cognitive radios, causal and non-causal C-IFC models attempt to capture the specifications of the cognitive radio technology \cite{Mit918}, which aims at developing communication systems with the capability of sensing the environment and then adapting to it. For this purpose, researchers focus mostly on the non-causal C-IFC models. Moreover, despite the complex nature of the CC-IFC model, it is unsuited to all scenarios. In fact, due to the cognitive user's cognitive ability, it may hear the primary user's transmitted signal earlier than do the receivers, and the cognitive user can utilize this extra information to cooperate in sending or to precode against the primary user's message.

The special features discussed above motivate us to define the Causal Cognitive Interference Channel With Delay (CC-IFC-WD) as an IFC where one of the transmitters can causally overhear the channel and its transmission depends on the $L$ future (noisy) received symbols as well as the past ones. This can be seen as the equivalent of the classical CC-IFC with $-L$ time units of delay on the cognitive user's received signal (or on the link between the transmitters). To physically motivate this channel model, we show that CC-IFC-WD (Fig.~\ref{fig:channel_graph}~(Right)) is equivalent to the CC-IFC with link delays shown in Fig.~\ref{fig:channel_graph}~(Left), where $L_{12}\leq L_{1r}$. As can be seen in Fig.~\ref{fig:channel_example}, this channel model may fit wireless networks where the transmitters are close to each other or there is a high-speed link between them. Since setting $L=0$ in the CC-IFC-WD model results in a classic causal model, CC-IFC-WD extends CC-IFC. Since, instead of the primary user's message, a noisy version of the primary user channel input is provided to the cognitive user (when the cognitive user has unlimited look-ahead and non-causally knows its entire received sequence), CC-IFC-WD also extends non-causal C-IFC. Therefore, CC-IFC-WD is a middle point between the genie-aided (non-causal) C-IFC and CC-IFC. In fact, a simple strategy which allows the users to cooperate instantaneously could be beneficial and could increase the channel capacity, as does the case in the Relay With Delay (RWD) channel \cite{Hass06}. The RWD channel has been vastly investigated in \cite{ElgaHasMam07, Hass06}, wherein different upper and lower bounds and some capacity results have been derived. The lower bounds are achieved based on the combination of cooperative strategies such as full or partial DF, instantaneous relaying (for $L>0$), where the relay sends a function of its current received symbol, and non-causal DF (for the unlimited look-ahead case), in which the relay pre-decodes part or all of the message at the beginning of the block and transmits the message to the receiver in cooperation with the source. A new general upper bound which holds for any arbitrary amount of delay has been derived in \cite{SalMirAref09} and is shown to be tighter in some cases than the previously established bounds. It has been shown that the capacity of the discrete memoryless RWD channel is strictly larger than that of the classical relay channel \cite{ElgaHasMam07, Hass06}.

\subsection{Main contributions and organization}\label{subsec:contributions}
In this paper, we study the IFC with causal and non-causal generalized feedback at the cognitive transmitter by defining the CC-IFC-WD. We derive new results regarding the capacity region of this channel for both discrete memoryless and Gaussian cases. Our contributions in the rest of the paper are organized as follows:

\begin{itemize}
\item We introduce the general CC-IFC-WD in Section~\ref{sec:definition}, where we also prove the equivalence of this channel model with \textit{CC-IFC with link delays}.

\item In Section~\ref{sec:OUTER}, we first derive an outer bound on the capacity region of the new channel model (CC-IFC-WD) for an arbitrary value of $L$. Based on the fact that the receivers cannot cooperate, we use the idea of \cite{Sato78} in providing the cognitive receiver with a side information which has the same marginal distribution as the primary receiver's signal and an arbitrary correlation with the cognitive receiver's signal. This idea has been utilized in \cite{TunITA10} to establish an outer bound on the capacity region of IFC-GF. We also make use of the techniques of \cite{SalMirAref09} to incorporate the amount of the delay ($L$). Next, we apply the strong interference condition at the primary receiver to the general outer bound in order to derive an outer bound under this condition, which is further tightened by setting the strong interference condition at the cognitive receiver.

\item To determine the achievable rate regions, we focus on three special cases in Section~\ref{sec:INNER}: 1)~Classical CC-IFC which corresponds to $L=0$, 2)~CC-IFC without delay ($L=1$), and 3)~CC-IFC with unlimited look-ahead.

\item A new inner bound for the classical CC-IFC is presented in Section~\ref{subsec:zerodelay}. This bound is based on the coding schemes which combine cooperative, collaborative and interference mitigating strategies. These strategies include rate splitting at both transmitters as in the Han-Kobayashi (HK) scheme \cite{HanKob81}, GP binning at the cognitive user, and Generalized block Markov coding (partial DF \cite{CoveElg79}). Next, we compare our scheme with the previous results and show that our scheme includes the scheme of \cite{SeyXinLian09} for CC-IFC, and the schemes of \cite{Tuni07}-\cite{YangTuni08} tailored to CC-IFC.

\item In Section~\ref{subsec:onedelay}, we consider the CC-IFC without delay ($L=1$), where the current received symbol (at the cognitive user) could also be utilized and present a new inner bound for this channel. Our coding scheme is based on the combination of the strategies of Section~\ref{subsec:zerodelay} with instantaneous relaying. This means that the cognitive user, having access to the current received symbol, sends a function of its current received symbol and the codeword obtained by other strategies.

\item CC-IFC with unlimited look-ahead, in which the cognitive user non-causally knows its entire received sequence, is investigated in Section~\ref{subsec:ndelay}. To obtain the achievable rate region, we employ non-causal partial DF strategy in which the cognitive user can contribute to the rate of the primary user by encoding a part of the primary user's message and cooperating with the primary user to transmit this decoded part of the message. We remark that using a coding scheme based on instantaneous relaying is feasible for this case. However, to compare this strategy with non-causal partial DF, we restrict the use of this scheme to $L=1$. When the cognitive link between transmitters is ideal, CC-IFC with unlimited look-ahead reduces to a non-causal C-IFC. Therefore, we compare our proposed scheme with the results in \cite{JoviVis09}-\cite{JiangXin08}, \cite{LiuMarGoldSha09} for non-causal C-IFC, and show that our scheme encompasses most of the previous results and all of the capacity achieving schemes of \cite{WuVishAra07} for weak interference, \cite{MariYatKra07} for strong interference, and \cite{LiuMarGoldSha09} for a class of Z cognitive channel.

\item In Section~\ref{sec:Cap}, we derive the capacity regions for the classes of degraded and semi-deterministic classical CC-IFC under strong interference conditions, where achievability proofs follow from the region of Section~\ref{subsec:zerodelay}, and for the converse parts, we evaluate the outer bound of Section~\ref{sec:OUTER} for $L=0$.

\item In Section~\ref{sec:Gaussian}, Gaussian CC-IFC-WD is investigated where we extend the achievable rate regions of Section~\ref{sec:INNER} to the Gaussian case. Providing some numerical examples for Gaussian CC-IFC-WD, we investigate the rate gain of the cognitive link for different delay values. In addition, we compare the strategies used in our coding schemes and show that instantaneous relaying and non-causal DF improve the rate region noticeably.

\item Finally, Section~\ref{sec:conclusion} concludes the paper.
\end{itemize}

\section{Channel Models and Preliminaries}\label{sec:definition}
Throughout this paper, the following notations are used: upper case letters, e.g. $X$, are used to denote Random Variables (RVs) and lower case letters, e.g. $x$, show their realizations. The probability mass function (p.m.f) of a Random Variable (RV) $X$ with alphabet set $\Xc$, is denoted by $p_X(x)$, where the subscript $X$ is occasionally omitted. Additionally, $|\Xc|$ denotes the cardinality of a finite discrete set $\Xc$. $A_\epsilon^n(X,Y)$ specifies the set of $\epsilon$-strongly, jointly typical sequences of length $n$ on $p(x,y)$, abbreviated by $A_\epsilon^n$ if it is clear from the context. The notation $X^j_i$ indicates a sequence of RVs $(X_i,X_{i+1},...,X_j)$, where we use $X^j$ instead of $X^j_1$ for the sake of brevity. $\Nc(0,\sigma^2)$ denotes the normal distribution with zero mean and variance $\sigma^2$.

\begin{figure}[tb]
  \centering
  \includegraphics[width=7.5cm]{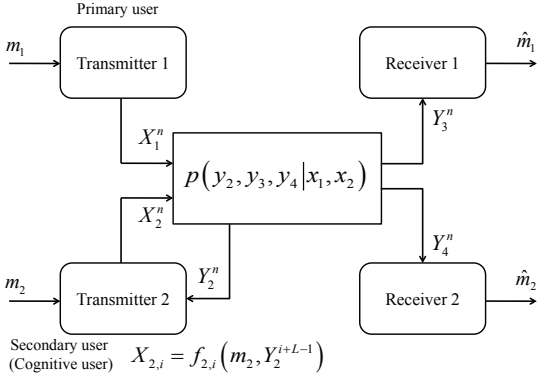}
  \caption{Causal Cognitive Interference Channel With Delay (CC-IFC-WD)}
  \label{fig:channelmodel}
\end{figure}

Consider the CC-IFC-WD in Fig.~\ref{fig:channelmodel} with finite input alphabets $\Xc_1,\Xc_2$, finite output alphabets $\Yc_2,\Yc_3,\Yc_4$, and a channel transition probability distribution $p(y_2,y_3,y_4|x_1,x_2)$, denoted by ($\Xc_1\times\Xc_2,p(y_2,y_3,y_4|x_1,x_2),$ $\Yc_2\times\Yc_3\times\Yc_4$), where $X_1\in\Xc_1$ and $X_2\in\Xc_2$ are inputs of Transmitter~1 (Tx1) and Transmitter~2 (Tx2), respectively, $Y_2\in\Yc_2$ is the secondary user output, $Y_3\in\Yc_3$ and $Y_4\in\Yc_4$ are channel outputs at the Receiver~1 (Rx1) and Receiver~2 (Rx2), respectively. In $n$ channel uses, each Transmitter~$u$ (Tx$u$) sends a message $m_u$ to the Receiver~$u$ (Rx$u$) for $u\in\{1,2\}$.
\begin{defi}\label{def:code}
A $(2^{nR_1},2^{nR_2},n,P_e^{(n)})$ code for CC-IFC-WD consists of (i) two message sets $\Mc_1=\{1,\ldots,2^{nR_1}\}$ and $\Mc_2=\{1,\ldots,2^{nR_2}\}$ for the primary and secondary users, respectively; (ii) an encoding function at the primary user $f_1:\Mc_1\mapsto\Xc_1^n$; (iii) a set of encoding functions at the secondary user $x_{2,i}=f_{2,i}(m_2,y_2^{i+L-1})$, for $1\leq i\leq n$ and $m_2\in\Mc_2$; (iv) two decoding functions at Rx1 and Rx2, $g_1: \Yc_3^n\mapsto\Mc_1$ and $g_2:\Yc_4^n\mapsto\Mc_2$. We assume that the channel is memoryless. Thus, for $m_1\in\Mc_1$, and $m_2\in\Mc_2$, the joint p.m.f of $\Mc_1\times\Mc_2\times\Xc_1\times\Xc_2\times\Yc_2\times\Yc_3\times\Yc_4$ is given by
\begin{IEEEeqnarray}{l}\label{eqn:pmf}
p(m_1,m_2,x_1^n,x_2^n,y_2^n,y_3^n,y_4^n)=p(m_1)p(m_2)\prod\limits_{i=1}^np(x_{1,i}|m_1)\nonumber\\
\times p(x_{2,i}|m_2,y_2^{i+L-1})p(y_{2,i}|x_{1,i})p(y_{3,i},y_{4,i}|x_{1,i},x_{2,i})
\end{IEEEeqnarray}
where we avoid instantaneous feedback from $X_2$ to $Y_2$, which may happen depending on the delay value ($L$) in the encoding
process of the cognitive user. The probability of error for this code is defined as $P_e^{(n)}=max\left(P_{e,1}^{(n)},P_{e,2}^{(n)}\right)$, where
for $u\in\{1,2\}$ we have
\begin{displaymath}
P_{e,u}^{(n)}=\frac{1}{2^{n(R_1+R_2)}}\sum\limits_{m_1,m_2}{Pr(g_u(Y^n_{u+2})\neq m_u | (m_1,m_2)\textrm{ sent})}
\end{displaymath}
\end{defi}

\begin{defi}\label{def:rate}
A rate pair $(R_1,R_2)$ is achievable if there exists a sequence of $(2^{nR_1},2^{nR_2},n,P_e^{(n)})$ codes with $P_e^{(n)}\rightarrow 0$ as $n\rightarrow\infty$. The capacity region $\Cc_L$ is the closure of the set of all achievable rates. Thus, $L=0$ corresponds to the \emph{classical CC-IFC} with capacity $\Cc_0$ and $L=1$ corresponds to the \textit{CC-IFC without delay} with capacity $\Cc_1$. Moreover, we define \emph{CC-IFC with unlimited look-ahead} as a CC-IFC-WD wherein the encoding functions at the secondary user take the form of $x_{2,i}=f_{2,i}(m_2,y_2^{n})$ and denote its capacity by $\Cc_n$.
\end{defi}

In the following Theorem, we physically motivate the CC-IFC-WD model, which is defined above.
\begin{theorem}\label{thm:model_equivalent}
A discrete memoryless CC-IFC-WD, using $L$ future symbols at the cognitive user, $x_{2,i}=f_{2,i}(m_2,y_2^{i-1+L})$, shown in Fig.~\ref{fig:channel_graph}~(Right), is equivalent to a discrete memoryless CC-IFC with link delays shown in Fig.~\ref{fig:channel_graph}~(Left).
\end{theorem}

\begin{IEEEproof}
The CC-IFC with link delays is defined as a channel $\Xc_1,\Xc_2\rightarrow \Yc_2,\Yc_3,\Yc_4$,
\begin{equation}\label{eqn:CC-IFC_with_link_delays}
\begin{array}{c}
\prod\limits_{i=1}^np(y_{2,i}|x_{1,i-L_{12}})p(y_{3,i},y_{4,i}|x_{1,i-L_{1r}},x_{2,i-L_{2r}})\\
x_{2,i}=f_{2,i}(m_2,y_{2,-L_{1r}+L_{12}+1},\ldots,y_{2,1},\ldots,y_{2,i-1})
\end{array}
\end{equation}

Now, we define
\begin{equation}\label{eqn:RVs_subs}
\begin{array}{c}
\tilde{y}_{3,i}=y_{3,i}\:,\:\tilde{y}_{4,i}=y_{4,i}\:,\:\tilde{y}_{2,i+L}=y_{2,i}\\
\tilde{x}_{1,i+L_{1r}}=x_{1,i}\:,\:\tilde{x}_{2,i+L_{2r}}=x_{2,i}
\end{array}
\end{equation}
where $L\doteq L_{1r}-L_{12}$. Substituting the RVs of (\ref{eqn:RVs_subs}) into (\ref{eqn:CC-IFC_with_link_delays}), the equivalent channel model of CC-IFC-WD is obtained as
\begin{equation}\label{eqn:CC-IFC-WD_equi}
\begin{array}{c}
\prod\limits_{i=1}^np(\tilde{y}_{2,i+L}|\tilde{x}_{1,i+L})p(\tilde{y}_{3,i},\tilde{y}_{4,i}|\tilde{x}_{1,i},\tilde{x}_{2,i})\\
\tilde{x}_{2,i}=\tilde{f}_{2,i}(m_2,\tilde{y}_{2,1},\ldots,\tilde{y}_{2,i},\ldots,\tilde{y}_{2,i+L-1}).
\end{array}
\end{equation}

Comparing (\ref{eqn:CC-IFC-WD_equi}) with (\ref{eqn:pmf}) completes the proof.
\end{IEEEproof}

\section{Outer Bounds on the Capacity Region of Discrete Memoryless CC-IFC-WD}\label{sec:OUTER}
In this section, we investigate the outer bounds on the capacity region of CC-IFC-WD. Since the receivers cannot cooperate, we give the cognitive receiver a side information with the same marginal distribution as the primary receiver's signal but an arbitrary correlation with the cognitive receiver's signal, as in \cite{Sato78, TunITA10}. Based on this idea, and using the techniques of \cite{SalMirAref09} for defining the auxiliary RVs, we first derive an outer bound on the capacity region of general CC-IFC-WD for arbitrary values of $L$.

\begin{theorem}\label{thm:outer_gen}
The capacity region of CC-IFC-WD with the joint p.m.f \eqref{eqn:pmf}, is contained in the region
\begin{IEEEeqnarray}{rl}
&\Rc_{out} =\bigcup\limits_{p(u,t)p(x_1|t)p(v|t),f'_2(v,u,y_2)}
\Big\{(R_1,R_2): R_1 \geq 0, R_2 \geq 0 \nonumber\\
&R_1 \leq I(X_1;Y_2|V,T)+I(X_1;Y_3|X_2,Y_2,T)\label{eqn:outer_genI}\\
&R_2 \leq I(V;Y_4|X_1,T)\label{eqn:outer_genII}\\
&R_2 \leq I(U,V;Y_3|X_1,T)+I(U,V,T;Y_4|X_1,Y'_3)\label{eqn:outer_genIII}\\
&R_1+R_2 \leq I(X_1,U,V,T;Y_3)+I(U,V,T;Y_4|X_1,Y'_3)\Big\}\IEEEeqnarraynumspace\label{eqn:outer_genIV}
\end{IEEEeqnarray}
where $x_2=f'_2(v,u,y_2)$. Also, $Y'_3$ has the same marginal distribution of $Y_3$; i.e., $p(y_2^n,y_3^{'n}|x_1^n,x_2^n)=p(y_2^n,y_3^n|x_1^n,x_2^n)$, but $p(y_3^{'n},y_4^n|x_1^n,x_2^n)$ is an arbitrary joint distribution. We remark that, the dependence on $L$ is through the input distribution.
\end{theorem}

\begin{remark} Nullifying $Y_4$ and setting $R_2=0$, $\Rc_{out}$ reduces to the capacity upper bound derived in \cite[Theorem~1]{SalMirAref09}
for the RWD channel.
\end{remark}

\begin{IEEEproof}
See Appendix~\ref{app:out_proof}.
\end{IEEEproof}

Now, we impose the strong interference condition at the primary receiver under which the interfering signal at Rx1 is strong enough that both messages can be decoded; we assume that the following strong interference condition hold
\begin{IEEEeqnarray}{rcl}
I(X_2;Y_4|X_1)&\leq& I(X_2;Y_3|X_1).\label{eqn:cond_str_rec1}
\end{IEEEeqnarray}

\begin{theorem}\label{thm:outer_str1}
The capacity region of CC-IFC-WD with the joint p.m.f (\ref{eqn:pmf}), satisfying (\ref{eqn:cond_str_rec1}), is contained in the region
\begin{IEEEeqnarray}{rl}
\Rc_{out}^{str_1} =&\bigcup\limits_{p(u,t)p(x_1|t)p(v|t),f'_2(v,u,y_2)}
\Big\{(R_1,R_2): R_1 \geq 0, R_2 \geq 0 \nonumber\\
&R_1 \leq I(X_1;Y_2|V,T)+I(X_1;Y_3|X_2,Y_2,T)\label{eqn:outer_str1I}\\
&R_2 \leq \min\{I(V;Y_4|X_1,T),I(U,V;Y_3|X_1,T)\}\label{eqn:outer_str1II}\\
&R_1+R_2 \leq I(X_1,U,V,T;Y_3)\Big\}\label{eqn:outer_str1III}
\end{IEEEeqnarray}
where $x_2=f'_2(v,u,y_2)$.
\end{theorem}
\begin{IEEEproof}
The strong interference condition at (\ref{eqn:cond_str_rec1}) implies that
\begin{IEEEeqnarray}{rcl}
I(X_2;Y_4|X_1,Y'_3)&\leq& I(X_2;Y_3|X_1,Y'_3).\label{eqn:cond_str_rec1_II}
\end{IEEEeqnarray}

Thus, we can compute the following mutual information term as
\begin{IEEEeqnarray*}{rcl}
I(U,V,T;Y_4|X_1,Y'_3)&\leq&I(U,V,T,X_2;Y_4|X_1,Y'_3)\\
&\stackrel{(a)}{=}&I(X_2;Y_4|X_1,Y'_3)\stackrel{(b)}{\leq} I(X_2;Y_3|X_1,Y'_3)
\end{IEEEeqnarray*}
where (a) follows from the memoryless property of the channel and (b) from (\ref{eqn:cond_str_rec1_II}). Now, by substituting $Y'_3=Y_3$, the region in Theorem~\ref{thm:outer_gen} ($\Rc_{out}$) reduces to $\Rc_{out}^{str_1}$.
\end{IEEEproof}

Next, we apply the strong interference condition at the cognitive receiver to further tighten the outer bound which we use to derive the capacity results in Section~\ref{sec:Cap}. Assume that the following strong interference condition at Rx2 hold
\begin{IEEEeqnarray}{rcl}
I(X_1;Y_3)&\leq& I(X_1;Y_4)\label{eqn:cond_str_rec2}
\end{IEEEeqnarray}

\begin{theorem}\label{thm:outer_str2}
The capacity region of CC-IFC-WD with the joint p.m.f (\ref{eqn:pmf}), satisfying (\ref{eqn:cond_str_rec1}) and (\ref{eqn:cond_str_rec2}), is contained in the region
\begin{IEEEeqnarray}{rl}
&\Rc_{out}^{str_2} =\bigcup\limits_{p(u,t)p(x_1|t)p(v|t),f'_2(v,u,y_2)}
\Big\{(R_1,R_2): R_1 \geq 0, R_2 \geq 0 \nonumber\\
&R_1 \leq I(X_1;Y_2|V,T)+I(X_1;Y_3|X_2,Y_2,T)\label{eqn:outer_str2I}\\
&R_2 \leq \min\{I(V;Y_4|X_1,T),I(U,V;Y_3|X_1,T)\}\label{eqn:outer_str2II}\\
&R_1+R_2 \leq\min\{I(X_1,U,V,T;Y_3),I(X_1,U,V,T;Y_4)\}\Big\}\label{eqn:outer_str2III}
\end{IEEEeqnarray}
where $x_2=f'_2(v,u,y_2)$.
\end{theorem}
\begin{IEEEproof}
See Appendix~\ref{app:out_proof}.
\end{IEEEproof}

\section{Inner Bounds on the Capacity Region of Discrete Memoryless CC-IFC-WD}\label{sec:INNER}
In this section, we consider the discrete memoryless CC-IFC-WD introduced in Section \ref{sec:definition} and concentrate on three special cases: 1)~Classical CC-IFC, which corresponds to $L=0$; 2)~CC-IFC without delay ($L=1$), where the current received symbol (at the cognitive user) can also be utilized; and 3)~CC-IFC with unlimited look-ahead, in which the cognitive user knows its entire received sequence non-causally. For all setups,
the inner bounds on the capacity region for the general discrete memoryless case are derived. For the first case, we utilize a coding scheme based on the combination of generalized block Markov superposition coding, rate splitting, and GP binning against part of the interference. In addition to the strategies used in the first case, we apply instantaneous relaying in the second setup due to the knowledge of the current received symbol at the cognitive user. Furthermore, we employ non-causal partial DF instead of generalized block Markov coding in the last case. The outline of the proofs are presented. Also, we compare our proposed schemes with the results in \cite{HanKob81}-\cite{JiangXin08}, \cite{SeyXinLian09}, \cite{Tuni07}-\cite{YangTuni08}, \cite{Hass06} and \cite{LiuMarGoldSha09} and show that some previously known rate regions are included in our achievable rate regions.

\subsection{Classical CC-IFC ($L=0$)}\label{subsec:zerodelay}
We present a new achievable rate region for this setup. In our coding scheme, we employ the following strategies:

\begin{itemize}
\item \textbf{Generalized block Markov coding} (partial DF \cite{CoveElg79}): In order to boost the rate of the primary user, the cognitive user can cooperate in sending the message of the primary user sent in the previous block.

\item \textbf{Rate splitting at both transmitters:} This allows the improvement of both rates through interference cancelation at both receivers as in the HK scheme \cite{HanKob81}. Also, the cognitive user can \emph{partially} decode the primary user's message due to the splitting.

The message of the primary user ($m_1$) is split into four parts, i.e., $m_1=(m_{1cd},m_{1cn},m_{1pd},m_{1pn})$. The \emph{private} parts $(m_{1pd},m_{1pn})$ can be decoded only at the intended receiver (Rx1), while the \emph{common} parts $(m_{1cd},m_{1cn})$ can be decoded at the non-intended receiver (Rx2) as well, allowing interference cancelation at Rx2. Note that, subscript $c$ (or $p$) refers to the common (or private) part of the message. Moreover, as Tx2 attempts to decode the primary user's message via overhearing the channel ($Y_2$), we further consider two parts for partial decoding at the cognitive user (Tx2), where subscript $d$ (or $n$) refers to the part of the primary user's message which can (or cannot) be decoded by the cognitive user. Therefore, $(m_{1cd},m_{1pd})$ can be decoded at Tx2, and we refer to them as \emph{cooperative} messages, while $(m_{1cn},m_{1pn})$ cannot be decoded at Tx2, and we refer to them as \emph{non-cooperative} messages.

The cognitive user splits its message ($m_2$) into two parts, i.e., $m_2=(m_{2c},m_{2p})$, for interference cancelation at Rx1, where $m_{2c}$ and $m_{2p}$ are the common and private messages, as in the HK scheme \cite{HanKob81}.

\item \textbf{GP binning at the cognitive user:} The cognitive user precodes its message against the part of the primary user's message which was sent in the previous block and decoded by the cognitive user. This approach improves the rate of the cognitive user by correlated codebooks (using block Markov coding). Moreover, since the common message should be decoded in both receivers, binning against $m_{1cd}$ provides no improvement. Therefore, Tx2 generates codewords for $m_{2c}$ and $m_{2p}$, superimposing on $m_{1cd}$ in order to support its transmission, and bins its codewords against $m_{1pd}$ to pre-cancel this part of the interference. Previous results generally focus on two binning techniques: in the first technique, two independent binning steps are applied for GP coding, as in \cite{JiangXin08} for non-causal C-IFC, while in the second technique, the second codeword is superimposed on the first binned one prior to the second binning step as in \cite{MariGolKraSha08} for non-causal C-IFC. Instead, we use joint binning, which brings potential improvements.
\end{itemize}

Consider auxiliary RVs, $T_c,T_p,U_{1c},U_{1p},V_{1c},V_{1p},U_{2c},U_{2p}$ and a time-sharing RV $Q$ defined on arbitrary finite sets $\Tc_c,\Tc_p,\Uc_{1c},\Uc_{1p},\Vc_{1c},\Vc_{1p},\Uc_{2c},\Uc_{2p}$ and $\Qc$, respectively. Let $Z_1=(Q,T_c,T_p,U_{1c},U_{1p},$ $V_{1c},V_{1p},U_{2c},U_{2p},X_1,X_2,Y_2,Y_3,Y_4)$, and $\mathcal{P}_1$ denote the set of all joint p.m.fs $p(.)$ on $Z_1$ that can be factored in the form of
\begin{IEEEeqnarray}{l}\label{eqn:pmfzerodelay}
p(z_1)=p(q)p(t_c|q)p(t_p|t_c,q)p(u_{1c}|t_c,q)p(u_{1p}|u_{1c},t_p,t_c,q)\nonumber\\
p(v_{1c}|t_c,q)p(v_{1p}|v_{1c},t_p,t_c,q)p(x_1|v_{1p},v_{1c},u_{1p},u_{1c},t_p,t_c,q)\nonumber\\
p(u_{2c},u_{2p}|t_p,t_c,q)p(x_2|u_{2c},u_{2p},t_p,t_c,q)p(y_2,y_3,y_4|x_1,x_2)\;\:\quad
\end{IEEEeqnarray}
\addtocounter{equation}{21}
Let $\Rc_1(Z_1)$ denote the set of all nonnegative rate pairs $(R_1,R_2)$ where $R_1=R_{1cd}+R_{1cn}+R_{1pd}+R_{1pn}$ and $R_2=R_{2c}+R_{2p}$, such that there exist nonnegative $(L_{2c},L_{2p})$ satisfying \eqref{eqn:zerodelay_Tx2_enc_begin}-\eqref{eqn:zerodelay_Tx2_dec_end}.
\newcounter{tempequationcounter}
\begin{figure*}[!t]
\normalsize
\setcounter{tempequationcounter}{\value{equation}}
\begin{IEEEeqnarray}{rcl}
\setcounter{equation}{19}
L_{2c}&\geq&I(U_{2c};T_p|T_cQ)\doteq I_1\label{eqn:zerodelay_Tx2_enc_begin}\\
L_{2p}&\geq&I(U_{2p};T_p|T_cQ)\doteq I_2\label{eqn:zerodelay_I2}\\
L_{2c}+L_{2p}&\geq&I\!(U_{2c};U_{2p}|T_cQ)\!+\!I(U_{2c}U_{2p};T_p|T_cQ)\doteq I_3\label{eqn:zerodelay_Tx2_enc_end}\\
R_{1pn}&\leq& I(V_{1p};Y_3|U_{2c}V_{1c}U_{1p}U_{1c}T_pT_cQ)\doteq
I_4\label{eqn:zerodelay_Rx1_begin}\\
R_{1cd}+R_{1cn}+R_{1pd}+R_{1pn}+L_{2c}+R_{2c}&\leq&
I_1+I(U_{2c}V_{1p}V_{1c}U_{1p}U_{1c}T_pT_c;Y_3|Q)\doteq
I_5\label{eqn:zerodelay_I5}\\
R_{1cn}+R_{1pd}+R_{1pn}&\leq&
I(V_{1p}V_{1c}U_{1p}T_p;Y_3U_{2c}|U_{1c}T_cQ)\doteq
I_6\label{eqn:zerodelay_I6}\\
R_{1pd}+R_{1pn}&\leq&
I(V_{1p}U_{1p}T_p;Y_3U_{2c}|V_{1c}U_{1c}T_cQ)\doteq I_7\label{eqn:zerodelay_I7}\\
R_{1pd}+R_{1pn}+L_{2c}+R_{2c}&\leq& I_1+I(U_{2c}V_{1p}U_{1p}T_p;Y_3|V_{1c}U_{1c}T_cQ)\doteq I_8\label{eqn:zerodelay_I8}\\
R_{1cn}+R_{1pn}&\leq&I(V_{1c}V_{1p};Y_3|U_{2c}U_{1p}U_{1c}T_pT_cQ)\doteq
I_9\label{eqn:zerodelay_I9}\\
R_{1pn}+L_{2c}+R_{2c}&\leq&I_1+I(V_{1p}U_{2c};Y_3|V_{1c}U_{1p}U_{1c}T_pT_cQ)\doteq
I_{10}\label{eqn:zerodelay_I10}\\
R_{1cn}+R_{1pd}+R_{1pn}+L_{2c}+R_{2c}&\leq&I_1+I(U_{2c}V_{1p}V_{1c}U_{1p}T_p;Y_3|U_{1c}T_cQ)\doteq
I_{11}\label{eqn:zerodelay_I11}\\
R_{1cn}+R_{1pn}+L_{2c}+R_{2c}&\leq&I_1+I(U_{2c}V_{1p}V_{1c};Y_3|U_{1p}U_{1c}T_pT_cQ)\doteq
I_{12}\label{eqn:zerodelay_Rx1_end}\\
L_{2c}+R_{2c}&\leq&I(U_{2c};Y_4U_{2p}|V_{1c}U_{1c}T_cQ)\doteq I_{13}\label{eqn:zerodelay_Rx2_begin}\\
L_{2p}+R_{2p}&\leq&I(U_{2p};Y_4U_{2c}|V_{1c}U_{1c}T_cQ)\doteq I_{14}\label{eqn:zerodelay_I14}\\
R_{1cd}+R_{1cn}+L_{2c}+R_{2c}+L_{2p}+R_{2p}&\leq& I(U_{2c}U_{2p}V_{1c}U_{1c}T_c;Y_4|Q)\doteq I_{15}\label{eqn:zerodelay_I15}\\
R_{1cn}+L_{2c}+R_{2c}&\leq&I(U_{2c}V_{1c};Y_4U_{2p}|U_{1c}T_cQ)\doteq
I_{16}\label{eqn:zerodelay_I16}\\
R_{1cn}+L_{2p}+R_{2p}&\leq&I(U_{2p}V_{1c};Y_4U_{2c}|U_{1c}T_cQ)\doteq
I_{17}\label{eqn:zerodelay_I17}\\
R_{1cn}+L_{2c}+R_{2c}+L_{2p}+R_{2p}&\leq& I(U_{2c}U_{2p}V_{1c};Y_4|U_{1c}T_cQ)\doteq I_{18}\label{eqn:zerodelay_I18}\\
L_{2c}+R_{2c}+L_{2p}+R_{2p}&\leq&I(U_{2c}U_{2p};Y_4|V_{1c}U_{1c}T_cQ)\doteq
I_{19}\label{eqn:zerodelay_Rx2_end}\\
R_{1pd}&\leq& I(U_{1p};Y_2|U_{2c}U_{2p}U_{1c}T_pT_cQ)\doteq I_{20}\label{eqn:zerodelay_Tx2_dec_begin}\\
R_{1cd}+R_{1pd}&\leq&I(U_{1c}U_{1p};Y_2|U_{2c}U_{2p}T_pT_cQ)\doteq
I_{21}\label{eqn:zerodelay_Tx2_dec_end}
\end{IEEEeqnarray}
\setcounter{equation}{\value{tempequationcounter}}
\hrulefill
\vspace*{4pt}
\end{figure*}

\begin{theorem}\label{thm:zerodelay}
For any $p(.)\in\mathcal{P}_1$, the region $\Rc_1(Z_1)$ is an achievable rate region for the discrete memoryless classical CC-IFC (CC-IFC-WD with $L=0$), i.e., $\bigcup_{Z_1\in\mathcal{P}_1}{\!\!\Rc_1(Z_1)}\subseteq\Cc_0$.
\end{theorem}

\begin{IEEEproof}[Outline of the Proof]
We propose the following random coding scheme, which contains regular generalized block Markov superposition coding, rate splitting and GP coding in the encoding part. For decoding at the receivers, we utilize backward decoding. As mentioned earlier, messages of the primary and cognitive users are split into four and two parts, respectively, i.e.: $m_1=(m_{1cd},m_{1cn},m_{1pd},m_{1pn})$ and $m_2=(m_{2c},m_{2p})$. Tx1 uses generalized block Markov superposition coding technique and creates $t_c^n,t_p^n$ codewords for cooperative messages of the previous block ($m_{1cd,b-1},m_{1pd,b-1}$), $u_{1c}^n,u_{1p}^n$ for cooperative messages of the current block ($m_{1cd,b},m_{1pd,b}$), and $v_{1c}^n,v_{1p}^n$ for non-cooperative messages of the current block ($m_{1cn,b},m_{1pn,b}$), where $c$ in the subscript refers to a codeword related to the common part of the message (to be decoded at both receivers) and $p$ refers to a codeword related to the private part of the message (to be decoded at the intended receiver only). At Tx1, all codewords related to the private messages are superimposed on the codewords related to the common messages. Note that, the cognitive user can decode $u_{1p}^n,u_{1c}^n$ using $t_c^n,t_p^n$, where $v_{1c}^n,u_{1c}^n$ are decoded at Rx2 and all of the above codewords are decoded at Rx1. Additionally, Tx2 encodes its split message with two codewords: joint binning against $t_p^n$ conditioned on $t_c^n$ is used to create $u_{2c}^n,u_{2p}^n$ for $m_{2c},m_{2p}$, respectively. We remark that, in order to establish a cooperative strategy, all codewords are correlated due to block Markov scheme. The encoding scheme and relation between RVs are graphically shown in Fig.~\ref{fig:RV}. Now, consider a block Markov encoding scheme with $B$ blocks of transmission, each of $n$ symbols.
\begin{figure*}[tb]
  \centering
  \includegraphics[width=17cm]{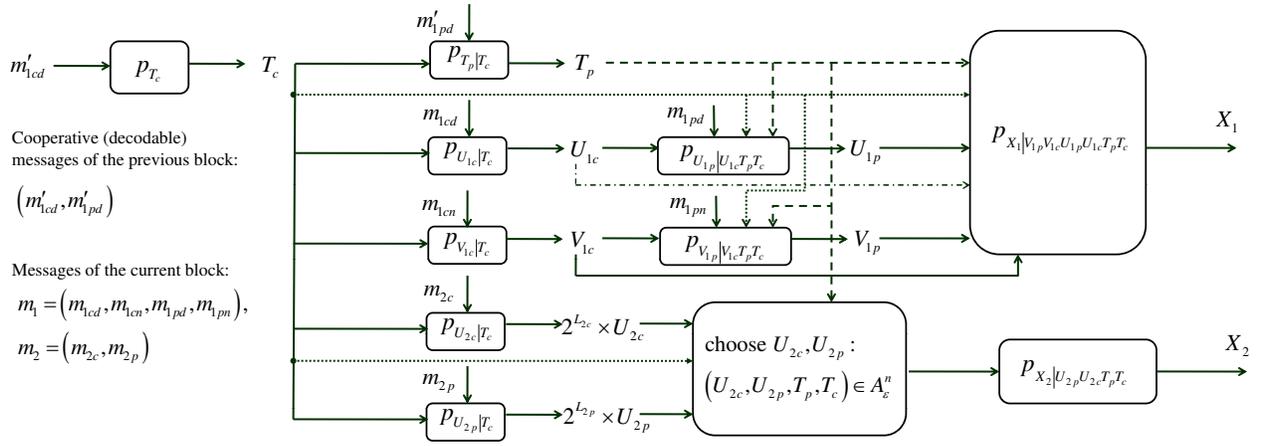}
  \caption{The encoding scheme for Theorem~\ref{thm:zerodelay}}
  \label{fig:RV}
\end{figure*}

\textit{\textbf{Codebook Generation}:} Let $q^n$ be a random sequence of $\Qc^n$ according to the probability $\prod\limits_{i=1}^np(q_{i})$ and fix a joint p.m.f as (\ref{eqn:pmfzerodelay}).

\emph{Primary User:}
\begin{enumerate}
\item Generate $2^{nR_{1cd}}$ independent and identically distributed (i.i.d) $t_c^n$ sequences, each with probability $\prod\limits_{i=1}^np(t_{c,i}|q_{i})$. Index them as $t_c^n(m'_{1cd})$ where $m'_{1cd}\in$ $[1,2^{nR_{1cd}}]$.

\item For each $t_c^n(m'_{1cd})$, generate $2^{nR_{1pd}}$ i.i.d $t_p^n$ sequences, according to $\prod\limits_{i=1}^np(t_{p,i}|t_{c,i},q_{i})$. Index them as $t_p^n(m'_{1pd},$ $m'_{1cd})$ where $m'_{1pd}\in[1,2^{nR_{1pd}}]$.

\item For each $t_c^n(m'_{1cd})$, generate $2^{nR_{1cd}}$ i.i.d $u_{1c}^n$ sequences, according to $\prod\limits_{i=1}^np(u_{1c,i}|t_{c,i},q_{i})$. Index them as $u_{1c}^n(m_{1cd},m'_{1cd})$ where $m_{1cd}\in[1,2^{nR_{1cd}}]$.

\item For each $t_c^n(m'_{1cd})$, generate $2^{nR_{1cn}}$ i.i.d $v_{1c}^n$ sequences, each with probability $\prod\limits_{i=1}^np(v_{1c,i}|t_{c,i},q_{i})$. Index them as $v_{1c}^n(m_{1cn},m'_{1cd})$ where $m_{1cn}\in[1,2^{nR_{1cn}}]$.

\item For each $(u_{1c}^n(m_{1cd},m'_{1cd}),t_p^n(m'_{1pd},m'_{1cd}),t_c^n(m'_{1cd}))$, generate $2^{nR_{1pd}}$ i.i.d $u_{1p}^n$ sequences, according to $\prod\limits_{i=1}^np(u_{1p,i}|u_{1c,i},t_{p,i},t_{c,i},q_{i})$. Index them as $u_{1p}^n(m_{1pd},$ $m_{1cd},m'_{1pd},m'_{1cd})$ where $m_{1pd}\in[1,2^{nR_{1pd}}]$.

\item For each $(v_{1c}^n(m_{1cn},m'_{1cd}),t_p^n(m'_{1pd},m'_{1cd}),t_c^n(m'_{1cd}))$, generate $2^{nR_{1pn}}$ i.i.d $v_{1p}^n$ sequences, according to $\prod\limits_{i=1}^np(v_{1p,i}|v_{1c,i},t_{p,i},t_{c,i},q_{i})$. Index them as $v_{1p}^n(m_{1pn},m_{1cn},m'_{1pd},m'_{1cd})$ where $m_{1pn}\in[1,2^{nR_{1pn}}]$.
\end{enumerate}

\emph{Cognitive User:} From the p.m.f in (\ref{eqn:pmfzerodelay}), compute the marginals $p(u_{2c}|t_c,q)$ and $p(u_{2p}|t_c,q)$ (drop the dependence on $t_p$).
\begin{enumerate}
\item  For each $t_c^n(m'_{1cd})$, generate $2^{n(R_{2c}+L_{2c})}$ i.i.d $u_{2c}^n$ sequences, each with probability $\prod\limits_{i=1}^np(u_{2c,i}|t_{c,i},q_{i})$. Index them as $u_{2c}^n([m_{2c},l_{2c}],m'_{1cd})$, where $m_{2c}\in[1,2^{nR_{2c}}]$ and $l_{2c}\in[1,2^{nL_{2c}}]$.
\item  For each $t_c^n(m'_{1cd})$, generate $2^{n(R_{2p}+L_{2p})}$ i.i.d $u_{2p}^n$ sequences, according to $\prod\limits_{i=1}^np(u_{2p,i}|t_{c,i},q_{i})$. Index them as $u_{2p}^n([m_{2p},l_{2p}],$ $m'_{1cd})$, where $m_{2p}\in[1,2^{nR_{2p}}]$ and $l_{2p}\in[1,2^{nL_{2p}}]$.
\end{enumerate}

\textit{\textbf{Encoding} (at the beginning of block $b$):}

\emph{Primary User (Transmitter 1):}

In order to transmit the message $m_{1,b}=(m_{1cd,b},m_{1cn,b},m_{1pd,b},m_{1pn,b})$, encoder~1 picks codewords $v_{1p}^n(m_{1pn,b},m_{1cn,b},m_{1pd,b-1},m_{1cd,b-1})$, $v_{1c}^n(m_{1cn,b},m_{1cd,b-1})$ , $u_{1p}^n(m_{1pd,b},m_{1cd,b},m_{1pd,b-1},m_{1cd,b-1})$, $u_{1c}^n(m_{1cd,b},m_{1cd,b-1})$, $t_p^n(m_{1pd,b-1},m_{1cd,b-1})$, and $t_c^n(m_{1cd,b-1})$. Then, it sends $x_1^n$ generated according to
$\prod\limits_{i=1}^np(x_{1,i}|v_{1p,i},v_{1c,i},u_{1p,i},u_{1c,i},$ $t_{p,i},t_{c,i},q_{i})$. We assume that in the first block, cooperative information is $(m_{1pd,b-1},m_{1cd,b-1})=(m_{1pd,0},$ $m_{1cd,0})=(0,0)$, and in the last block, a previously known message $(m_{1pd,b},m_{1cd,b})=(m_{1pd,B},m_{1cd,B})=(1,1)$ is transmitted.

\begin{table*}
\renewcommand{\arraystretch}{1.3}
\caption{Corresponding Codewords for Theorem~\ref{thm:zerodelay}}
\label{tbl:codeword} \centering
\begin{tabular}{|c|c|c|c|}
\hline
 & block 1  & block $b$, $b=2,\ldots,B-1$  &  block $B$\\
\hline \hline
& $t_c^n(0)$  & $t_c^n(m_{1cd,b-1})$  & $t_c^n(m_{1cd,B-1})$\\
\cline{2-4}
& $t_p^n(0,0)$  & $t_p^n(m_{1pd,b-1},m_{1cd,b-1})$  & $t_p^n(m_{1pd,B-1},m_{1cd,B-1})$\\
\cline{2-4}
Tx1& $u_{1c}^n(m_{1cd,1},0)$  & $u_{1c}^n(m_{1cd,b},m_{1cd,b-1})$  & $u_{1c}^n(1,m_{1cd,B-1})$\\
\cline{2-4}
& $u_{1p}^n(m_{1pd,1},m_{1cd,1},0,0)$ & $u_{1p}^n(m_{1pd,b},m_{1cd,b},m_{1pd,b-1},m_{1cd,b-1})$ & $u_{1p}^n(1,1,m_{1pd,B-1},m_{1cd,B-1})$\\
\cline{2-4}
& $v_{1c}^n(m_{1cn,1},0)$ & $v_{1c}^n(m_{1cn,b},m_{1cd,b-1})$ & $v_{1c}^n(m_{1cn,B},m_{1cd,B-1})$\\
\cline{2-4}
& $v_{1p}^n(m_{1pn,1},m_{1cn,1},0,0)$ & $v_{1p}^n(m_{1pn,b},m_{1cn,b},m_{1pd,b-1},m_{1cd,b-1})$ & $v_{1p}^n(m_{1pn,B},m_{1cn,B},m_{1pd,B-1},m_{1cd,B-1})$\\
\hline
Tx2& $u_{2c}^n([m_{2c,1},l_{2c,1}],0)$ & $u_{2c}^n([m_{2c,b},l_{2c,b}],m_{1cd,b-1})$ & $u_{2c}^n([m_{2c,B},l_{2c,B}],m_{1cd,B-1})$\\
\cline{2-4}
& $u_{2p}^n([m_{2p,1},l_{2p,1}],0)$ & $u_{2p}^n([m_{2p,b},l_{2p,b}],m_{1cd,b-1})$ & $u_{2p}^n([m_{2p,B},l_{2p,B}],m_{1cd,B-1})$\\
\hline
\end{tabular}
\end{table*}
\emph{Cognitive User (Transmitter 2):}

Tx2 at the beginning of block $b$ knows $\tilde{m}_{1cd,b-1}$ and $\tilde{m}_{1pd,b-1}$, which are estimates of the parts of the common and private messages sent by Tx1 in the previous block and can be decoded by the cognitive user. In order to send $m_{2,b}=(m_{2c,b},m_{2p,b})$, encoder~2, knowing the codewords $t_p^n(\tilde{m}_{1pd,b-1},\tilde{m}_{1cd,b-1})$ and $t_c^n(\tilde{m}_{1cd,b-1})$, seeks an index pair $(l_{2c,b},l_{2p,b})$ such that
\begin{IEEEeqnarray}{c}
(u_{2c}^n([m_{2c,b},l_{2c,b}],\tilde{m}_{1cd,b-1}),u_{2p}^n([m_{2p,b},l_{2p,b}],\tilde{m}_{1cd,b-1}),\nonumber\\
t_p^n(\tilde{m}_{1pd,b-1},\tilde{m}_{1cd,b-1}),t_c^n(\tilde{m}_{1cd,b-1}),q^n)\nonumber\\
\in A_\epsilon^n\left(U_{2c},U_{2p},T_p,T_c,Q\right)
\end{IEEEeqnarray}

If there is more than one such index pair, Tx2 picks the smallest. If there are no such codewords, it declares an error. Using mutual covering lemma \cite{ElgKim10}, it can be shown that there exist such indices $l_{2c,b}$ and $l_{2p,b}$ with a sufficiently high probability, if $n$ is sufficiently large and (\ref{eqn:zerodelay_Tx2_enc_begin})-(\ref{eqn:zerodelay_Tx2_enc_end}) hold. Then, Tx2 sends $x_2^n$ generated according to $\prod\limits_{i=1}^np(x_{2,i}|u_{2p,i},u_{2c,i},t_{p,i},t_{c,i},q_{i})$. The codewords at Tx1 and Tx2 used in transmission are listed in Table~\ref{tbl:codeword}.

\textit{\textbf{Decoding}:}

\emph{Cognitive User (Transmitter 2):} Tx2 at the end of block $b$ wants to correctly recover $m_{1pd,b}$ and $m_{1cd,b}$. Hence, it looks for a unique pair $(\tilde{m}_{1pd,b},\tilde{m}_{1cd,b})$, such that
\begin{IEEEeqnarray}{c}
(y_2^n(b),u_{1p}^n(\tilde{m}_{1pd,b},\tilde{m}_{1cd,b},m_{1pd,b-1},m_{1cd,b-1}),\nonumber\\
u_{1c}^n(\tilde{m}_{1cd,b},m_{1cd,b-1}),u_{2c}^n([m_{2c,b},l_{2c,b}],m_{1cd,b-1}),\nonumber\\
u_{2p}^n([m_{2p,b},l_{2p,b}],m_{1cd,b-1}),t_p^n(m_{1pd,b-1},m_{1cd,b-1}),\nonumber\\
t_c^n(m_{1cd,b-1}),q^n)\in A_\epsilon^n\left(Y_2,U_{2c},U_{2p},U_{1p},U_{1c},T_p,T_c,Q\right)\IEEEeqnarraynumspace
\end{IEEEeqnarray}

This step can be accomplished with small enough probability of error, i.e., $(\tilde{m}_{1pd,b},\tilde{m}_{1cd,b})=(m_{1pd,b},m_{1cd,b})$, for sufficiently large $n$ if (\ref{eqn:zerodelay_Tx2_dec_begin})-(\ref{eqn:zerodelay_Tx2_dec_end}) hold.

Backward decoding is used at Rx1 and Rx2, hence they begin to decode after all $B$ blocks are received.

\emph{Receiver 1:} In block $b$, Rx1 looks for a unique quadruple $(m_{1pn,b},m_{1cn,b},m_{1pd,b-1},m_{1cd,b-1})$ and some pair $(m_{2c,b},l_{2c,b})$ such that
\begin{IEEEeqnarray}{c}
(y_3^n(b),u_{2c}^n([m_{2c,b},l_{2c,b}],m_{1cd,b-1}),v_{1c}^n(m_{1cn,b},m_{1cd,b-1}),\nonumber\\
v_{1p}^n(m_{1pn,b},m_{1cn,b},m_{1pd,b-1},m_{1cd,b-1}),\nonumber\\
\!\!\!\!\!\!u_{1p}^n(m_{1pd,b},m_{1cd,b},m_{1pd,b-1},m_{1cd,b-1}),u_{1c}^n(m_{1cd,b},m_{1cd,b-1}),\nonumber\\
t_p^n(m_{1pd,b-1},m_{1cd,b-1}),t_c^n(m_{1cd,b-1}),q^n)\nonumber\\
\in A_\epsilon^n\left(Y_3,U_{2c},V_{1p},V_{1c},U_{1p},U_{1c},T_p,T_c,Q\right)
\end{IEEEeqnarray}
where $(m_{1pd,b},m_{1cd,b})$ were decoded in the previous step of backward decoding (i.e., block $b+1$). Here, for large enough $n$, the probability of error can be made sufficiently small if (\ref{eqn:zerodelay_Rx1_begin})-(\ref{eqn:zerodelay_Rx1_end}) hold.

\emph{Receiver 2:} In block $b$, Rx2 finds a unique triple $(m_{2c,b},m_{2p,b},m_{1cd,b-1})$ and some triple $(l_{2c,b},l_{2p,b},m_{1cn,b})$ such that
\begin{IEEEeqnarray}{c}
(u_{2c}^n([m_{2c,b},l_{2c,b}],m_{1cd,b-1}),u_{2p}^n([m_{2p,b},l_{2p,b}],m_{1cd,b-1}),\nonumber\\
v_{1c}^n(m_{1cn,b},m_{1cd,b-1}),u_{1c}^n(m_{1cd,b},m_{1cd,b-1}),t_c^n(m_{1cd,b-1}),\nonumber\\
q^n,y_4^n(b))\in A_\epsilon^n\left(Y_4,U_{2c},U_{2p},V_{1c},U_{1c},T_c,Q\right)
\end{IEEEeqnarray}
where $m_{1cd,b}$ was decoded in the previous step of backward decoding (i.e., block $b+1$). Note that, since $m_{1cd}$ plays a fundamental role in the backward decoding, it is necessary for Rx2 to correctly decode $m_{1cd,b-1}$. However, this causes no additional constraint on the rate region. With an arbitrarily high probability, no error occurs in the second receiver if $n$ is sufficiently large and (\ref{eqn:zerodelay_Rx2_begin})-(\ref{eqn:zerodelay_Rx2_end}) hold. In Appendix~\ref{app:error}, we provide the complete error analysis.
\end{IEEEproof}

\begin{remark}[Comparison with existing results]\label{remark:compare_zerodelay}
Now, we compare the scheme of Theorem~\ref{thm:zerodelay} with the known results for CC-IFC and special cases of this channel and show that Theorem~\ref{thm:zerodelay} includes the rate regions of the following schemes:
\end{remark}
\subsubsection{The HK region \cite{HanKob81}}
Consider the case where the cognitive user cannot overhear the channel, i.e., $Y_2=\emptyset$. If we set $T_c=T_p=U_{1c}=U_{1p}=\emptyset$ and
$L_{2c}=L_{2p}=R_{1cd}=R_{1pd}=0$, rename $V_{1p}=X_1$, and define $X_2$ as a deterministic function of $U_{2c}$ and $U_{2p}$, then the derived rate region reduces to the HK region.

\subsubsection{The relay channel}
If we omit Rx2, i.e., $Y_4=\emptyset$, and the cognitive user has no message to transmit, i.e., $R_2=0$, then the model reduces to the relay channel. By setting $T_c=T_p=U_{1p}=V_{1p}=U_{2p}=\emptyset$ and $L_{2c}=L_{2p}=R_{1pn}=R_{1pd}=R_2=0$, and re-defining $U_{2c}=X_2$, the rate region reduces to the partial DF rate for the relay channel \cite{CoveElg79}, which includes the capacity regions of the degraded \cite{CoveElg79} and semi-deterministic relay channels \cite{Aref1982}. Note that (\ref{eqn:zerodelay_I8}), (\ref{eqn:zerodelay_I10}), and (\ref{eqn:zerodelay_I17})-(\ref{eqn:zerodelay_Rx2_end}) can be dropped, because these bounds correspond to the decoding of the common message from the non-intended transmitter. Hence, these events cause no error unless another intended message is incorrectly decoded.

\subsubsection{The region in \cite{SeyXinLian09} for CC-IFC ($\Rc_{SJXW}$)}
Scheme of \cite{SeyXinLian09} to achieve $\Rc_{SJXW}$ differs from our scheme to achieve $\Rc_1$ in the followings:
\begin{itemize}
\item The message of the primary user in $\Rc_{SJXW}$ is fully decoded by the cognitive user; therefore, $m_1$ is split into two parts. While in $\Rc_1$, we use partial DF and split $m_1$ into four parts, in which we can achieve the scheme of $\Rc_{SJXW}$ by nullifying extra parts. By introducing two extra parts that are sent directly to the receivers, we aim to achieve a reasonable rate region (no less than IFC) even when the condition of the cognitive link is poor.

\item In $\Rc_{SJXW}$, the codewords conveying the private and common messages are generated independently. However, we use superposition encoding on the codewords related to the private messages by using codewords related to the common messages as cloud centers. Thus, we derive a potentially larger achievable rate region with a simpler description.

\item The codewords of Tx2 in $\Rc_{SJXW}$ are generated independently and binnned against all codewords of Tx1. However, in $\Rc_1$ we generate the codewords of Tx2 ($U_{2c},U_{2p}$) by superimposing them on the common cooperative codeword of Tx1 ($U_{1c}$) and then binning them against the private cooperative codeword of Tx1 ($U_{1p}$) conditioned on $U_{1c}$. Thus, $\Rc_1$ can be reduced to $\Rc_{SJXW}$ if $U_{2c},U_{2p}$ and $U_{1c}$ are generated independently. Note that, since common message should be decoded by both receivers, binning against the common message provides no improvement. A similar result has been concluded in \cite{LiuMarGoldSha09} for the cognitive Z-IFC.
\end{itemize}

By setting $V_{1c}=V_{1p}=\emptyset$ and $R_{1cn}=R_{1pn}=0$, $\Rc_1$ reduces to $\Rc'_1\subseteq\Rc_1$. Note that, in this scenario (\ref{eqn:zerodelay_I10}) and (\ref{eqn:zerodelay_Rx1_end}) can be dropped, since they correspond to the incorrect decoding of the common message from the non-intended transmitter. Now, in the scheme of $\Rc_{SJXW}$, generate $U_1$ and $X_{11}$ conditioned on $U_2$ and $X_{12},U_1$, respectively. Then, bounds (3), (9) and (12) in \cite{SeyXinLian09} can be dropped and $\Rc_{SJXW}$ is enlarged to a region $\Rc'_{SJXW}$ as a result of removing these rate constraints ($\Rc_{SJXW}\subseteq\Rc'_{SJXW}$). By redefining $T_{p}=U_{1}$, $U_{1p}=X_{11}$, $T_{c}=U_{2}$, $U_{1c}=X_{12}$, $U_{2c}=V_{1}$, and $U_{2p}=V_{2}$ in $\Rc'_1$, one gets $\Rc'_{SJXW}\subseteq\Rc'_1$. Therefore, $\Rc_{SJXW}\subseteq\Rc_1$.

\subsubsection{The region in \cite{CaoChen07} tailored to CC-IFC ($\Rc_{CC}$)}
The region in \cite{CaoChen07} has been derived for IFC-GF and can be reduced to a region for CC-IFC. In order to perform this reduction, assume that Tx1 is the cognitive user and set $\tilde{Y}_1=G_{1}=H_{1}=W_1=\emptyset$ and $R_{13}=0$ in the region of \cite{CaoChen07} to obtain $\Rc_{CC}$. Note that, indices 1 and 2 are switched, due to the positions of the primary and cognitive users being switched in this model. $\Rc_{CC}$ is different from $\Rc_1$ in that
\begin{itemize}
\item In $\Rc_{CC}$, the primary user splits its message into three parts. In fact, the cooperative message is private and is not decoded at the cognitive user's receiver. This means that the cognitive user cannot decode the common message of the primary user.

\item The scheme of \cite{CaoChen07} is based on the irregular encoding/successive decoding technique, while we use the regular encoding/backward decoding \cite{KraGasGup05}. The latter results in fewer RVs and a simpler scheme.

\item The binning in $\Rc_{CC}$ is done sequentially, in contrast to the joint binning technique employed in $\Rc_1$, which brings potential improvement.
\end{itemize}

By setting $U_{1c}=T_{c}=\emptyset$ and $R_{1cd}=0$, and redefining $U_{2c}=N_{1}$, $U_{2p}=M_{1}$, $T_{p}=S_{2}$, $U_{1p}=W_{2}$, $V_{1c}=U_{2}$, and $V_{1p}=V_{2}$, $\Rc_1$ reduces to a region which includes $\Rc_{CC}$ as a subset.

\subsubsection{The region in \cite{Tuni07} tailored to CC-IFC ($\Rc_{T}$)}
Similar to the previous case, we reduce the region in \cite{Tuni07} to CC-IFC, which has been originally derived for IFC-GF. We assume that Tx2 is the cognitive user and set $V_{2}=V_0$ and $R_{2c}=0$ in the region of \cite{Tuni07} to obtain the reduced region $\Rc_{T}$ for CC-IFC. $\Rc_1$ includes $\Rc_{T}$ as a special case, because
\begin{itemize}
\item The message of the primary user is split into three parts in $\Rc_{T}$, i.e., the cognitive user only decodes a part of the common message and there is no cooperative private message.

\item There is no binning in the $\Rc_{T}$ scheme and the cognitive user acts simply as a relay for the primary user's message.
\end{itemize}

Applying the assignments, $R_{1pd}=L_{2c}=L_{2p}=0$, $T_p=U_{1p}=\emptyset$, $T_c=V_0$, $U_{1c}=V_1$, $V_{1c}=U_1$, $V_{1p}=X_1$, $U_{2c}=U_2$ and $U_{2p}=X_2$, $\Rc_1$ reduces to $\Rc_{T}$.

\subsubsection{The region in \cite{YangTuni08} tailored to CC-IFC ($\Rc_{YT}$)}
Considering Tx2 as the cognitive user, we reduce the region in \cite{YangTuni08} to CC-IFC by setting $V_{2}=S_{2}=Z_{2}=\emptyset$ and $R_{20c}=R_{22c}=R'_{22c}=0$ in the region of \cite{YangTuni08} to obtain the reduced region $\Rc_{YT}$ for CC-IFC. Moreover, by nullifying $S_{2}$, one can set $R'_{11c}=0$ since the first binning step in \cite{YangTuni08} can be omitted. The scheme of $\Rc_{YT}$ is different from $\Rc_{1}$ in the following aspects:
\begin{itemize}
\item In $\Rc_{YT}$, binning is done sequentially and conditionally, while $\Rc_{1}$ utilizes joint binning technique. Therefore, our scheme achieves a potentially larger rate region compared to $\Rc_{YT}$.

\item We use joint backward decoding at the receivers, while two-step decoding is used for $\Rc_{YT}$. Joint decoding cannot have worse performance than the sequential ones.
\end{itemize}

By setting $T_c=Q$, $T_p=S_1$, $U_{1c}=V_1$, $U_{1p}=Z_1$, $V_{1c}=U_1$, $V_{1p}=T_1$, $U_{2c}=U_2$ and $U_{2p}=T_2$, $\Rc_1$ reduces to a region which includes $\Rc_{YT}$ as a result of the above differences.

Now, in order to understand the shape of the achievable rate region, we give a compact expression for $\Rc_1(Z_1)$ which is easier to compute.

\begin{corollary}\label{cor:zerodelay}
The region $\Rc_1(Z_1)$, after Fourier-Motzkin elimination \cite{ElgKim10}, can be expressed as
\begin{IEEEeqnarray*}{rl}
R_1\leq \min\Big(&\min(I_{21}+I_4+I_{16},I_5)-I_1,\\
&I_{21}+\min(I_4+I_{17}-I_2,I_9)\Big)\\
R_2\leq \min\Big(&I_{19},I_{14}+\min(I_{10},I_{13})\Big)-I'_1\\
R_1+R_2\leq \min\Big(&I_{14}+I_5,I_{15}+\min(I_{7},I_{8}-I_{1}),\\
&I_{21}+I_{17}+\min(I_{10},I_{4}+I_{13}),\\
&I_{4}+\min(I_{21}+I_{18},I_{20}+I_{15}),\\
I_{21}+I_{14}+&\min(I_{12},I_{4}+I_{16},I_{10}+I_{17}-I_{2})\Big)-I'_1\\
2R_1+R_2\leq\min\Big(&I_{4}+I_{15}+\min(I_{6},I_{11}-I_1),\\
&I_{21}+2I_{4}+I_{17}+I_{16},\\
&I_{4}+I_{17}+\min(I_{21}+I_{12},I_{5})\Big)+I_{21}-I'_1\\
R_1+2R_2\leq \min\Big(&I_{21}+I_{10}+I_{14}+\min(I_{14}+I_{16},I_{18}),\\
&I_{14}+I_{15}+\min(I_{20}+I_{10},I_{8})\Big)-2I'_1\\
2R_1+2R_2\leq \min\Big(&I_4+\min(I_{14}+I_{11},I_{17}+I_{8}),\\
I_{10}+I_{14}+&\min(I_{6},I_{11}-I_1)\Big)+I_{21}+I_{15}-2I'_1\\
2R_1+3R_2\leq \,I_{21}\,+\,&I_{10}+2I_{14}+I_{11}+I_{15}-3I'_1\\
3R_1+2R_2\,\leq \,2\,I_{21}\,&+2I_{4}+I_{11}+I_{17}+I_{15}-2I'_1\\
\textrm{subject to}\:\:\:&I_1\leq I_{16} \qquad\textrm{ and
}\qquad I_2\leq I_{17},
\end{IEEEeqnarray*}
where $\{I_i,i=1,\ldots,21\}$ are defined in (\ref{eqn:zerodelay_Tx2_enc_begin})-(\ref{eqn:zerodelay_Tx2_dec_end}), and $I'_1\doteq max(I_1+I_2,I_3)$.
\end{corollary}

\subsection{CC-IFC without delay ($L=1$)}\label{subsec:onedelay}
In this case, the cognitive user can utilize the current received symbol as well as the past ones in order to cooperate with the primary user or reduce the interference effect. Note that the derived inner bound in Theorem \ref{thm:zerodelay} is an inner bound on the capacity region for the CC-IFC without delay. However, knowledge of the present received symbol may lead to the expectation of achieving higher rates using this additional information. Instantaneous relaying is a cooperative scheme which exploits only the current received symbol. In general, the cognitive user may need to utilize both the current and the past received symbols to obtain an optimal coding scheme for the CC-IFC without delay. Hence, we establish an achievable rate region based on a scheme which involves the superposition of the scheme used in Theorem~\ref{thm:zerodelay} with instantaneous relaying. In fact, the cognitive user, knowing the current received symbol, sends a function of the codeword obtained by the scheme of Theorem~\ref{thm:zerodelay} and its current received symbol. This method can improve the rate region by allowing the primary and cognitive users to to cooperate instantaneously.

Consider auxiliary RVs $T_c,T_p,U_{1c},U_{1p},V_{1c},V_{1p},U_{2c},U_{2p},$ $V_2$ and a time-sharing RV $Q$ defined on arbitrary finite sets $\Tc_c,\Tc_p,\Uc_{1c},\Uc_{1p},\Vc_{1c},\Vc_{1p},\Uc_{2c},\Uc_{2p},\Vc_{2}$ and $\Qc$, respectively. Let $Z_2=(Q,T_c,T_p,U_{1c},U_{1p},V_{1c},V_{1p},U_{2c},U_{2p},V_2,X_1,X_2,$ $Y_2,Y_3,Y_4)$, and $\mathcal{P}_2$ be the set of all joint p.m.fs $p(.)$ on $Z_2$ that can be factored in the form of
\begin{IEEEeqnarray}{l}\label{eqn:pmfwithoutdelay}
p(z_2)=p(q)p(t_c|q)p(t_p|t_c,q)p(u_{1c}|t_c,q)p(u_{1p}|u_{1c},t_p,t_c,q)\nonumber\\
p(v_{1c}|t_c,q)p(v_{1p}|v_{1c},t_p,t_c,q)p(x_1|v_{1p},v_{1c},u_{1p},u_{1c},t_p,t_c,q)\nonumber\\
p(u_{2c},u_{2p}|t_p,t_c,q)p(v_2|u_{2c},u_{2p},t_p,t_c,q)p(x_2|v_2,y_2,q).
\end{IEEEeqnarray}
In fact $x_2=f'_2(v_2,y_2,q)$, where $f'_2(.)$ is an arbitrary deterministic function.

Let $\Rc_2(Z_2)$ be the set of all nonnegative rate pairs $(R_1,R_2)$ where $R_1=R_{1cd}+R_{1cn}+R_{1pd}+R_{1pn}$ and $R_2=R_{2c}+R_{2p}$, such that there exist nonnegative $(L_{2c},L_{2p})$ which satisfy (\ref{eqn:zerodelay_Tx2_enc_begin})-(\ref{eqn:zerodelay_Tx2_dec_end}).

\begin{theorem}\label{thm:withoutdelay}
For any $p(.)\in\mathcal{P}_2$, the region $\Rc_2(Z_2)$ is an achievable rate region for the discrete memoryless CC-IFC without delay (CC-IFC-WD with $L=1$), i.e., $\bigcup_{Z_2\in\mathcal{P}_2}{\Rc_2(Z_2)}\subseteq\Cc_1$.
\end{theorem}

\begin{IEEEproof}[Proof]
The achievability proof follows by combining the scheme used in Theorem~\ref{thm:zerodelay} and instantaneous relaying. Encoding and decoding follow the same lines as in Theorem~\ref{thm:zerodelay}. Hence, we only highlight the differences for the sake of brevity. The main difference is that during the codebook generation at the cognitive user (Tx2), $v_2^n$ (instead of $x_2^n$ in Theorem \ref{thm:zerodelay}) is generated according to
$\prod\limits_{i=1}^np(v_{2,i}|u_{2p,i},u_{2c,i},t_{p,i},t_{c,i},q_{i})$, and in the encoding session, Transmitter 2 at time $i$ and upon receiving $y_{2,i}$, sends a deterministic function of $y_{2,i}$ and $v_{2,i}$, i.e., $x_{2,i}=f'_{2,i}(v_{2,i},y_{2,i},q_{i})$ where $f'_2(v_2,y_2,q)$ has been fixed at the beginning of codebook generation.
\end{IEEEproof}

\begin{remark}
This scheme is analogous to \textit{Shannon's strategy} for the state-dependent channel with causal channel state information at the transmitter \cite{Shan58, MirAkhAref09}. Shannon found the capacity of this channel by considering an extended input alphabet. Therefore, if we assume that $\Vc_2$ has an extended alphabet of size $|\Xc_2|^{|\Yc_2|}$, and the code for this channel is constructed over the alphabet of all mappings from
$\Yc_2$ to $\Xc_2$, then this scheme will be similar to Shannon's strategies. However, for continuous alphabets, e.g., Gaussian channels, where $\Yc_2$ is infinite in limit, optimal codes cannot be constructed over extended input alphabets. Hence, we consider linear mapping for the Gaussian CC-IFC without delay in Section~\ref{sec:Gaussian}.
\end{remark}

\begin{remark}
Here, Tx2 sends a deterministic function of $Y_2$. Therefore, a function (not necessarily deterministic) of $X_1$ is transmitted by Tx2, which interferes at Rx2. This scheme can boost $R_2$ as it allows Rx2 to decode the unwanted message and to cancel the interference. Hence, we can refer to this scheme as \emph{instantaneous} interference forwarding according to \cite{DabMarGol08}.
\end{remark}

\begin{remark}
This scheme is feasible for any $L\geq1$. Moreover, $f'_2(.)$ can be extended to $x_{2,i}=f'_{2,i}(v_{2,i},y_{2,i},...,y_{2,i+L-1},q_{i})$.
\end{remark}

\begin{remark} Nullifying $Y_4$, $T_c$, $T_p$, $U_{1p}$, $V_{1p}$ and $U_{2p}$, and setting $R_2=L_{2c}=L_{2p}=R_{1pn}=R_{1pd}=0$ and $U_{2c}=V_2$, the model and the achievable rate region reduce to the model and the achievable rate based on partial DF and instantaneous relaying for the RWD channel \cite[Theorem 2.5]{Hass06}, which achieves all known capacity results for discrete memoryless RWD \cite{ElgaHasMam07, Hass06}. We remark that, as discussed for the relay channel in Remark~\ref{remark:compare_zerodelay}, (\ref{eqn:zerodelay_I8}), (\ref{eqn:zerodelay_I10}), (\ref{eqn:zerodelay_I17})-(\ref{eqn:zerodelay_Rx2_end}) can be dropped in Theorem~\ref{thm:withoutdelay} for this scenario.
\end{remark}

\subsection{CC-IFC with unlimited look-ahead}\label{subsec:ndelay}
Now, we investigate the CC-IFC with unlimited look-ahead, defined in Section~\ref{sec:definition}. This means that the cognitive user non-causally knows its entire received sequence. We derive an achievable rate region using a coding scheme based on combining non-causal partial DF, rate splitting and GP binning against part of the interference.

\begin{itemize}
\item \textbf{Non-causal partial DF:} The cognitive user can contribute to the rate of the primary user by encoding a part of the primary user's message and cooperating with the primary user to transmit the decoded part. This is possible only when the cognitive user non-causally has knowledge of the entire received sequence, as in the \emph{unlimited look-ahead} case.

\item \textbf{Rate splitting:} Similar to the scheme used in Theorem~\ref{thm:zerodelay}, rate splitting is employed at both transmitters and the messages of the primary and cognitive users are split into four and two parts, respectively, i.e.: $m_1=(m_{1cd},m_{1cn},m_{1pd},m_{1pn})$ and $m_2=(m_{2c},m_{2p})$. In fact, common (subscript~$c$) and private (subscript~$p$) parts are used for interference cancelation at the non-intended receivers as in the HK scheme \cite{HanKob81}, and cooperative (subscript~$d$) and non-cooperative (subscript~$n$) parts account for non-causal partial DF strategy. Moreover, Tx2 jointly bins its codewords against the cooperative private part of $m_1$, (i.e., $m_{1pd}$) to pre-cancel this part of the interference at Rx2.

\item \textbf{GP binning at the cognitive user:} The cognitive user can partially decode the primary user's message in a non-causal manner, and its rate is improved by precoding against the (partially) known interference.

\end{itemize}

Consider auxiliary RVs $U_{1c},U_{1p},V_{1c},V_{1p},U_{2c},U_{2p}$ and a time-sharing RV $Q$ defined on arbitrary finite sets $\Uc_{1c},\Uc_{1p},\Vc_{1c},\Vc_{1p},\Uc_{2c},\Uc_{2p}$ and $\Qc$, respectively. Let $Z_3=(Q,U_{1c},U_{1p},V_{1c},V_{1p},U_{2c},U_{2p},X_1,X_2,$
$Y_2,Y_3,Y_4)$, and $\mathcal{P}_3$ denote the set of all joint p.m.fs $p(.)$ on $Z_3$ that can be factored in the form of (\ref{eqn:pmfzerodelay}) with $(t_p,t_c)=(u_{1p},u_{1c})$. Let $\Rc_3(Z_3)$ be the set of all nonnegative rate pairs $(R_1,R_2)$ where $R_1=R_{1cd}+R_{1cn}+R_{1pd}+R_{1pn}$ and $R_2=R_{2c}+R_{2p}$, such that there exist nonnegative $(L_{2c},L_{2p})$ which satisfy (\ref{eqn:zerodelay_Tx2_enc_begin})-(\ref{eqn:zerodelay_Rx2_end}) with $(T_p,T_c)=(U_{1p},U_{1c})$ and
\begin{IEEEeqnarray}{rCl}
R_{1pd}&\leq& I(U_{1p};Y_2|U_{1c}Q)\label{eqn:blockdelay_Tx2_dec_begin}\\
R_{1cd}+R_{1pd}&\leq&I(U_{1c}U_{1p};Y_2|Q).\label{eqn:blockdelay_Tx2_dec_end}
\end{IEEEeqnarray}

\begin{theorem}\label{thm:blockdelay}
For any $p(.)\in\mathcal{P}_3$, the region $\Rc_3(Z_3)$ is an achievable rate region for the discrete memoryless CC-IFC with unlimited look-ahead, i.e., $\bigcup_{Z_3\in\mathcal{P}_3}{\Rc_3(Z_3)}\subseteq\Cc_n$.
\end{theorem}

\begin{remark}
For the unlimited look-ahead case, using a coding scheme based on instantaneous relaying is feasible. However, to compare this strategy with non-causal partial DF, we restrict the use of this scheme to $L=1$. In fact, applying the strategy of Theorem~\ref{thm:withoutdelay} to the CC-IFC with unlimited look-ahead without using non-causal partial DF will achieve $\Rc_2$ and in order to utilize the extra information in this case we must employ a \emph{non-causal} strategy. Moreover, a strategy based on non-causal partial DF and instantaneous relaying achieves a region which encompasses the ones for other values of $L$, wherein eliminating the non-causal partial DF will result in $\Rc_2$ and deleting instantaneous relaying will result in $\Rc_3$. Thus, to compare the non-causal partial DF and instantaneous relaying strategies, we must consider $\Rc_2$ and $\Rc_3$. Therefore, to reduce complexity, we prefer to exclude the instantaneous relaying in the scheme of Theorem~\ref{thm:blockdelay}.
\end{remark}

\begin{IEEEproof}
The proof of Theorem~\ref{thm:blockdelay} is similar to that of Theorem~\ref{thm:zerodelay}, with the exception that there is no dependence on the previous block messages and the transmitters non-causally cooperate using correlated codewords. For this reason, Tx1 uses superposition coding with four codewords: $u_{1c}^n,u_{1p}^n$ for cooperative messages ($m_{1cd},m_{1pd}$), and $v_{1c}^n,v_{1p}^n$ for $m_{1cn},m_{1pn}$, where all codewords related to the private messages are superimposed on the codewords related to the common messages and codewords conveying non-cooperative information are superimposed on the cooperative codewords. Using joint binning against $u_{1p}^n$, Tx2 creates $u_{2c}^n,u_{2p}^n$ for its own messages, while in order to relay $m_{1cd}$, these codewords are generated conditioned on $u_{1c}^n$. Due to the non-causal cooperative scheme, simultaneous joint decoding is used instead of backward decoding at the receivers. Thus, the proof follows the same lines as in Theorem~\ref{thm:zerodelay} and is omitted here for the sake of brevity.
\end{IEEEproof}

\begin{remark}\label{remark:compare_ndelayI}
As mentioned earlier, CC-IFC with unlimited look-ahead generalizes the non-causal C-IFC and can reduce to this channel model when $p(y_2|x_2)$ is ideal, i.e., the cognitive link between the transmitters is noise-free. To obtain an achievable rate region for this case, we use the region $\Rc_3$ of Theorem~\ref{thm:blockdelay} and assume that the cognitive user can fully decode the message of the primary user ($m_1$). Therefore, in $\Rc_3$, we set $V_{1c}=V_{1p}=\emptyset$ and $R_{1cn}=R_{1pn}=0$, drop (\ref{eqn:zerodelay_Tx2_dec_begin}) and (\ref{eqn:zerodelay_Tx2_dec_end}) due to the elimination of the cognitive link, and drop (\ref{eqn:zerodelay_I10}) and (\ref{eqn:zerodelay_Rx1_end}) because they correspond to the incorrect decoding of the common message from the non-intended transmitter, and derive $\Rc_{NC}$ for non-causal C-IFC. Now, we compare $\Rc_{NC}$ with the known results for non-causal C-IFC:
\end{remark}

\subsubsection{The region in \cite{MariGolKraSha08} ($\Rc_{MGKS}$)}
\begin{itemize}
\item In $\Rc_{MGKS}$, the binning is done sequentially and conditionally in two steps, while we utilize the joint binning technique in $\Rc_{NC}$ with potential improvement.

\item In $\Rc_{MGKS}$, the message of the primary user is split into two parts; however, the non-intended receiver decodes none of these parts.
\end{itemize}

Noting that the positions of the primary and cognitive users are switched in $\Rc_{MGKS}$, setting $R_{1cd}=0$ and $U_{1c}=\emptyset$, redefining $U_{1p}=(X_{2a},X_{2b})$, $U_{2c}=U_{1c}$, and $U_{2p}=U_{1a}$, and considering the above discussion, reduces $\Rc_{NC}$ to a region which includes $\Rc_{MGKS}$.

\subsubsection{The region in \cite{JiangXin08} ($\Rc_{JX}$)}
\begin{itemize}
\item There is no rate splitting for the message of the primary user in $\Rc_{JX}$.

\item In $\Rc_{JX}$, the binning is done independently in contrast with our joint binning technique in $\Rc_{NC}$.
\end{itemize}

Thus, if we set $R_{1cd}=0$ and $U_{1c}=\emptyset$, and redefine $U_{1p}=W$, $U_{2c}=U$, and $U_{2p}=V$ in $\Rc_{NC}$, our region is reduced to one which includes $\Rc_{JX}$ as a subset.

\subsubsection{Weak interference in \cite[Proposition~3.1]{WuVishAra07} ($\Rc_{WVA}$)}
By switching the position of the primary and cognitive users in \cite{WuVishAra07}, we assume that the second transmitter is cognitive. Now, set $R_{1cd}=R_{2c}=L_{2c}=0$ and $U_{1c}=U_{2c}=\emptyset$, and redefine $U_{1p}=(X_1,U)$ and $U_{2p}=V$ in $\Rc_{NC}$. Since, there is no common message to be decoded at Rx1, drop (\ref{eqn:zerodelay_I16}). Applying these assignments, $\Rc_{NC}$ reduces to $\Rc_{WVA}$.

\subsubsection{Strong interference in \cite[Theorem~5]{MariYatKra07} ($\Rc_{MYK}$)}
By setting $R_{1pd}=R_{2p}=L_{2c}=L_{2p}=0$ and $U_{1p}=U_{2p}=\emptyset$, redefining $U_{1c}=X_1$ and $U_{2c}=X_2$ and dropping (\ref{eqn:zerodelay_I8}), (\ref{eqn:zerodelay_I11}) and (\ref{eqn:zerodelay_I17}) which are due to the incorrect decoding of the common message at the non-intended receivers, $\Rc_{NC}$ reduces to the capacity region of non-causal C-IFC with strong interference, also referred to as strong interference channel with unidirectional cooperation, derived in \cite[Theorem~5]{MariYatKra07}.

\subsubsection{The region in \cite{LiuMarGoldSha09} ($\Rc_{LMGS}$)}
Noting that the first transmitter is cognitive in \cite{LiuMarGoldSha09}, set $R_{2p}=0$ and $U_{2p}=\emptyset$; redefine $U_{1c}=V$, $U_{2c}=U$, and $U_{1p}=X_2$ in $\Rc_{NC}$; and drop (\ref{eqn:zerodelay_I17}). Then, it can be easily shown that our region reduces to $\Rc_{LMGS}$ which achieves the capacity for a class of the cognitive Z-IFCs.

\subsubsection{The regions in \cite{RiniIZS10} and \cite{RiniIT11} ($\Rc_{RTD}$)}
The region in \cite{RiniIT11} is the largest known achievable rate region for the non-causal C-IFC, which has some differences in the binning technique with the one in \cite{RiniIZS10}. Our scheme does not include these regions. The reason is as follows: In $\Rc_{RTD}$, a part of the primary user's message is sent \emph{only} by the cognitive user based on using Marton coding \cite{Mart79}. In fact, this scheme is possible because the cognitive user knows the primary user's message by a genie. However, in our proposed model, i.e., the CC-IFC with unlimited look-ahead, the cognitive user must \emph{decode} the message of the primary user in a non-causal manner. Therefore, the entire message must be sent by the primary user and our scheme cannot include the method of $\Rc_{RTD}$.

\subsubsection{The region in \cite[Theorem~4.1]{JiaMarGolCui09} for non-causal C-IFC ($\Rc_{JMGC}$)}
The broadcast channel with two cognitive relays is considered in \cite{JiaMarGolCui09}, which is reduced to non-causal C-IFC by removing one of the relays. Our scheme and the one used to achieve $\Rc_{JMGC}$ differ in the binning technique in the cognitive user. In $\Rc_{JMGC}$, Marton coding is used for sending the private parts of the primary and cognitive user's messages. However, we use GP binning for the common and private parts of the cognitive user's message against the private message of the primary user. It appears that no subset relation can be established between $\Rc_{JMGC}$ and $\Rc_{NC}$.

\subsubsection{The regions in \cite{JiaMarGolShaCui10} for non-causal cognitive Z-IFC}
In \cite{JiaMarGolShaCui10}, simple and easily computable rate regions have been derived for non-causal cognitive Z-IFC, which are also achievable for non-causal C-IFC. The region in \cite[Proposition~3.1]{JiaMarGolShaCui10} is based on \cite[Theorem~4.1]{JiaMarGolCui09}, which was discussed
above. By setting $R_{2p}=0$ and $U_{2p}=\emptyset$, $\Rc_{NC}$ includes the regions in \cite[Corollary~3.2]{JiaMarGolShaCui10} and \cite[Proposition~3.2]{JiaMarGolShaCui10}.

\begin{remark}\label{remark:compare_ndelayII}
The region $\Rc_3$ of Theorem~\ref{thm:blockdelay} achieves the capacity region of the partially-cognitive IFC under strong interference conditions characterized in \cite[Theorem~5]{MariGolKraSha07}. Setting $(m_{1cd},m_{1pd})=W_0$, $(m_{1cn},m_{1pn})=W_1$ and $m_2=W_2$ in the scheme of CC-IFC with unlimited look-ahead results in the model of the partially-cognitive IFC, also referred to as IFC with partial unidirectional cooperation. In order to derive the region of \cite[Theorem~5]{MariGolKraSha07}, set $R_{1pd}=R_{1pn}=R_{2p}=L_{2c}=L_{2p}=0$ and
$U_{1p}=V_{1p}=U_{2p}=\emptyset$; rename $R_{1cd}=R_0$, $R_{1cn}=R_1$, $R_{2c}=R_2$, $U_{1c}=U$, $V_{1c}=X_1$, and $U_{2c}=X_2$; and drop (\ref{eqn:zerodelay_I8}), (\ref{eqn:zerodelay_I10}) and (\ref{eqn:zerodelay_I17}) in $\Rc_3$. Note that the events corresponding to these bounds cause no error in this case.
\end{remark}

\section{Capacity Results for Two Special Cases of the Classical CC-IFC}\label{sec:Cap}
In this section, we investigate the classical CC-IFC (CC-IFC-WD with $L=0$) with joint p.m.f $p^*$, given by (\ref{eqn:pmf}) with $L=0$. We find the capacity regions for the classes of degraded and semi-deterministic classical CC-IFC under strong interference conditions, where we use the achievable rate region in Theorem~\ref{thm:zerodelay} for the achievability of these regions and the outer bound of Theorem~\ref{thm:outer_str2} for the converse parts.

\subsection{Degraded classical CC-IFC}
We define degraded classical CC-IFC as a classical CC-IFC (CC-IFC-WD with $L=0$) where the degradedness condition for the Tx1-Rx1 pair with the cognitive user as a relay holds for every $p^*$. More precisely,
\begin{IEEEeqnarray}{rcl}
p(y_3|x_1,x_2,y_2)=p(y_3|x_2,y_2),\label{eqn:cond_deg}
\end{IEEEeqnarray}
i.e., $X_1\rightarrow (X_2,Y_2)\rightarrow Y_3$ forms a Markov chain. Next, we assume that the strong interference conditions (\ref{eqn:cond_str_rec1}) and (\ref{eqn:cond_str_rec2}) at Rx1 and Rx2 hold for every distribution $p^*$, where under these conditions the interfering signals are strong enough to decode both messages at both receivers.

\begin{theorem}\label{thm:cap_str_deg}
The capacity region of the degraded classical CC-IFC with the joint p.m.f $p^*$, satisfying conditions (\ref{eqn:cond_str_rec1}) and
(\ref{eqn:cond_str_rec2}), is given by
\begin{IEEEeqnarray}{rl}
\Cc_0^{d} =&\bigcup_{p(t)p(x_1|t)p(x_2|t)} \Big\{
(R_1,R_2): R_1 \geq 0, R_2 \geq 0 \nonumber\\
&R_1 \leq (X_1;Y_2|X_2,T)\label{eqn:cap_str_deg1}\\
&R_2 \leq I(X_2;Y_4|X_1,T)\label{eqn:cap_str_deg2}\\
&R_1+R_2 \leq\min\{I(X_1,X_2;Y_3),I(X_1,X_2;Y_4)\}\Big\}.\quad\label{eqn:cap_str_deg3}
\end{IEEEeqnarray}
\end{theorem}

\begin{remark}\label{remark:cap_zerodelay_deg}
The message of the cognitive user ($m_2$) can be decoded at Rx1 under condition (\ref{eqn:cond_str_rec1}). Hence, the bound in (\ref{eqn:cap_str_deg1}) and the first bound of (\ref{eqn:cap_str_deg3}) give the capacity region of the degraded relay channel of (\ref{eqn:cond_deg}) \cite{CoveElg79} with a private message $m_2$ from the relay to the receiver. Note that, due to the degradedness condition, the cognitive user is able to decode the message of the primary user ($m_1$). Moreover, $m_1$ can be decoded at Rx2 under condition (\ref{eqn:cond_str_rec2}). Therefore, we have a MAC with common information at Rx2, where $R_1$ is the common rate, $R_2$ is the private rate for the second transmitter, and the private rate for the first transmitter is zero. The bound in (\ref{eqn:cap_str_deg2}) and the second bound of (\ref{eqn:cap_str_deg3}) give the capacity region for this MAC \cite{SleWol73}.
\end{remark}

\begin{IEEEproof}
\underline{Achievability:} For this part, we use the region $\Rc_1$ in Theorem~\ref{thm:zerodelay} (or Corollary~\ref{cor:zerodelay}) and ignore the time-sharing RV $Q$. Let, $T_p=U_{1p}=V_{1p}=U_{2p}=\emptyset$ and $R_{2p}=R_{1pn}=R_{1pd}=0$, which negate the private parts of both messages, making the messages common to both receivers. Furthermore, assume that the cognitive user (Tx2) fully decodes the message of the primary user ($m_1$). Consequently, it is necessary to set $R_{1cn}=0$ and $V_{1c}=\emptyset$. In order to omit the GP coding, we set $L_{2c}=L_{2p}=0$ as well. Note that (\ref{eqn:zerodelay_I8}), (\ref{eqn:zerodelay_I10})-(\ref{eqn:zerodelay_Rx1_end}) and (\ref{eqn:zerodelay_I17}) can be dropped, because these bounds correspond to the decoding of the common message from the non-intended transmitter. Redefining $T_c=T$, $U_{2c}=X_2$, and $U_{1c}=X_1$ completes the proof for the achievability.

\underline{Converse:} To prove the converse part, we evaluate $\Rc_{out}^{str_2}$ of Theorem~\ref{thm:outer_str2} for $L=0$ (classical CC-IFC). Considering (\ref{eqn:RV_U}),
\begin{IEEEeqnarray}{c}
U_{i}=Y_{2,i+1}^{i+L-1}|_{L=0}=\emptyset\label{eqn:RV_U_zeroL}
\end{IEEEeqnarray}

Moreover, in this case Definition~\ref{def:code} provides
\begin{IEEEeqnarray}{c}
X_{2,i}=f_{2,i}(M_2,y_2^{i-1+L})=f_{2,i}(M_2,y_2^{i-1})=f_{2,i}(V_{i})\label{eqn:X2_zeroL}
\end{IEEEeqnarray}

Therefore, the p.m.f in Theorem~\ref{thm:outer_str2} reduces to the one in Theorem~\ref{thm:cap_str_deg}. Based on (\ref{eqn:outer_str2I}),
\begin{IEEEeqnarray*}{rcl}
R_1 &\leq& I(X_1;Y_2|V,T)+I(X_1;Y_3|X_2,Y_2,T)\\
&\stackrel{(a)}{=}&I(X_1;Y_2|V,T,X_2)+I(X_1;Y_3|X_2,Y_2,T)\\
&\stackrel{(b)}{\leq}&H(Y_2|T,X_2)-H(Y_2|X_1,V,T,X_2)+I(X_1;Y_3|X_2,Y_2,T)\\
&\stackrel{(c)}{=}&I(X_1;Y_2|X_2,T)+I(X_1;Y_3|X_2,Y_2,T)\yesnumber\label{eqn:cap_degI}
\end{IEEEeqnarray*}
where (a) is obtained using (\ref{eqn:X2_zeroL}), (b) is due to the fact that conditioning does not increase the entropy, and (c) follows from the joint p.m.f $p^*$, given by (\ref{eqn:pmf}) with $L=0$. Subsequently, applying condition (\ref{eqn:cond_deg}) to (\ref{eqn:cap_degI}) results in (\ref{eqn:cap_str_deg1}).
Similarly, we utilize the first bound in (\ref{eqn:outer_str2II}) to obtain (\ref{eqn:cap_str_deg2}) as follows:
\begin{IEEEeqnarray*}{rcl}
R_2 &\leq& I(V;Y_4|X_1,T)\\
&\stackrel{(a)}{=}&I(V,X_2;Y_4|X_1,T)=H(Y_4|X_1,T)-H(Y_4|X_1,T,V,X_2)\\
&\stackrel{(b)}{=}&I(X_2;Y_4|X_1,T)\yesnumber\label{eqn:cap_degII}
\end{IEEEeqnarray*}
where for (a) we use (\ref{eqn:X2_zeroL}) and (b) is obtained from the joint p.m.f $p^*$. In a similar manner, we derive the first bound in (\ref{eqn:cap_str_deg3}) by using the first bound of (\ref{eqn:outer_str2III}),
\begin{IEEEeqnarray*}{rcl}
R_1+R_2 &\leq& I(X_1,V,T;Y_3)\\
&\stackrel{(a)}{=}&I(X_1,V,T,X_2;Y_3)=H(Y_3)-H(Y_3|X_1,V,T,X_2)\\
&\stackrel{(b)}{=}&I(X_1,X_2;Y_3)\yesnumber\label{eqn:cap_degIII}
\end{IEEEeqnarray*}
where (a) and (b) are obtained with the same reasons as that used in (\ref{eqn:cap_degII}). Finally, similar to (\ref{eqn:cap_degIII}), the second bound in (\ref{eqn:cap_str_deg3}) can be easily obtained from the second bound of (\ref{eqn:outer_str2III}). This completes the converse proof.
\end{IEEEproof}

If we consider the following condition
\begin{IEEEeqnarray}{rcl}
I(X_1,X_2;Y_3)&\leq& I(X_1,X_2;Y_4)\label{eqn:cond_str_rec1_cor}
\end{IEEEeqnarray}
instead of (\ref{eqn:cond_str_rec2}), the capacity region is given by the following corollary:

\begin{corollary}\label{cor:cap_str_deg}
The capacity region of the degraded classical CC-IFC with the joint p.m.f $p^*$, satisfying conditions (\ref{eqn:cond_str_rec1}) and (\ref{eqn:cond_str_rec1_cor}), is given by
\begin{IEEEeqnarray*}{rl}
\Cc_0^{*} =\bigcup_{p(t)p(x_1|t)p(x_2|t)} \Big\{&
(R_1,R_2): R_1 \geq 0, R_2 \geq 0 \nonumber\\
&R_1 \leq (X_1;Y_2|X_2,T)\yesnumber\label{eqn:cap_str_deg_cor1}\\
&R_2 \leq I(X_2;Y_4|X_1,T)\yesnumber\label{eqn:cap_str_deg_cor2}\\
&R_1+R_2 \leq I(X_1,X_2;Y_3)\Big\}.\yesnumber\label{eqn:cap_str_deg_cor3}
\end{IEEEeqnarray*}
\end{corollary}

\begin{remark}
If we assume that the cognitive link between the transmitters is ideal, then the cognitive user can decode the message of the primary user without any rate constraint and the bound in (\ref{eqn:cap_str_deg_cor1}) will be dropped. In this case, by setting $T=\emptyset$, $\Cc_0^{*}$ coincides with the capacity region of the strong interference channel with unidirectional cooperation (or non-causal C-IFC), satisfying (\ref{eqn:cond_str_rec1}) and (\ref{eqn:cond_str_rec1_cor}), which has been characterized in \cite[Theorem 5]{MariYatKra07}, \cite{MarYatKra06}.
\end{remark}

\begin{remark}[Comparison of two sets of conditions]
We can write (\ref{eqn:cond_str_rec1_cor}) as
\begin{IEEEeqnarray*}{rcl}
I(X_1;Y_3)+\underbrace{[I(X_2;Y_3|X_1)-I(X_2;Y_4|X_1)]}_{I_{diff}}&\leq&I(X_1;Y_4)
\end{IEEEeqnarray*}

Considering (\ref{eqn:cond_str_rec1}), it can be seen that $I_{diff}\geq 0$. Hence, the conditions of Corollary~\ref{cor:cap_str_deg} imply those of Theorem~\ref{thm:cap_str_deg}. Therefore, the strong interference conditions of Theorem~\ref{thm:cap_str_deg} are weaker compared to the conditions obtained in \cite{MariYatKra07, MarYatKra06}.
\end{remark}

\subsection{Semi-deterministic classical CC-IFC}
Here, we consider classical CC-IFC (CC-IFC-WD with $L=0$) with the deterministic component for the channel output of the cognitive transmitter, i.e., the received signal at the cognitive user (Tx2) is a deterministic function of the primary user's input signal:
\begin{IEEEeqnarray}{l}
Y_2=h_2(X_1)\label{eqn:cond_semi}
\end{IEEEeqnarray}

Assume that for every distribution $p^*$, this semi-deterministic classical CC-IFC satisfies (\ref{eqn:cond_str_rec1}), (\ref{eqn:cond_str_rec2}) and the following additional condition:
\begin{IEEEeqnarray}{rcl}
I(X_1;Y_3|Y_2,X_2)&\leq&I(X_1;Y_4|Y_2,X_2)\label{eqn:cond_str_rec1_semi}
\end{IEEEeqnarray}

\begin{theorem}\label{thm:cap_str_semi}
The capacity region of the semi-deterministic classical CC-IFC, defined by (\ref{eqn:cond_semi}) with the joint p.m.f $p^*$, satisfying conditions (\ref{eqn:cond_str_rec1}), (\ref{eqn:cond_str_rec2}) and (\ref{eqn:cond_str_rec1_semi}), is given by
\begin{IEEEeqnarray}{rl}
\Cc_0^s =&\bigcup\limits_{p(t)p(x_1|t)p(x_2|t)} \Big\{
(R_1,R_2): R_1 \geq 0, R_2 \geq 0  \nonumber\\
&R_1 \leq H(Y_2|X_2,T)+I(X_1;Y_3|Y_2,X_2,T)\label{eqn:cap_str_semi_R1}\\
&R_2 \leq I(X_2;Y_4|X_1,T)\label{eqn:cap_str_semi_R2}\\
&R_1+R_2 \leq\min\{I(X_1,X_2;Y_3),I(X_1,X_2;Y_4)\}\Big\}.\quad\label{eqn:cap_str_semi_sum}
\end{IEEEeqnarray}
\end{theorem}

\begin{remark}\label{remark:cap_zerodelay_semi}
Similar to Remark~\ref{remark:cap_zerodelay_deg}, the above channel model can be seen as a semi-deterministic relay channel of (\ref{eqn:cond_semi}) \cite{Aref1982} with a private message $m_2$ from the relay to the receiver and a MAC with common information at Rx2 \cite{SleWol73}.
\end{remark}

\begin{IEEEproof}
\underline{Achievability:} Similar to Theorem~\ref{thm:cap_str_deg}, we specialize the region $\Rc_1$ in Theorem~\ref{thm:zerodelay} with $Q=\emptyset$. In order to cancel the private parts of the messages, let $T_p=U_{1p}=V_{1p}=U_{2p}=\emptyset$ and $R_{2p}=R_{1pn}=R_{1pd}=0$. Moreover, ignore GP coding by setting $L_{2c}=L_{2p}=0$. Also, redefine $U_{2c}=X_2$, $V_{1c}=X_1$, $U_{1c}=U$ and $T_c=T$. Thus, $\Rc_1$ reduces to
\begin{IEEEeqnarray}{rcl}
R_1 &\leq& I(U;Y_2|X_2,T)+I(X_1;Y_3|U,X_2,T)\label{eqn:ach_str_semi_R1_1}\\
R_1 &\leq& I(U;Y_2|X_2,T)+I(X_1;Y_4|U,X_2,T)\label{eqn:ach_str_semi_R1_2}\\
R_2 &\leq& I(X_2;Y_3|X_1,T)\label{eqn:ach_str_semi_R2_1}\\
R_2 &\leq& I(X_2;Y_4|X_1,T)\label{eqn:ach_str_semi_R2_2}\\
R_1+R_2 &\leq& \min\{I(X_1,X_2;Y_3),I(X_1,X_2;Y_4)\}\label{eqn:ach_str_semi_sum1}\\
R_1+R_2 &\leq& I(U;Y_2|X_2,T)+\label{eqn:ach_str_semi_sum2}\\
&&\min\{I(X_1,X_2;Y_3|U,T),I(X_1,X_2;Y_4|U,T)\}\nonumber
\end{IEEEeqnarray}
where $\mathcal{P}_1$ in (\ref{eqn:pmfzerodelay}) becomes
\begin{IEEEeqnarray}{c}
p(t)p(x_1,u|t)p(x_2|t).\label{eqn:pmf_ach_cap_str_semi}
\end{IEEEeqnarray}

Due to condition (\ref{eqn:cond_str_rec1}), the bound in (\ref{eqn:ach_str_semi_R2_1}) is redundant. Now, in the above region let $U=Y_2$, which is feasible because the primary user knows $Y_2=h_2(X_1)$. Then, due to condition (\ref{eqn:cond_str_rec1_semi}), the bound in (\ref{eqn:ach_str_semi_R1_2}) becomes redundant and (\ref{eqn:ach_str_semi_R1_1}), (\ref{eqn:ach_str_semi_R2_2}) and (\ref{eqn:ach_str_semi_sum1}) reduce to (\ref{eqn:cap_str_semi_R1}), (\ref{eqn:cap_str_semi_R2}) and (\ref{eqn:cap_str_semi_sum}), respectively. Moreover, p.m.f in (\ref{eqn:pmf_ach_cap_str_semi}) becomes
\begin{IEEEeqnarray}{c}
p(t)p(x_1|t)p(x_2|t).\label{eqn:pmf_cap_str_semi}
\end{IEEEeqnarray}

Hence, due to the conditional independence of $X_2$ and $X_1$ given $T$, the following equations are obtained:
\begin{IEEEeqnarray}{l}
\!\!\!I(X_2;Y_3|X_1,T)=I(X_2;Y_3|X_1,Y_2,T)=I(X_2;Y_3|Y_2,T)\;\quad\label{eqn:red_semi1}\\
\!\!\!I(X_2;Y_4|X_1,T)=I(X_2;Y_4|X_1,Y_2,T)=I(X_2;Y_4|Y_2,T)\;\quad\label{eqn:red_semi2}
\end{IEEEeqnarray}

Combining (\ref{eqn:red_semi1}) and (\ref{eqn:cond_str_rec1}), the first bound in (\ref{eqn:ach_str_semi_sum2}) becomes redundant. In a similar manner, (\ref{eqn:red_semi2}) and (\ref{eqn:cond_str_rec1_semi}) make the second bound in (\ref{eqn:ach_str_semi_sum2}) redundant. This completes the proof for achievability.

\underline{Converse:} For this part, we use the bounds derived in the converse proof of Theorem~\ref{thm:cap_str_deg}. Bounds in (\ref{eqn:cap_str_semi_R2}) and (\ref{eqn:cap_str_semi_sum}) are obtained directly from (\ref{eqn:cap_degII}) and (\ref{eqn:cap_degIII}). For (\ref{eqn:cap_str_semi_R1}), we use (\ref{eqn:cap_degI}) to obtain
\begin{IEEEeqnarray*}{ll}
R_1&\leq I(X_1;Y_2|X_2,T)+I(X_1;Y_3|X_2,Y_2,T)\\
&=H(Y_{2}|X_{2},T)+I(X_{1};Y_{3}|X_{2},Y_{2},T)\yesnumber\label{eqn:fanoI_ti_sh_semi}
\end{IEEEeqnarray*}
where (\ref{eqn:cond_semi}) has been used for (\ref{eqn:fanoI_ti_sh_semi}).
\end{IEEEproof}

\section{Gaussian Causal Cognitive Interference Channel With Delay}\label{sec:Gaussian}
In this section, we consider Gaussian CC-IFC-WD and extend the achievable rate regions $\Rc_1(Z_1)$, $\Rc_2(Z_2)$, and $\Rc_3(Z_3)$, derived for the discrete memoryless classical CC-IFC ($L=0$), the discrete memoryless CC-IFC without delay ($L=1$), and the discrete memoryless CC-IFC with unlimited look-ahead, respectively, to the Gaussian case. Moreover, we present some numerical examples in order to investigate the effects of the delay and the rate gain of the cognitive link in this channel. Thus, we compare the strategies which are used for achieving the above rate regions.

\begin{figure}[tb]
  \centering
  \includegraphics[width=8cm]{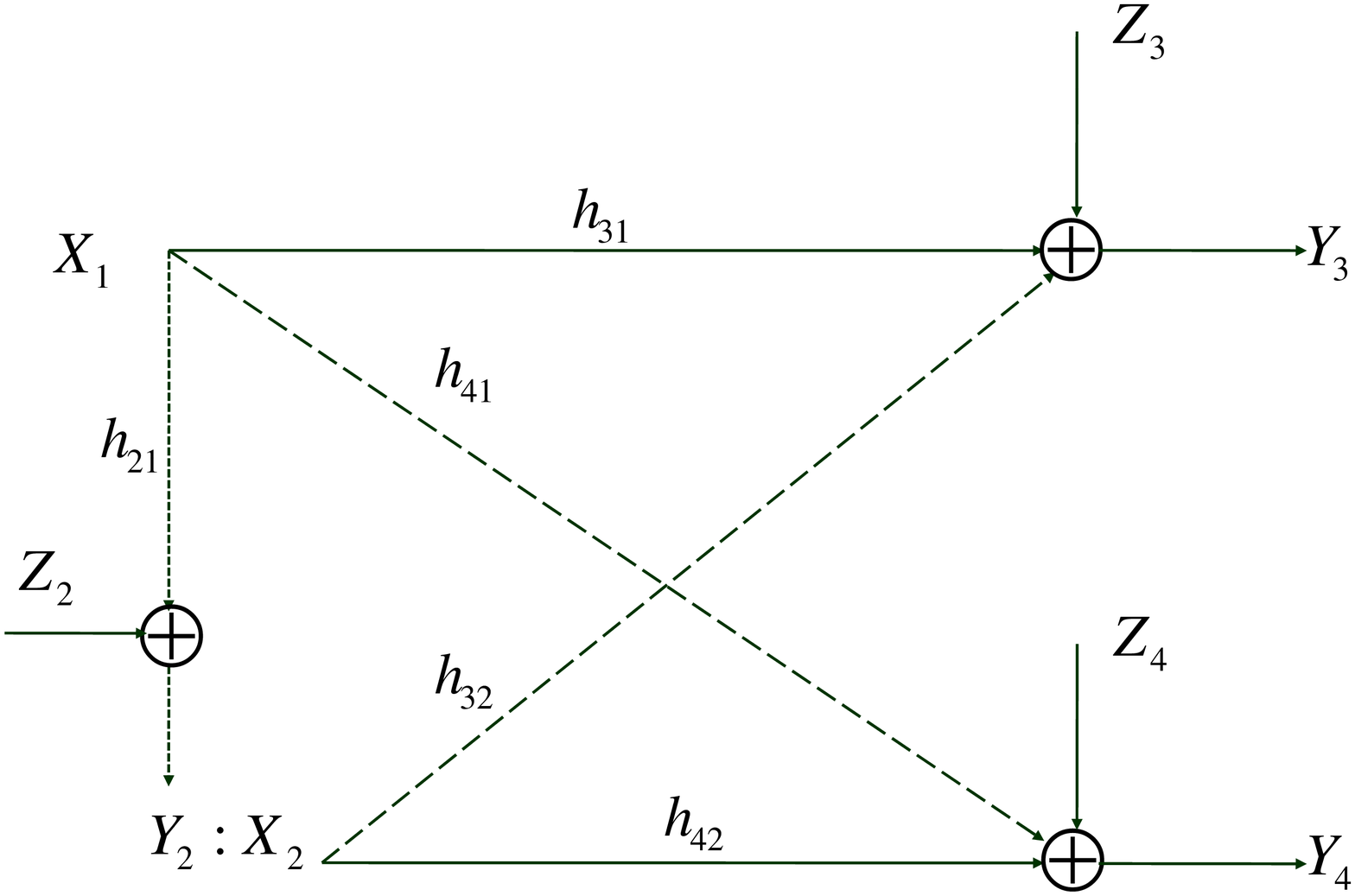}
  \caption{Gaussian Causal Cognitive Interference Channel With Delay (Gaussian CC-IFC-WD).}
  \label{fig:Gauschannelmodel}
\end{figure}
\subsection{Channel Model for the Gaussian CC-IFC-WD}\label{subsec:Gaussian_model}
Gaussian CC-IFC-WD, as depicted in
Fig.~\ref{fig:Gauschannelmodel}, at time $i=1,\ldots,n$ can be modeled mathematically as
\begin{IEEEeqnarray}{rcl}
Y_{2,i}&\:=\:&h_{21}X_{1,i}+Z_{2,i}\nonumber\\
Y_{3,i}&\:=\:&h_{31}X_{1,i}+h_{32}X_{2,i}+Z_{3,i}\label{eqn:Gaussian_model}\\
Y_{4,i}&\:=\:&h_{41}X_{1,i}+h_{42}X_{2,i}+Z_{4,i}\nonumber
\end{IEEEeqnarray}
where $h_{21}$, $h_{31}$, $h_{32}$, $h_{41}$ and $h_{42}$ are known channel gains. Additionally, $X_{1,i}$ and $X_{2,i}$ are input signals with average power constraints
\begin{IEEEeqnarray}{rcl}
\frac{1}{n}\sum\limits_{i=1}^n(x_{u,i})^2\leq P_u\label{eqn:power_cons}
\end{IEEEeqnarray}
for $u\in\{1,2\}$. Also, $Z_{2,i}$, $Z_{3,i}$ and $Z_{4,i}$ are i.i.d and independent zero mean Gaussian noise components with powers $N_2$, $N_3$ and $N_4$, respectively. Note that, at the cognitive user, we have a set of encoding functions $x_{2,i}=f_{2,i}(m_2,y_2^{i-1+L})$ for $i=1,\ldots,n$ and $m_2\in\Mc_2$.

\subsection{Achievable Rate Regions for the Gaussian CC-IFC-WD}\label{subsec:Gaussian_rate}
To simplify notation, we define
\begin{IEEEeqnarray}{rcl}
\theta(x)\doteq \frac{1}{2}\,log(1+x)
\end{IEEEeqnarray}

\begin{figure*}[!t]
\normalsize
\setcounter{tempequationcounter}{\value{equation}}{\small
\begin{IEEEeqnarray}{rcl}
\setcounter{equation}{79}
I_1^*&\doteq&\theta\left(\frac{\alpha_1^2h_{41}^2\beta_3P_1}{\gamma_1P_2}\right)\label{eqn:Gaus_zerodelay_Tx2_enc_begin}\\
I_2^*&\doteq&
\theta\left(\frac{\alpha_2^2h_{41}^2\beta_3P_1}{(\gamma_2+\alpha_2^2h_{42}^2\gamma_1)P_2}\right)\\
I_3^*&\doteq&I_1^*+\theta\left(\frac{\alpha_2^2h_{41}^2\beta_3(1-\alpha_1h_{42})^2\gamma_1P_1P_2}{A\gamma_2P_2}\right)+\underbrace{\theta\left(\frac{\alpha_2^2(\alpha_1h_{41}^2\beta_3P_1+h_{42}\gamma_1P_2)^2}{A\gamma_2P_2+C\alpha_2^2}\right)}_{I_3'^*}\label{eqn:Gaus_zerodelay_Tx2_enc_end}\\
I_4^*&\doteq&\theta\left(\frac{h_{31}^2\beta'_1P_1}{h_{32}^2\gamma_2P_2+N_3}\right)\label{eqn:Gaus_zerodelay_Rx1_begin}\\
I_5^*&\doteq&I_1^*+\theta\left(\frac{h_{31}^2P_1+h_{32}^2(\gamma_1+\gamma_3)P_2+2h_{31}h_{32}\sqrt{\beta_4\gamma_3P_1P_2}}{h_{32}^2\gamma_2P_2+N_3}\right)\\
I_6^*&\doteq&I_1^*+\theta\left(P_1\frac{Ah_{31}^2(\beta'_1+\beta_1+\beta'_2+\beta_2+\beta_3)+\alpha_1^2h_{41}^2\beta_3(h_{32}^2\gamma_2P_2-h_{31}^2\beta_3P_1)-2\alpha_1h_{31}h_{32}h_{41}\beta_3\gamma_1P_2}{A}\right)\\
I_7^*&\doteq&I_1^*+\theta\left(P_1\frac{Ah_{31}^2(\beta'_1+\beta'_2+\beta_2+\beta_3)+\alpha_1^2h_{41}^2\beta_3(h_{32}^2\gamma_2P_2-h_{31}^2\beta_3P_1)-2\alpha_1h_{31}h_{32}h_{41}\beta_3\gamma_1P_2}{A}\right)\\
I_8^*&\doteq&I_1^*+\theta\left(\frac{h_{31}^2(\beta'_1+\beta'_2+\beta_3)P_1+h_{32}^2\gamma_1P_2}{h_{32}^2\gamma_2P_2+N_3}\right)\\
I_9^*&\doteq&\theta\left(\frac{h_{31}^2(\beta'_1+\beta_1)P_1}{h_{32}^2\gamma_2P_2+N_3}\right)\\
I_{10}^*&\doteq&I_1^*+\theta\left(\frac{h_{31}^2\beta'_1P_1+h_{32}^2\gamma_1P_2}{h_{32}^2\gamma_2P_2+N_3}\right)\\
I_{11}^*&\doteq&I_1^*+\theta\left(\frac{h_{31}^2(\beta'_1+\beta_1+\beta'_2+\beta_3)P_1+h_{32}^2\gamma_1P_2}{h_{32}^2\gamma_2P_2+N_3}\right)\\
I_{12}^*&\doteq&I_1^*+\theta\left(\frac{h_{31}^2(\beta'_1+\beta_1)P_1+h_{32}^2\gamma_1P_2}{h_{32}^2\gamma_2P_2+N_3}\right)\label{eqn:Gaus_zerodelay_Rx1_end}\\
I_{13}^*&\doteq&I_3'^*+\theta\left(\gamma_2P_2^2\frac{Ah_{41}^2\beta_3\gamma_2(1-2\alpha_2h_{42})P_1+C\gamma_2(2\alpha_1h_{42}-1)+h_{42}^2\gamma_1(\gamma_1\gamma_2(1-\alpha_2h_{42})^2P_2^2-\alpha_2^2)}{B\Big((N_4+h_{41}^2(\beta'_1+\beta'_2)P_1)(A\gamma_2P_2+C\alpha_2^2)+C(1-\alpha_2h_{42})^2\gamma_2P_2\Big)}\right)\label{eqn:Gaus_zerodelay_Rx2_begin}\\
I_{14}^*&\doteq&I_3'^*+\theta\left(\frac{\frac{C}{\gamma_1P_2}(\alpha_2^2h_{41}^2\beta_3P_1+2\alpha_2h_{42}\gamma_2(\alpha_1^2+\gamma_1P_2)P_2)+A^2h_{42}^2\gamma_2P_2}{A(N_4+h_{41}^2(\beta'_1+\beta'_2)P_1)(A\gamma_2P_2+C\alpha_2^2)+C(1-\alpha_2h_{42})^2\gamma_2P_2}\right)\\
I_{15}^*&\doteq&\theta\left(\frac{(A\gamma_2P_2+C\alpha_2^2)((\beta_1+\beta_2+\beta_4)h_{41}^2P_1+h_{42}^2\gamma_3P_2+2h_{41}h_{42}\sqrt{\beta_4\gamma_3P_1P_2})+F}{(N_4+h_{41}^2(\beta'_1+\beta'_2)P_1)(A\gamma_2P_2+C\alpha_2^2)+C(1-\alpha_2h_{42})^2\gamma_2P_2}\right)\\
I_{16}^*&\doteq&I_{13}^*+\theta\left(\frac{Bh_{41}^2\beta_1P_1}{B(N_4+h_{41}^2(\beta'_1+\beta'_2)P_1)+h_{41}^2\beta_3\gamma_2(1-2\alpha_2h_{42})P_1P_2+h_{42}^2\gamma_1\gamma_2(1-\alpha_2h_{42})^2P_2^2}\right)\\
I_{17}^*&\doteq&I_{14}^*+\theta\left(\frac{Ah_{41}^2\beta_1P_1}{A(N_4+h_{41}^2(\beta'_1+\beta'_2)P_1+h_{41}^2\gamma_2P_2)+C}\right)\\
I_{18}^*&\doteq&\theta\left(\frac{h_{41}^2\beta_1P_1(A\gamma_2P_2+C\alpha_2^2)+F}{(N_4+h_{41}^2(\beta'_1+\beta'_2)P_1)(A\gamma_2P_2+C\alpha_2^2)+C(1-\alpha_2h_{42})^2\gamma_2P_2}\right)\\
I_{19}^*&\doteq&\theta\left(\frac{F}{(N_4+h_{41}^2(\beta'_1+\beta'_2)P_1)(A\gamma_2P_2+C\alpha_2^2)+C(1-\alpha_2h_{42})^2\gamma_2P_2}\right)\label{eqn:Gaus_zerodelay_Rx2_end}\\
I_{20}^*&\doteq&\theta\left(\frac{h_{21}^2\beta'_2P_1}{h_{21}^2(\beta'_1+\beta_1)P_1+N_2}\right)\label{eqn:Gaus_zerodelay_Tx2_dec_begin}\\
I_{21}^*&\doteq&\theta\left(\frac{h_{21}^2(\beta'_2+\beta_2)P_1}{h_{21}^2(\beta'_1+\beta_1)P_1+N_2}\right)\label{eqn:Gaus_zerodelay_Tx2_dec_end}
\end{IEEEeqnarray}}\setcounter{equation}{\value{tempequationcounter}}
\hrulefill
\vspace*{4pt}
\end{figure*}
\addtocounter{equation}{21}
First, we consider the Gaussian classical CC-IFC ($L=0$). For certain $\left\{0\leq\beta_r\leq1,r\in\{1,2,3,4\}\right\}$, $\left\{0\leq\beta'_s\leq1,s\in\{1,2\}\right\}$ and $\left\{0\leq\gamma_t\leq1,t\in\{1,2,3\}\right\}$ with $\beta'_1+\beta_1+\beta'_2+\beta_2+\beta_3+\beta_4\leq1$ and $\gamma_1+\gamma_2+\gamma_3\leq1$, we define $I_i^*\,,\,i=1,\ldots,21$ as (\ref{eqn:Gaus_zerodelay_Tx2_enc_begin})-(\ref{eqn:Gaus_zerodelay_Tx2_dec_end}) at the top of the following page, where
\begin{IEEEeqnarray}{rcl}
\alpha_1&\doteq&\frac{h_{42}\gamma_1P_2}{h_{42}^2\gamma_1P_2+D+(h_{41}\sqrt{\beta_4P_1}+h_{42}\sqrt{\gamma_3P_2})^2}\label{eqn:alpha_1}\\
\alpha_2&\doteq&\frac{h_{42}\gamma_2P_2}{D+(h_{41}\sqrt{\beta_4P_1}+h_{42}\sqrt{\gamma_3P_2})^2}\label{eqn:alpha_2}\\
A&\doteq&\gamma_1P_2+\alpha_1^2h_{41}^2\beta_3P_1\\
B&\doteq&\alpha_2^2h_{41}^2\beta_3P_1+(\gamma_2+h_{42}^2\alpha_2^2\gamma_1)P_2\\
C&\doteq&h_{41}^2\beta_3\gamma_1P_1P_2(1-\alpha_1h_{42})^2\\
D&\doteq&N_4+h_{41}^2(\beta'_1+\beta_1+\beta'_2+\beta_2)P_1+h_{42}^2\gamma_2P_2\\
F&\doteq&(h_{41}^2\beta_3\alpha_1P_1+2h_{42}\gamma_1P_2)h_{41}^2\beta_3\alpha_1\gamma_2P_1P_2\nonumber\\
&&+C\alpha_2(\alpha_2h_{42}^2\gamma_1P_2+\alpha_2h_{41}^2\beta_3P_1+2h_{42}\gamma_2P_2)\nonumber\\
&&+(A\gamma_2+\gamma_1^2P_2)h_{42}^2\gamma_2P_2^2
\end{IEEEeqnarray}

Now, replacing each term in (\ref{eqn:zerodelay_Tx2_enc_begin})-(\ref{eqn:zerodelay_Tx2_dec_end}) with the corresponding term from (\ref{eqn:Gaus_zerodelay_Tx2_enc_begin})-(\ref{eqn:Gaus_zerodelay_Tx2_dec_end}) (replacing $I_i$ with $I_i^*$ for $i=1,\ldots,21$), we obtain the Gaussian counterpart of $\Rc_1$, namely $\Rc_1^*$.

\begin{theorem}\label{thm:Gaus_zerodelay}
For the Gaussian classical CC-IFC (CC-IFC-WD with $L=0$), defined in Section \ref{subsec:Gaussian_model}, the convex closure of the region
$\bigcup\limits_{\substack{\{\beta_r,\beta'_s,\gamma_t\}\in[0,1]\\\beta'_1+\beta_1+\beta'_2+\beta_2+\beta_3+\beta_4\leq1\\ \gamma_1+\gamma_2+\gamma_3\leq1}}\Rc_1^*$, where $r\in\{1,2,3,4\}$, $s\in\{1,2\}$ and $t\in\{1,2,3\}$, is an achievable rate region.
\end{theorem}

\begin{IEEEproof}
The achievable rate region $\Rc_1$ in Theorem~\ref{thm:zerodelay} (or Corollary~\ref{cor:zerodelay}) can be extended to the discrete-time Gaussian memoryless case with continuous alphabets by standard arguments \cite{CovTho06}. Hence, it is sufficient to evaluate (\ref{eqn:zerodelay_Tx2_enc_begin})-(\ref{eqn:zerodelay_Tx2_dec_end}) with an appropriate choice of input distribution to reach (\ref{eqn:Gaus_zerodelay_Tx2_enc_begin})-(\ref{eqn:Gaus_zerodelay_Tx2_dec_end}). We constrain all the inputs to be Gaussian and set the time-sharing RV $Q=\emptyset$.

For certain $\left\{0\leq\beta_r\leq1,r\in\{1,2,3,4\}\right\}$, $\{0\leq\beta'_s\leq1,$ $s\in\{1,2\}\}$, and $\{0\leq\gamma_t\leq1,t\in\{1,2,3\}\}$, consider the following mapping ($MAP_1$) for the codebook generated in Theorem~\ref{thm:zerodelay} with respect to the p.m.f (\ref{eqn:pmfzerodelay}), which contains the Gaussian version of the generalized block Markov superposition coding, rate splitting, and GP coding:
\begin{IEEEeqnarray}{lll}
T_c\sim\Nc(0,\beta_4P_1)&&\label{eqn:map_zerodelay_tc}\\
T_p=T'_p+T_c&\mbox{where}&\: T'_p\sim\Nc(0,\beta_3P_1)\IEEEeqnarraynumspace\;\\
U_{1c}=U'_{1c}+T_c&\mbox{where}&\: U'_{1c}\sim\Nc(0,\beta_2P_1)\\
U_{1p}=U'_{1p}+U'_{1c}+T'_p+T_c&\mbox{where}&\: U'_{1p}\sim\Nc(0,\beta'_2P_1)\\
V_{1c}=V'_{1c}+T_c&\mbox{where}&\: V'_{1c}\sim\Nc(0,\beta_1P_1)\\
V_{1p}=V'_{1p}+V'_{1c}+T'_p+T_c&\mbox{where}&\: V'_{1p}\sim\Nc(0,\beta'_1P_1)\\
X_1=V'_{1p}\,+\,V'_{1c}\,+\,U'_{1p}\,+&U'_{1c}+T'_p&+T_c\\
U_{2c}=U'_{2c}+\alpha_1S_1&\mbox{where}&\: U'_{2c}\sim\Nc(0,\gamma_1P_2)\\
U_{2p}=U'_{2p}+\alpha_2S_2&\mbox{where}&\: U'_{2p}\sim\Nc(0,\gamma_2P_2)\label{eqn:map_zerodelay_u2p}\\
X_2=U'_{2p}+U'_{2c}+\sqrt{\frac{\gamma_3P_2}{\beta_4P_1}}&T_c&
\end{IEEEeqnarray}
where $\alpha_1$ and $\alpha_2$ are defined in (\ref{eqn:alpha_1}) and (\ref{eqn:alpha_2}), respectively, and
\begin{IEEEeqnarray}{rcl}
S_1&=&h_{41}T'_p\label{eqn:map_zerodelay_s1}\\
S_2&=&h_{41}T'_p+h_{42}U'_{2c}.\label{eqn:map_zerodelay_s2}
\end{IEEEeqnarray}

Parameters $\beta_4$ and $\beta_3$ determine the amounts of $P_1$ which are dedicated for constructing the basis of cooperation for sending common and private messages, respectively. Parameter $\beta_2$ specifies the amount of $P_1$ which is allocated for relaying through the cognitive user for sending the common message. Parameter $\beta'_2$ indicates the amount of $P_1$ which enables the cognitive user to perform GP coding. The remaining parts of $P_1$, distinguished with parameters $\beta_1$ and $\beta'_1$, are sent directly to Rx1. Parameters $\gamma_1$, $\gamma_2$, and $\gamma_3$ determine the amounts of $P_2$ which are dedicated for sending the common message, the private message and relaying, respectively. To execute GP coding, parameters $\alpha_1$ and $\alpha_2$ are utilized. In fact, optimal values for $\alpha_1$, $\alpha_2$, $S_1$, and $S_2$ can be determined by optimizing the rate region for these parameters. However, this method is cumbersome, so we use the modified version of Costa's dirty paper coding (DPC) results \cite{Cost83}.

Applying the power constraints in (\ref{eqn:power_cons}) to $MAP_1$ yields
\begin{IEEEeqnarray*}{c}
\beta'_1+\beta_1+\beta'_2+\beta_2+\beta_3+\beta_4\leq1\\
\gamma_1+\gamma_2+\gamma_3\leq1.
\end{IEEEeqnarray*}

Using the above mapping ($MAP_1$) with the channel model in (\ref{eqn:Gaussian_model}), the remainder of the proof is straightforward.
\end{IEEEproof}

Next, we investigate the Gaussian CC-IFC without delay ($L=1$). First, we modify $I_{i}^*, i=1,\ldots,21$ in (\ref{eqn:Gaus_zerodelay_Tx2_enc_begin})-(\ref{eqn:Gaus_zerodelay_Tx2_dec_end}), by replacing $h_{u1}$ with $h'_{u1}$, $h_{u2}$ with $h'_{u2}$, and
$N_u$ with $N'_u$ for $u\in\{3,4\}$, and refer to them as $I_{i}^{**}$ for $i=1,\ldots,21$.

Consider the channel model in Fig.~\ref{fig:Gauschannelmodel} and (\ref{eqn:Gaussian_model}) with $L=1$, i.e., $X_{2}=f_{2}(m_2,Y_2^{i})$. In order to obtain the Gaussian counterpart of $\Rc_2$, namely $\Rc_2^*$, we replace each term $\{I_{i}, i=1,\ldots,21\}$ in (\ref{eqn:zerodelay_Tx2_enc_begin})-(\ref{eqn:zerodelay_Tx2_dec_end}) with its corresponding term $\{I_{i}^{**}, i=1,\ldots,21\}$, for certain $\left\{0\leq\beta_r\leq1,r\in\{1,2,3,4\}\right\}$, $\left\{0\leq\beta'_s\leq1,s\in\{1,2\}\right\}$, $\left\{0\leq\gamma_t\leq1,t\in\{1,2,3\}\right\}$, and
\begin{IEEEeqnarray*}{rcl}
h'_{u1}&=&h_{u1}+h\beta h_{21}h_{u2}\\
h'_{u2}&=&h(1-\beta)h_{u2}\\
N'_u&=&N_u+h^2\beta^2 h_{u2}^2N_2
\end{IEEEeqnarray*}
for $u\in\{3,4\}$, where $0\leq \beta \leq 1$, $h$ is a normalizing parameter and the following inequalities hold:{\small
\begin{IEEEeqnarray*}{c}
\beta'_1+\beta_1+\beta'_2+\beta_2+\beta_3+\beta_4\leq1\yesnumber\label{eqn:Gaus_constraint_withoutdelay}\\
\frac{P_2}{h^2}\geq P'\doteq h_{21}^2\beta^2(\beta'_1+\beta_1+\beta'_2+\beta_2+\beta_3)P_1+\\
\big(h_{21}\beta\sqrt{\beta_4P_1}+(1-\beta)\sqrt{\gamma_3P_2}\big)^2+\beta^2N_2+(1-\beta)^2(\gamma_1+\gamma_2)P_2
\end{IEEEeqnarray*}}
\begin{theorem}\label{thm:Gaus_withoutdelay}
For the Gaussian CC-IFC without delay (CC-IFC-WD with $L=1$), defined in Section~\ref{subsec:Gaussian_model}, the convex closure of the region
$\bigcup\limits_{\substack{h,\{\beta,\beta_r,\beta'_s,\gamma_t\}\in[0,1]\\\beta'_1+\beta_1+\beta'_2+\beta_2+\beta_3+\beta_4\leq1\\ h^2P'\leq P_2}}\Rc_2^*$, where $r\in\{1,2,3,4\}$, $s\in\{1,2\}$, and $t\in\{1,2,3\}$, is an achievable rate region.
\end{theorem}

\begin{IEEEproof}
The proof of Theorem~\ref{thm:Gaus_withoutdelay} is similar to that of Theorem~\ref{thm:Gaus_zerodelay}. Considering Theorem~\ref{thm:withoutdelay}, $V_2$ is generated according to $\prod\limits_{i=1}^np(v_{2,i}|u_{2p,i},u_{2c,i},t_{p,i},t_{c,i},q_{i})$, and $x_{2,i}=f'_{2,i}(v_{2,i},y_{2,i},q_{i})$. For Gaussian inputs and $Q=\emptyset$, appropriate mapping ($MAP_2$) for the codebook generated in Theorem~\ref{thm:withoutdelay}, with respect to the p.m.f $\mathcal{P}_2$ defined in (\ref{eqn:pmfwithoutdelay}), consists of (\ref{eqn:map_zerodelay_tc})-(\ref{eqn:map_zerodelay_u2p}), (\ref{eqn:map_zerodelay_s1}), (\ref{eqn:map_zerodelay_s2}) and
\begin{IEEEeqnarray}{l}
V_2=U'_{2p}+U'_{2c}+\sqrt{\frac{\gamma_3P_2}{\beta_4P_1}}T_c\\
X_2=h(\beta Y_2+(1-\beta)V_2)
\end{IEEEeqnarray}
where $0\leq \beta \leq 1$ and $h$ is a normalizing parameter. In fact, the cognitive user sends a linear function of its received symbol and the codeword $V_2$, where $h\beta$ determines the amount of $P_2$ which is dedicated for instantaneous relaying by the cognitive user. Also,
(\ref{eqn:Gaus_constraint_withoutdelay}) is obtained by applying the power constraints in (\ref{eqn:power_cons}) to $MAP_2$.

\begin{figure}[tb]
  \centering
  \includegraphics[width=7.5cm]{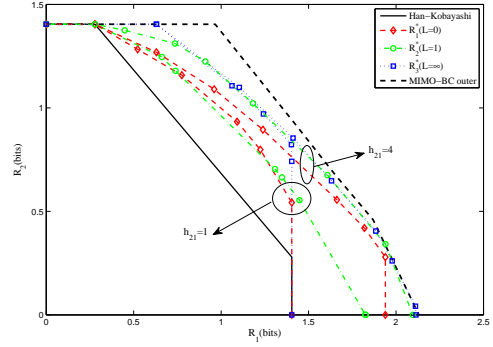}
  \caption{Comparison among $\Rc_1^*$, $\Rc_2^*$, $\Rc_3^*$ and HK region \cite{HanKob81}. $P_1=P_2=6$, $h_{31}=h_{42}=1$, $h_{32}=h_{41}=\sqrt{0.55}$, and $N_2=N_3=N_4=1$.}
  \label{fig:Gaus_Case1}
\end{figure}
Evaluating $\{I_{i}, i=1,\ldots,21\}$ in (\ref{eqn:zerodelay_Tx2_enc_begin})-(\ref{eqn:zerodelay_Tx2_dec_end}), using the above mapping ($MAP_2$) and (\ref{eqn:Gaussian_model}), results in $\{I_{i}^{**}, i=1,\ldots,21\}$. Considering Theorem~\ref{thm:withoutdelay}, the remainder of the proof is straightforward.
\end{IEEEproof}

Finally, we derive the rate region for the Gaussian CC-IFC with unlimited look-ahead. Let $I_{i}^{***}=I_{i}^*\,,\,i=1,\ldots,19$ and{\small
\begin{IEEEeqnarray}{rcl}
I_{20}^{***}&\doteq&\theta\left(\frac{h_{21}^2\beta_3P_1}{h_{21}^2(\beta'_1+\beta_1)P_1+N_2}\right)\label{eqn:Gaus_blockdelay_Tx2_dec_begin}\\
I_{21}^{***}&\doteq&\theta\left(\frac{h_{21}^2(\beta_3+\beta_4)P_1}{h_{21}^2(\beta'_1+\beta_1)P_1+N_2}\right).\label{eqn:Gaus_blockdelay_Tx2_dec_end}
\end{IEEEeqnarray}}
Now, replacing each term in (\ref{eqn:zerodelay_Tx2_enc_begin})-(\ref{eqn:zerodelay_Rx2_end}), (\ref{eqn:blockdelay_Tx2_dec_begin}) and (\ref{eqn:blockdelay_Tx2_dec_end}) with its corresponding term from (\ref{eqn:Gaus_zerodelay_Tx2_enc_begin})-(\ref{eqn:Gaus_zerodelay_Rx2_end}),
(\ref{eqn:Gaus_blockdelay_Tx2_dec_begin}) and (\ref{eqn:Gaus_blockdelay_Tx2_dec_end}), i.e., replacing $I_i$ with $I_i^{***}$ for $i=1,\ldots,21$, yields the Gaussian counterpart of $\Rc_3$, to which we refer as $\Rc_3^*$.

\begin{theorem}\label{thm:Gaus_blockdelay}
For the Gaussian CC-IFC with unlimited look-ahead, defined in Section~\ref{subsec:Gaussian_model}, the convex closure of the region $\bigcup\limits_{\substack{\{\beta_r,\beta'_1,\gamma_t\}\in[0,1]\\\beta'_2=\beta_2=0\\\beta'_1+\beta_1+\beta_3+\beta_4\leq1\\\gamma_1+\gamma_2+\gamma_3\leq1}}\Rc_3^*$, where $r\in\{1,3,4\}$ and $t\in\{1,2,3\}$, is an achievable rate region.
\end{theorem}
\begin{IEEEproof}
The proof follows the same lines as that of Theorem~\ref{thm:Gaus_zerodelay}, except that according to Theorem~\ref{thm:blockdelay} there is no dependence on the previous block messages. Therefore, it is possible to set $T_c=U_{1c}$ and $T_p=U_{1p}$, or equivalently $\beta'_2=0$ and $\beta_2=0$ in $MAP_1$, to obtain $MAP_3$.
\end{IEEEproof}
\begin{figure}[!b]
  \centering
  \includegraphics[width=7.5cm]{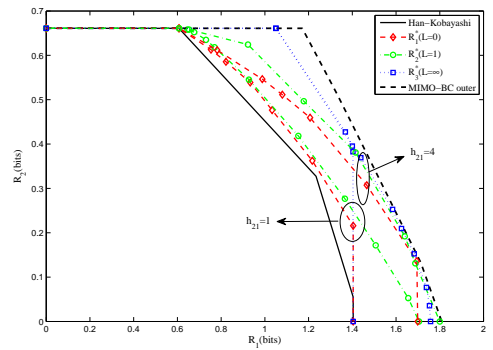}
  \caption{Comparison among $\Rc_1^*$, $\Rc_2^*$, $\Rc_3^*$ and HK region \cite{HanKob81}. $P_1=6$, $P_2=1.5$, $h_{31}=h_{42}=1$, $h_{32}=h_{41}=\sqrt{0.55}$, and $N_2=N_3=N_4=1$.}
  \label{fig:Gaus_Case2}
\end{figure}
\subsection{Numerical Examples for the Gaussian CC-IFC-WD}\label{subsec:Gaussian_example}
Here, we provide some numerical examples of the rate regions $\Rc_1^*$ in Theorem~\ref{thm:Gaus_zerodelay}, $\Rc_2^*$ in Theorem~\ref{thm:Gaus_withoutdelay} and $\Rc_3^*$ in Theorem~\ref{thm:Gaus_blockdelay}. Comparing the strategies used to achieve the above rate regions, we investigate the effects of the delay, cooperation, and interference cancelation in this channel. First, we consider the rate gain of the cognitive link for different strategies.

Fig.~\ref{fig:Gaus_Case1} compares the regions $\Rc_1^*$, $\Rc_2^*$, and $\Rc_3^*$ with the HK region in \cite{HanKob81}, where the overheard information is neglected, for $P_1=P_2=6$, $h_{31}=h_{42}=1$, $h_{32}=h_{41}=\sqrt{0.55}$ and $N_2=N_3=N_4=1$. Moreover, an outer bound on the capacity region of CC-IFC-WD is provided by intersecting the capacity region of the Gaussian MIMO broadcast channel (MIMO-BC) \cite{WeiSteSha06} with the rate of the Tx2-Rx2 interference-free channel, i.e., $R_2\leq\theta(\frac{P_2}{N_4})$. These regions are shown in Fig.~\ref{fig:Gaus_Case2} for $P_1=6$ and $P_2=1.5$. Due to the cooperative strategies, $\Rc_1^*$, $\Rc_2^*$, and $\Rc_3^*$ outperform the HK region. Especially when the cognitive link is sufficiently strong, i.e. $h_{21}=4$, $\Rc_2^*$ and $\Rc_3^*$ achieve rates close to the outer bound for a small $R_2$, because the cognitive user can decode and cooperate more effectively and can allocate more power for simultaneous cooperation. Due to instantaneous relaying and non-causal DF schemes, larger regions are obtained for $L=1$ and unlimited look-ahead ($L=\infty$) cases than for $L=0$.
\begin{figure*}[htb]%
\centering
\parbox{8cm}{\includegraphics[width=8cm]{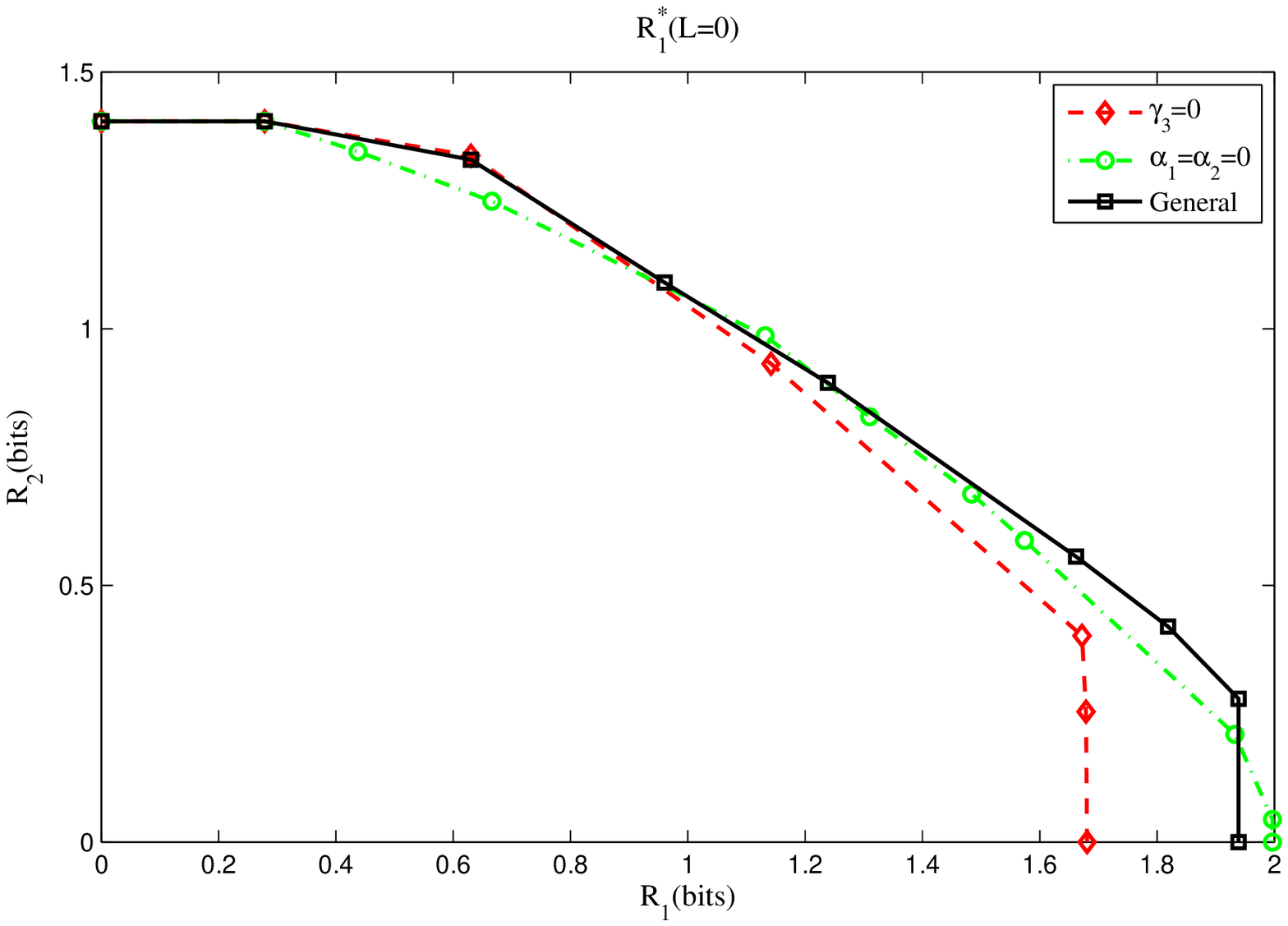}}%
\,
\begin{minipage}{8.3cm}%
\includegraphics[width=8cm]{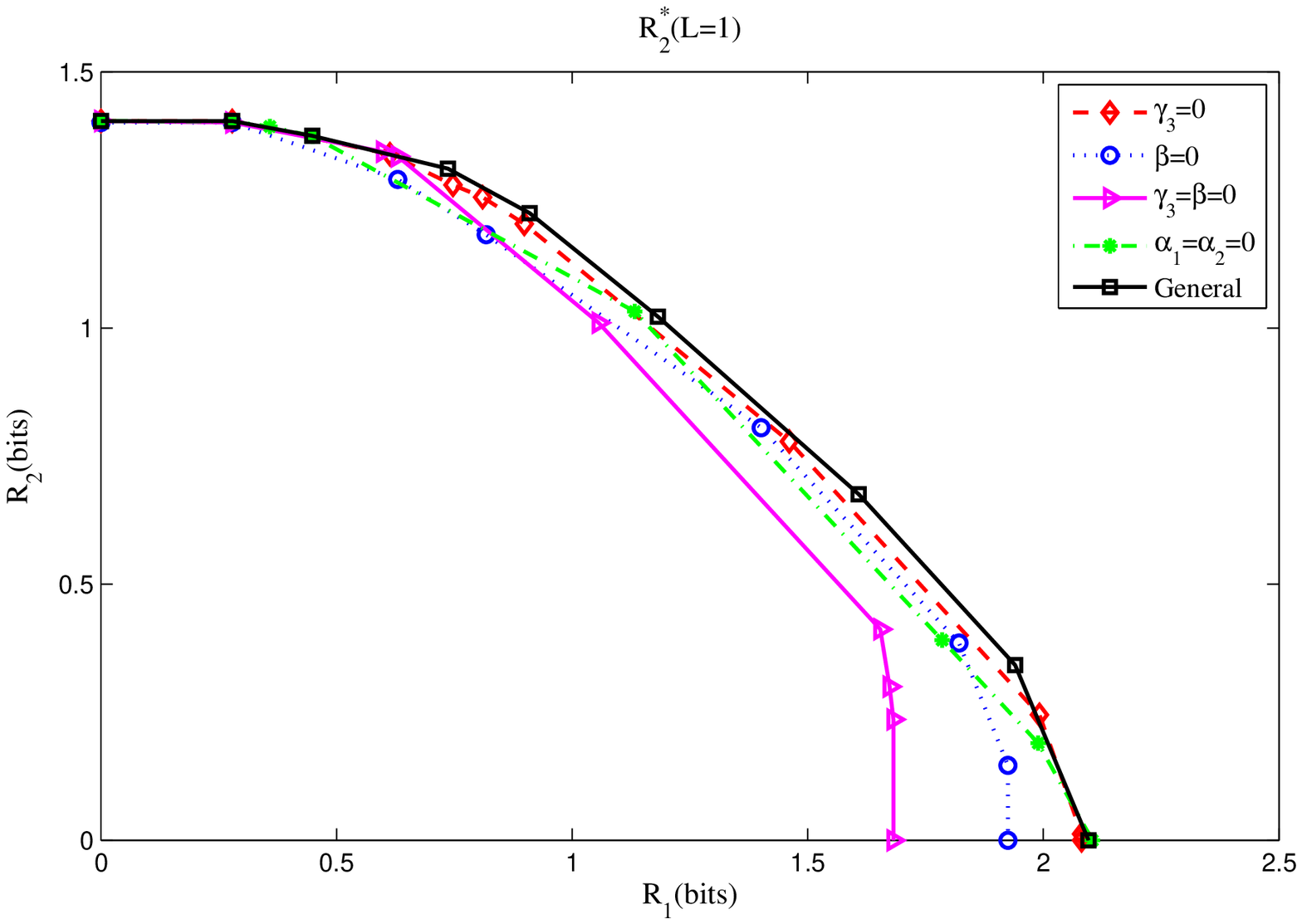}
\end{minipage}\\
\quad\includegraphics[width=8cm]{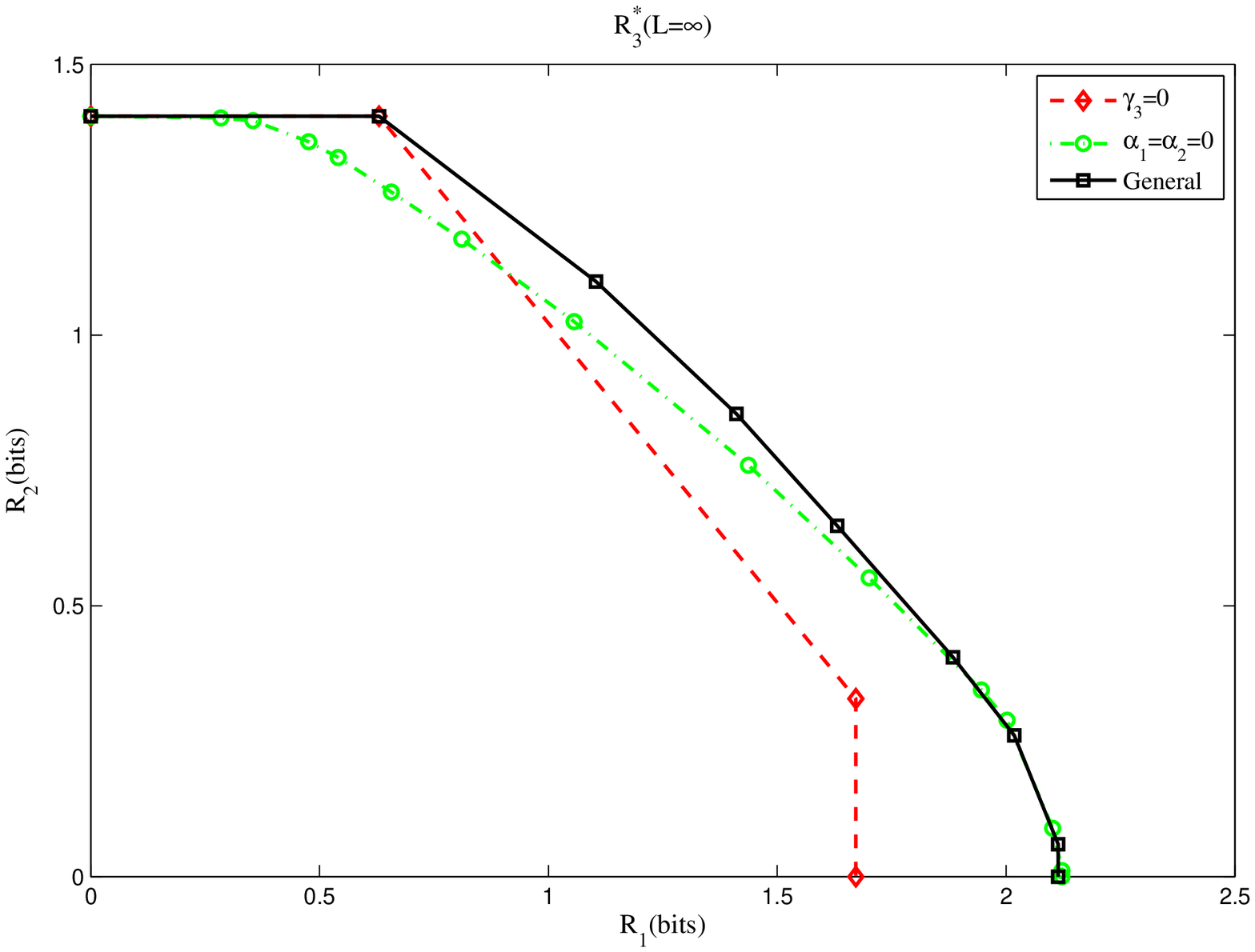}
\caption{Impacts of partial DF relaying ($\gamma_3$), instantaneous relaying ($\beta$) and interference cancelation by DPC ($\alpha_1$ and $\alpha_2$). Setting each of the above parameters to zero, eliminates the related strategy.}%
\label{fig:Gaus_Case4}
\end{figure*}

To compare the non-causal DF ($\Rc_3^*$) and instantaneous relaying ($\Rc_2^*$) based on Fig.~\ref{fig:Gaus_Case1} and Fig.~\ref{fig:Gaus_Case2}, one must consider the condition of the cognitive link. For a strong cognitive link ($h_{21}=4$), the performance of the non-causal DF scheme is better (especially when the cognitive user sends at higher rates), and allowing sufficient time for the cognitive user to decode increases the rates that can be achieved. However, when $h_{21}=1$, instantaneous relaying outperforms DF for small $R_2$. In fact, when poor conditions exist for the cognitive link, instantaneous relaying is the only scheme that can outperform the HK scheme for the primary user ($R_1$) when the cognitive user sends at lower rates. We remark that, since an instantaneous relaying scheme is feasible for every $L\geq 1$, the convex hull of the regions $\Rc_2^*$ and $\Rc_3^*$ is achievable for CC-IFC with unlimited look-ahead ($L=\infty$) using a coding scheme based on a combination of instantaneous relaying with non-causal DF strategies.

Fig.~\ref{fig:Gaus_Case4} portrays the impacts of partial DF relaying ($\gamma_3$), instantaneous relaying ($\beta$), and interference cancelation by DPC ($\alpha_1$ and $\alpha_2$) for $P_1=P_2=6$, $h_{31}=h_{42}=1$, $h_{32}=h_{41}=\sqrt{0.55}$, and $N_2=N_3=N_4=1$. Considering $\Rc_1^*$ ($L=0$) and $\Rc_3^*$ ($L=\infty$), we see that when $R_2$ is large, setting $\gamma_3=0$ (no DF relaying) performs better. This more efficient performance means that in this case interference cancelation by DPC is a better strategy. However, when the cognitive user sends at lower rates and can allocate more power for relaying, DPC provides less improvement. It is worth noting that the region related to $\alpha_1=\alpha_2=0$, can also be obtained by the general scheme if the rate region is optimized for these parameters instead of using (\ref{eqn:alpha_1}) and (\ref{eqn:alpha_2}). A similar argument can be made for $\Rc_2^*$ ($L=1$). However, the performance improvement in the latter case is due mostly to the instantaneous relaying, especially when $R_2$ is small.

\begin{figure*}[tb]%
\centering
\parbox{8cm}{\includegraphics[width=8cm]{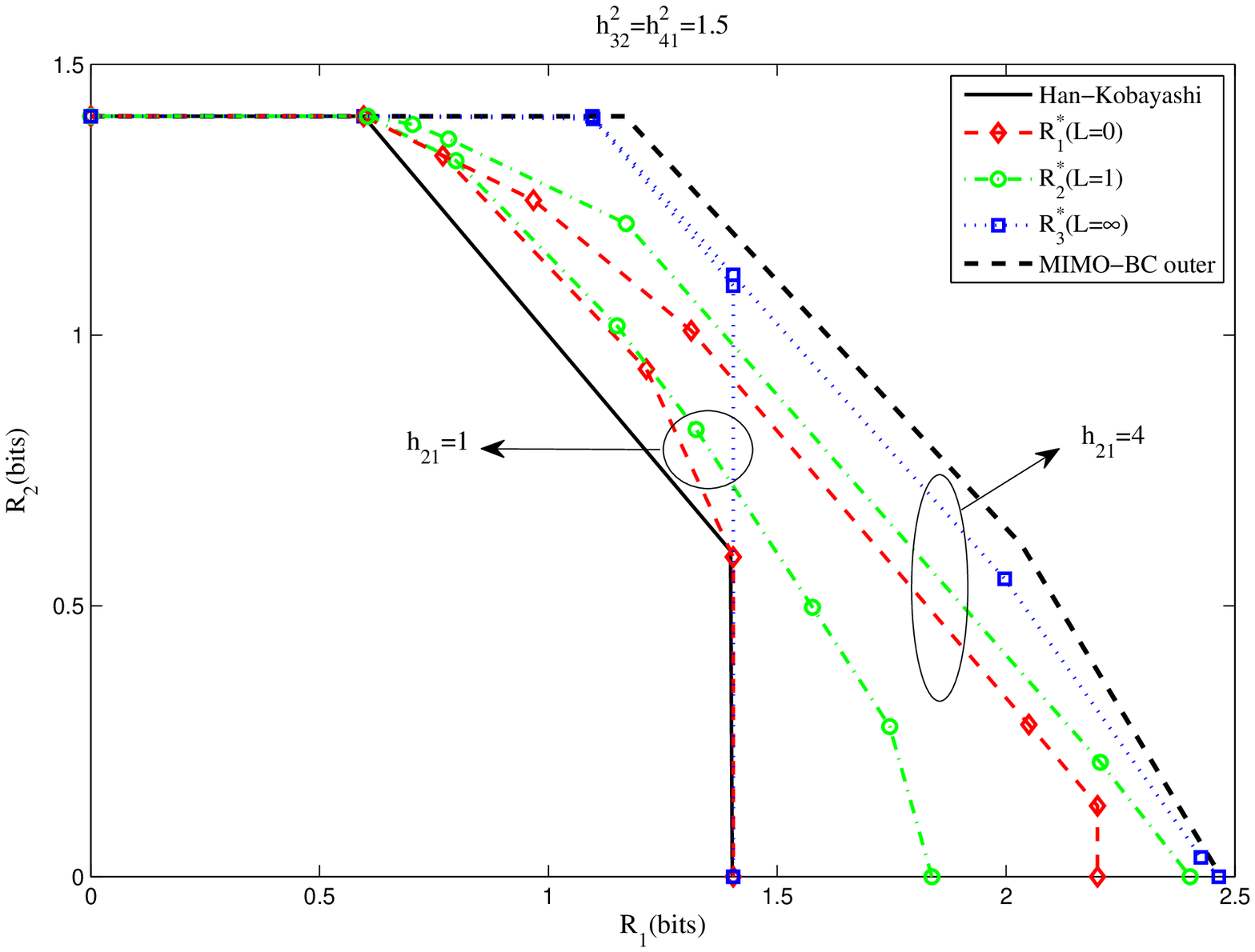}}%
\,
\begin{minipage}{8.3cm}%
\includegraphics[width=8cm]{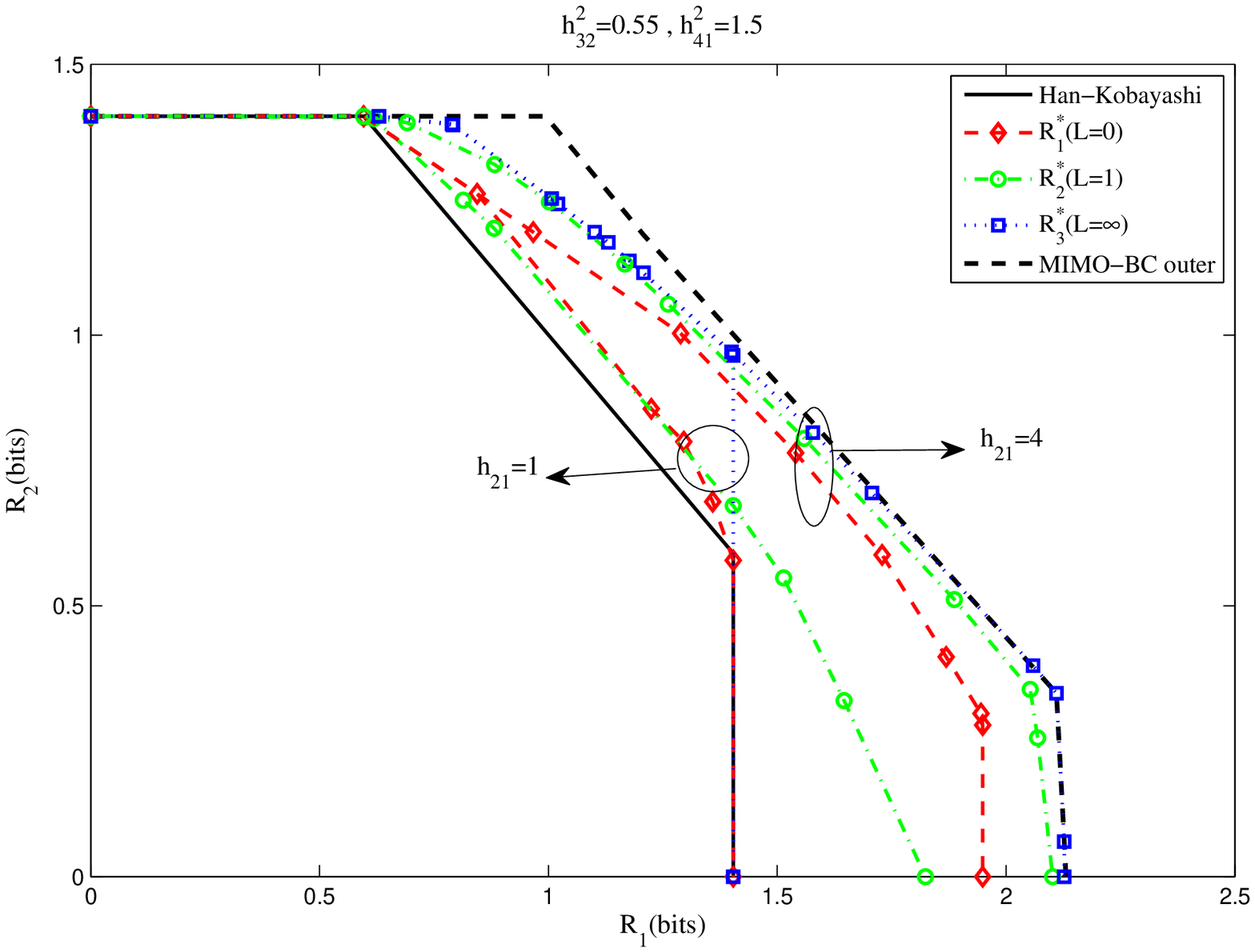}
\end{minipage}
\caption{Comparison among $\Rc_1^*$, $\Rc_2^*$, $\Rc_3^*$ and HK region \cite{HanKob81}. $P_1=P_2=6$, $h_{31}=h_{42}=1$, and $N_2=N_3=N_4=1$.}%
\label{fig:Gaus_Case4_5}%
\end{figure*}

Fig.~\ref{fig:Gaus_Case4_5} compares the regions $\Rc_1^*$, $\Rc_2^*$ and $\Rc_3^*$, with the HK region for $P_1=P_2=6$, $h_{31}=h_{42}=1$, $N_2=N_3=N_4=1$, and different values of $h_{32}$ and $h_{41}$, where results similar to those depicted in Fig.~\ref{fig:Gaus_Case1} can be concluded at the strong interference ($h_{32}=h_{41}=\sqrt{1.5}$) and the mixed interference ($h_{32}=\sqrt{0.55}$, $h_{41}=\sqrt{1.5}$) regimes.

\begin{figure}[tb]
  \centering
  \includegraphics[width=8cm]{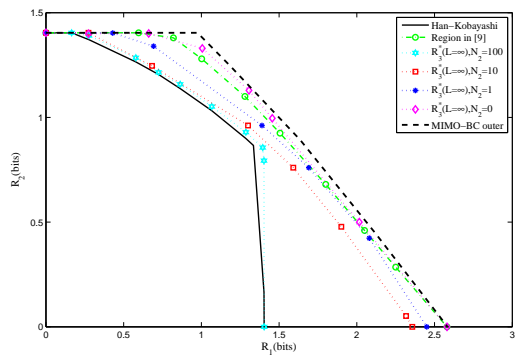}
  \caption{Comparison among $\Rc_3^*$ and the non-causal scheme of \cite{MariGolKraSha08} for different values of $N_2$. $P_1=P_2=6$, $h_{31}=h_{42}=1$, $h_{32}=\sqrt{2}$, $h_{41}=\sqrt{0.3}$, and $N_3=N_4=1$.}
  \label{fig:Gaus_Case6}
\end{figure}

In Fig.~\ref{fig:Gaus_Case6}, in order to investigate the effect of the noise in the channel between the transmitters (the cognitive link), we compare the region $\Rc_3^*$ for the unlimited look-ahead case ($L=\infty$) with the non-causal scheme of \cite{MariGolKraSha08} for $P_1=P_2=6$, $h_{31}=h_{42}=1$, $h_{32}=\sqrt{2}$, $h_{41}=\sqrt{0.3}$, $N_3=N_4=1$, and different values of $N_2$. We see that, when poor conditions exist for the cognitive link, i.e., $N_2=100$, one cannot gain very much using the strategy of $\Rc_3^*$ in comparison with the HK scheme. As $N_2$ decreases, the performance approaches the rates achieved in the non-causal scheme of \cite{MariGolKraSha08} as well as the outer bound. For $N_2=0$, our rate region outperforms that in \cite{MariGolKraSha08} in agreement with the discussion in part~1 of Remark~\ref{remark:compare_ndelayI}.

\section{Conclusion}\label{sec:conclusion}
We introduced the Causal Cognitive Interference Channel With Delay (CC-IFC-WD) and investigated its capacity region. We derived a general outer bound on the capacity region for an arbitrary value of $L$ and specialized it to the strong interference case. We tightened the outer bound under strong interference conditions. We also obtained achievable rate regions for three special cases: 1)~Classical CC-IFC, 2)~CC-IFC without delay, and
3)~CC-IFC with unlimited look-ahead. Coding schemes were based on the generalized block Markov superposition coding, rate splitting and Gel'fand-Pinsker (GP) binning. Moreover, instantaneous relaying and non-causal partial Decode-and-Forward (DF) were employed in the second and third cases, respectively. Furthermore, using the derived inner and outer bounds, we characterized the capacity regions for the classes of the degraded
and semi-deterministic classical CC-IFC under strong interference conditions. We showed that these channel models can be seen as a combination of the degraded or semi-deterministic relay channel with private message from the relay to the receiver and the MAC with common information.

Also, we investigated Gaussian CC-IFC-WD by extending our achievable rate regions to the Gaussian case and providing some numerical examples in order to examine the rate gain of the cognitive link. We compared different strategies which we have used in the coding schemes and showed that instantaneous relaying and non-causal DF improve the rate region noticeably and achieve rates close to the outer bound for a strong cognitive link, especially when the rate of the cognitive user is small. In addition, comparing the partial (causal or non-causal) DF, instantaneous relaying and DPC (GP binning) strategies, we attempted to identify the cases wherein each strategy is dominant. The results showed that when the cognitive user sends at higher rates interference cancelation by DPC is a better strategy. However, when the cognitive user sends at lower rates and can dedicate more power to cooperating with the primary user, DPC provides less improvement.

\appendices
\section{Proof of the outer bounds}\label{app:out_proof}
\begin{IEEEproof}[Proof of Theorem~\ref{thm:outer_gen}]
Consider a $(2^{nR_1},2^{nR_2},n,P_e^{(n)})$ code with an average error probability $P_e^{(n)}\rightarrow 0$, which implies that $P_{e,1}^{(n)}\rightarrow 0$ and $P_{e,2}^{(n)}\rightarrow 0$. Applying Fano's inequality \cite{CovTho06} results in
\begin{IEEEeqnarray}{rcl}
H(M_1|Y_3^n)\leq P_{e,1} log(2^{nR_1} - 1) + h(P_{e,1}^{(n)})\leq n\delta_{1n}\quad\label{eqn:Fano_deltaI}\\
H(M_2|Y_4^n)\leq P_{e,2} log(2^{nR_2} - 1) + h(P_{e,2}^{(n)})\leq n\delta_{2n}\quad\label{eqn:Fano_deltaII}
\end{IEEEeqnarray}
where $\delta_{un}\rightarrow 0$ as $P_{e,u}^{(n)}\rightarrow 0$ for $u\in\{1,2\}$. Now, the following RVs are defined for $i=1,\ldots,n$:
\begin{IEEEeqnarray}{c}
T_{i}=Y_2^{i-1}\label{eqn:RV_T}\\
V_{i}=(M_2,Y_2^{i-1})=(M_2,T_{i})\label{eqn:RV_V}\\
U_{i}=Y_{2,i+1}^{i+L-1}\label{eqn:RV_U}
\end{IEEEeqnarray}

Regarding the joint p.m.f (\ref{eqn:pmf}), we remark that $X_{1,i}\rightarrow T_i\rightarrow V_i$ forms a Markov chain. Moreover, it is noted that $X_{2,i}=f'_{2,i}(V_{i},U_{i},Y_{2,i})$. Thus, these choices of auxiliary RVs satisfy the p.m.f in Theorem~\ref{thm:outer_gen}. First, we provide some useful lemmas which we need in the proof of this theorem.

\begin{lemma}\label{lemma:MarkovI}
$(M_1,Y_u^{i-1})\rightarrow(X_{1,i},V_i,U_i)\rightarrow(Y_{u,i},Y_{2,i})$ forms a Markov chain, where $u\in \{3,4\}$.
\end{lemma}

\begin{IEEEproof} Noting (\ref{eqn:pmf}), consider $p(m_1,m_2,x_{1,i},y_u^{i},y_2^{i+L-1})$ which can be written as
\begin{IEEEeqnarray*}{l}
p(m_1,m_2,x_{1,i},y_u^{i},y_2^{i-1},y_{2,i+1}^{i+L-1},y_{2,i})\\
\stackrel{(a)}{=}p(m_1,x_{1,i},y_u^{i},v_i,u_i,y_{2,i})\\
=p(m_1,y_u^{i-1},x_{1,i},v_i,u_i)p(y_{u,i},y_{2,i}|m_1,y_u^{i-1},x_{1,i},v_i,u_i)\\
=p(m_1,y_u^{i-1},x_{1,i},v_i,u_i)p(y_{2,i}|m_1,y_u^{i-1},x_{1,i},v_i,u_i)\\
\quad\times p(y_{u,i}|m_1,y_u^{i-1},x_{1,i},v_i,u_i,y_{2,i})\\
\stackrel{(b)}{=}p(m_1,y_u^{i-1},x_{1,i},v_i,u_i)p(y_{2,i}|x_{1,i},v_i,u_i)\\
\quad\times p(y_{u,i}|m_1,y_u^{i-1},x_{1,i},v_i,u_i,y_{2,i},x_{2,i})\\
\stackrel{(c)}{=}p(m_1,y_u^{i-1},x_{1,i},v_i,u_i)p(y_{2,i}|x_{1,i},v_i,u_i)\\
\quad\times p(y_{u,i}|x_{1,i},v_i,u_i,y_{2,i},x_{2,i})\\
\stackrel{(d)}{=}p(m_1,y_u^{i-1},x_{1,i},v_i,u_i)p(y_{2,i}|x_{1,i},v_i,u_i)\\
\quad\times p(y_{u,i}|x_{1,i},v_i,u_i,y_{2,i})\\
=p(m_1,y_u^{i-1})p(x_{1,i},v_i,u_i|m_1,y_u^{i-1})p(y_{2,i},y_{u,i}|x_{1,i},v_i,u_i)
\end{IEEEeqnarray*}
where we use (\ref{eqn:RV_V}) and (\ref{eqn:RV_U}) for (a), and (b) to (d) follow from the joint p.m.f (\ref{eqn:pmf}) and the fact that $X_{2,i}$ is a deterministic function of $V_{i}$, $U_{i}$ and $Y_{2,i}$.
\end{IEEEproof}

\begin{lemma}\label{lemma:MarkovII}
For $u\in \{3,4\}$, $X_{2,i}\rightarrow(X_{1,i},V_i,U_i)\rightarrow(Y_{u,i},Y_{2,i})$ forms a Markov chain.
\end{lemma}

\begin{IEEEproof}
Note the joint p.m.f in (\ref{eqn:pmf}) and consider $p(m_2,x_{1,i},x_{2,i},y_2^{i+L-1},y_{u,i})$ which can be written as
\begin{IEEEeqnarray*}{l}
p(m_2,x_{1,i},x_{2,i},y_2^{i-1},y_{2,i+1}^{i+L-1},y_{2,i},y_{u,i})\\
\stackrel{(a)}{=}p(x_{1,i},x_{2,i},v_i,u_i,y_{2,i},y_{u,i})\\
=p(x_{2,i},x_{1,i},v_i,u_i)p(y_{2,i},y_{u,i}|x_{1,i},x_{2,i},v_i,u_i)\\
\stackrel{(b)}{=}p(x_{2,i},x_{1,i},v_i,u_i)p(y_{2,i}|x_{1,i},v_i,u_i)\\
\quad\times p(y_{u,i}|x_{1,i},x_{2,i},v_i,u_i,y_{2,i})\\
\stackrel{(c)}{=}p(x_{2,i},x_{1,i},v_i,u_i)p(y_{2,i}|x_{1,i},v_i,u_i)p(y_{u,i}|x_{1,i},v_i,u_i,y_{2,i})\\
=p(x_{2,i})p(x_{1,i},v_i,u_i|x_{2,i})p(y_{2,i},y_{u,i}|x_{1,i},v_i,u_i)
\end{IEEEeqnarray*}
where we use (\ref{eqn:RV_V}) and (\ref{eqn:RV_U}) for (a), (b) is due to the joint p.m.f given by (\ref{eqn:pmf}), and (c) follows from the fact that $X_{2,i}$ is a deterministic function of $V_{i}$, $U_{i}$ and $Y_{2,i}$.
\end{IEEEproof}

Now, using Fano's inequality, we derive the bounds in Theorem~\ref{thm:outer_gen}. For the first bound,
\begin{IEEEeqnarray*}{l}
nR_1=H(M_1)\stackrel{(a)}{=}H(M_1|M_2)\yesnumber\label{eqn:outer_fanoI_before}\\
\!\!=I(M_1;Y_3^n|M_2)+H(M_1|Y_3^n,M_2)\stackrel{(b)}{\leq} I(M_1;Y_3^n|M_2)+n\delta_{1n}
\end{IEEEeqnarray*}
where (a) follows since $M_1$ and $M_2$ are independent and (b) holds due to (\ref{eqn:Fano_deltaI}) and the fact that conditioning does not increase the entropy. Hence, we obtain
\begin{IEEEeqnarray*}{ll}
&nR_1- n\delta_{1n}\leq I(M_1;Y_3^n|M_2)\stackrel{(a)}{\leq} I(M_1;Y_3^n,Y_2^n|M_2)\\
&\stackrel{(b)}{=}I(M_1;Y_2^n|M_2)+I(M_1;Y_3^n|M_2,Y_2^n,X_2^n)\\
&\stackrel{(c)}{=}\sum\limits_{i=1}^{n}\Big\{I(M_1;Y_{2,i}|Y_2^{i-1},M_2)+\\
&\qquad\quad I(M_1;Y_{3,i}|Y_3^{i-1},M_2,Y_2^n,X_2^n)\Big\}\\
&\stackrel{(d)}{\leq}\sum\limits_{i=1}^{n}\Big\{I(M_1,X_{1,i};Y_{2,i}|Y_2^{i-1},M_2)+\\
&\qquad\quad I(M_1,X_{1,i};Y_{3,i}|Y_3^{i-1},M_2,Y_2^n,X_2^n)\Big\}\\
&\stackrel{(e)}{\leq}\sum\limits_{i=1}^{n}\Big\{H(Y_{2,i}|V_i,T_i)-H(Y_{2,i}|V_i,T_i,M_1,X_{1,i})\Big\}\\
&\:\:\:\:+\sum\limits_{i=1}^{n}\Big\{H(Y_{3,i}|X_{2,i},Y_{2,i},T_i)-\\
&\qquad\quad H(Y_{3,i}|M_1,X_{1,i},Y_3^{i-1},V_i,T_i,Y_{2,i}^n,X_2^n)\Big\}\\
&\stackrel{(f)}{=}\sum\limits_{i=1}^{n}\Big\{I(X_{1,i};Y_{2,i}|V_i,T_i)+I(X_{1,i};Y_{3,i}|X_{2,i},Y_{2,i},T_i)\Big\}\\
&\stackrel{(g)}{=}n\Big\{I(X_{1Q};Y_{2Q}|V_Q,T_Q,Q)+\\
&\qquad I(X_{1Q};Y_{3Q}|X_{2Q},Y_{2Q},T_Q,Q)\Big\}\\
&=n\Big\{I(X_{1};Y_{2}|V,T)+I(X_{1};Y_{3}|X_{2},Y_{2},T)\Big\}\yesnumber\label{eqn:outer_fanoI}
\end{IEEEeqnarray*}
where (a) and (d) are due to the non-negativity of mutual information, (b) is due to the fact that $X_{2}^n$ is a deterministic function of $M_2$ and $Y_2^n$, (c) is obtained from the chain rule, (e) follows from applying (\ref{eqn:RV_T}) and (\ref{eqn:RV_V}) and the fact that conditioning does not increase the entropy, (f) follows from the fact that the channel is memoryless with the joint p.m.f (\ref{eqn:pmf}), and (g) is obtained by using a standard time-sharing argument, where $Q$ is a time-sharing RV, independent of all other RVs and uniformly distributed over $\{1,2,\ldots,n\}$, and we define $X_{1Q}=X_1$, $X_{2Q}=X_2$, $Y_{2Q}=Y_2$, $Y_{3Q}=Y_3$, $V_{Q}=V$ and $(T_{Q},Q)=T$.

Now, as a result of applying Fano's inequality in (\ref{eqn:Fano_deltaII}) and the independence of the messages, we can bound $R_2$ as
\begin{IEEEeqnarray*}{ll}
nR_2- n\delta_{2n}&\leq I(M_2;Y_4^n|M_1)\\
&\stackrel{(a)}{\leq} I(M_2;Y_4^n,Y_2^n|M_1)\yesnumber\label{eqn:outer_fanoII}\\
&\stackrel{(b)}{=}\sum\limits_{i=1}^{n}I(M_2;Y_{4,i},Y_{2,i}|Y_4^{i-1},Y_2^{i-1},M_1)\\
&\stackrel{(c)}{=}\sum\limits_{i=1}^{n}I(M_2;Y_{4,i},Y_{2,i}|Y_4^{i-1},Y_2^{i-1},M_1,X_{1,i})\\
&\stackrel{(d)}{\leq}\sum\limits_{i=1}^{n}\Big\{I(M_2,Y_2^{i-1};Y_{4,i}|Y_4^{i-1},Y_2^{i-1},M_1,X_{1,i})+\\
&\qquad I(M_2;Y_{2,i}|Y_4^{i},Y_2^{i-1},M_1,X_{1,i})\Big\}\\
&\stackrel{(e)}{=}\sum\limits_{i=1}^{n}I(M_2,Y_2^{i-1};Y_{4,i}|Y_4^{i-1},M_1,Y_2^{i-1},X_{1,i})\\
&\stackrel{(f)}{=}\sum\limits_{i=1}^{n}I(V_{i};Y_{4,i}|X_{1,i},T_i)\\
&\stackrel{(g)}{=}n I(V_{Q};Y_{4Q}|X_{1Q},T_Q,Q)=n I(V;Y_{4}|X_1,T)
\end{IEEEeqnarray*}
where (a) and (d) are obtained from the non-negativity of the mutual information, (b) is based on the chain rule, (c) obtains since $X_{1,i}$ is a deterministic function of $M_1$, (e) holds since the channel is memoryless with the joint p.m.f (\ref{eqn:pmf}), (f) follows from (\ref{eqn:RV_T}), (\ref{eqn:RV_V}), and Lemma \ref{lemma:MarkovI} for $u=4$, and for (g) we use the time-sharing argument of (\ref{eqn:outer_fanoI}-g) and $Y_{4Q}=Y_4$.

Now, let $Y'_3$ be any RV with the same marginal distribution of $Y_3$, i.e., $p(y_2^n,y_3^{'n}|x_1^n,x_2^n)=p(y_2^n,y_3^n|x_1^n,x_2^n)$, but with an arbitrary joint distribution $p(y_3^{'n},y_4^n|x_1^n,x_2^n)$. Subsequently, the second bound on $R_2$ can be derived as
\begin{IEEEeqnarray*}{ll}
&nR_2- n\delta_{2n}\leq I(M_2;Y_4^n|M_1)\\
&\stackrel{(a)}{\leq}I(M_2,X_2^n;Y_3^{'n},Y_4^n|M_1,X_1^n)\yesnumber\label{eqn:outer_fanoIII}\\
&{\leq}I(M_2,X_2^n;Y_3^{'n},Y_2^n|M_1,X_1^n)\!+\!I(M_2,X_2^n;Y_4^n|M_1,X_1^n,Y_3^{'n})\\
&\stackrel{(b)}{=}\sum\limits_{i=1}^{n}I(M_2,X_2^n;Y'_{3,i}|Y_3^{'i-1},Y_2^{i-1},M_1,X_1^n)+\\
&\quad I(M_2,X_2^n;Y_{4,i}|Y_3^{'n},Y_4^{i-1},M_1,X_1^n)\\
&\stackrel{(c)}{\leq}\sum\limits_{i=1}^{n}H(Y_{3,i}|X_{1,i},Y_2^{i-1})+H(Y_{4,i}|Y'_{3,i},X_{1,i})\\
&\quad\qquad-H(Y_{3,i}|X_{1,i},Y_2^{i-1},M_2,Y_{2,i+1}^{i+L-1},X_{2,i})\\
&\quad\qquad-H(Y_{4,i}|Y'_{3,i},X_{1,i},M_2,Y_2^{i-1},Y_{2,i+1}^{i+L-1},X_{2,i})\\
&\stackrel{(d)}{=}\sum\limits_{i=1}^{n}I(U_{i},V_{i};Y_{3,i}|X_{1,i},T_i)+I(U_{i},V_{i},T_i;Y_{4,i}|Y'_{3,i},X_{1,i})\\
&{=}n I(U_{Q},V_{Q};Y_{3Q}|X_{1Q},T_Q,Q)+\\
&\quad\qquad n I(U_{Q},V_{Q},T_Q;Y_{4Q}|Y'_{3Q},X_{1Q},Q)\\
&{\leq}n I(U_{Q},V_{Q};Y_{3Q}|X_{1Q},T_Q,Q)+\\
&\quad\qquad n I(U_{Q},V_{Q},T_Q,Q;Y_{4Q}|Y'_{3Q},X_{1Q})\\
&\stackrel{(e)}{=}n I(U,V;Y_{3}|X_1,T)+n I(U,V,T;Y_{4}|Y'_{3},X_1)
\end{IEEEeqnarray*}
where (a) is based on the facts that $X_{1}^n$ is a deterministic function of $M_1$ and mutual information is non-negative, (b) is obtained from the chain rule and the memoryless property of the channel with the joint p.m.f (\ref{eqn:pmf}) , (c) is true due to the memoryless property of the channel, the definition of $Y'_3$, and the fact that conditioning does not increase the entropy, (d) follows from (\ref{eqn:RV_T})-(\ref{eqn:RV_U}) and Lemma~\ref{lemma:MarkovII} for $u=3$, and for (e) we use the defined time-sharing argument and $Y'_{3Q}=Y'_3$, $U_{Q}=U$.

Next, we bound $R_1+R_2$ as
\begin{IEEEeqnarray*}{ll}
&n(R_1+R_2)- n(\delta_{1n}+\delta_{2n})\leq I(M_1;Y_3^n)+I(M_2;Y_4^n|M_1)\\
&{\leq}I(M_1;Y_3^n)+I(M_2;Y_3^{'n},Y_4^n|M_1)\yesnumber\label{eqn:outer_fanoIV}\\
&{=}H(Y_3^n)-H(Y_3^n|M_1)+H(Y_3^{'n}|M_1)\\
&\qquad -H(Y_3^{'n}|M_1,M_2)+I(M_2;Y_4^n|M_1,Y_3^{'n})\\
&\stackrel{(a)}{\leq} H(Y_3^n)-H(Y_3^{'n}|M_1,M_2,X_1^n,X_2^n)+\\
&\qquad I(M_2,X_2^n;Y_4^n|M_1,X_1^n,Y_3^{'n})\\
&\stackrel{(b)}{\leq} I(X_1^n,X_2^n;Y_3^n)+I(M_2,X_2^n;Y_4^n|M_1,X_1^n,Y_3^{'n})\\
&\stackrel{(c)}{\leq}\sum\limits_{i=1}^{n}I(X_{1,i},X_{2,i};Y_{3,i})+I(U_{i},V_{i},T_i;Y_{4,i}|Y'_{3,i},X_{1,i})\\
&\stackrel{(d)}{\leq}\sum\limits_{i=1}^{n}I(X_{1,i},U_i,V_i,T_i;Y_{3,i})+I(U_{i},V_{i},T_i;Y_{4,i}|Y'_{3,i},X_{1,i})\\
&=n I(X_{1Q},U_Q,V_{Q},T_Q;Y_{3Q}|Q)+\\
&\quad\qquad nI(U_{Q},V_{Q},T_Q;Y_{4Q}|Y'_{3Q},X_{1Q},Q)\\
&\leq n I(X_{1Q},U_Q,V_{Q},T_Q,Q;Y_{3Q})+\\
&\quad\qquad nI(U_{Q},V_{Q},T_Q,Q;Y_{4Q}|Y'_{3Q},X_{1Q})\\
&=n I(X_{1},U,V,T;Y_{3})+n I(U,V,T;Y_{4}|Y'_{3},X_1)
\end{IEEEeqnarray*}
where (a) follows from the definition of $Y'_3$ and the fact that conditioning does not increase the entropy, (b) is true due to the memoryless property of the channel and the definition of $Y'_3$, (c) follows from the steps (b) to (d) in (\ref{eqn:outer_fanoIII}) and the fact that the channel is memoryless, and (d) follows from Lemma~\ref{lemma:MarkovII} for $u=3$ and the fact that mutual information is non-negative. This completes the proof.
\end{IEEEproof}

\begin{IEEEproof}[Proof of Theorem~\ref{thm:outer_str2}]
The bounds in (\ref{eqn:outer_str2I}) and (\ref{eqn:outer_str2II}) and the first bound in (\ref{eqn:outer_str2III}) follow from (\ref{eqn:outer_str1I})-(\ref{eqn:outer_str1III}). Therefore, we need to prove the second sum-rate bound in (\ref{eqn:outer_str2III}). Consider a code with the properties of that in the proof of Theorem~\ref{thm:outer_gen}. First, we state the following lemma:
\begin{lemma}\label{lemma:cond_str^n}
If (\ref{eqn:cond_str_rec2}) holds, then
\begin{IEEEeqnarray}{rcl}
I(X_1^n;Y_3^n)&\leq& I(X_1^n;Y_4^n).\label{eqn:cond_str_rec2^n}
\end{IEEEeqnarray}
\end{lemma}
\begin{IEEEproof}
The proof relies on the result in \cite[Proposition~1]{KorMar77} and follows the same lines as in \cite[Lemma~5]{MariYatKra07} and \cite[Lemma]{CosElg79}.
\end{IEEEproof}

Before proceeding to bound the sum-rate, we need to state the following inequalities:
\begin{IEEEeqnarray*}{ll}
I(M_1;Y_3^n)&\stackrel{(a)}{=} I(M_1,X_1^n;Y_3^n)\yesnumber\label{eqn:fanoIV_bef1}\\
&\stackrel{(b)}{=} I(X_1^n;Y_3^n)+H(M_1|X_1^n)-H(M_1|X_1^n,Y_3^n)\\
&\stackrel{(c)}{=} I(X_1^n;Y_3^n)-H(M_1|X_1^n,Y_3^n)\stackrel{(d)}{\leq}I(X_1^n;Y_3^n)
\end{IEEEeqnarray*}
where (a) and (c) follow from the deterministic relation between
$X_1^n$ and $M_1$, (b) is due to the chain rule, and (d) holds due
to the non-negativity of the entropy.
\begin{IEEEeqnarray*}{l}
I(M_2;Y_4^n|M_1){\leq}I(M_2,X_2^n;Y_4^n|M_1,X_1^n)\yesnumber\label{eqn:fanoIV_bef2}\\
\qquad=H(Y_4^n|M_1,X_1^n)-H(Y_4^n|M_1,X_1^n,M_2,X_2^n)\\
\qquad\stackrel{(a)}{\leq} H(Y_4^n|X_1^n)-H(Y_4^n|X_1^n,X_2^n)=I(X_2^n;Y_4^n|X_1^n)
\end{IEEEeqnarray*}
where (a) is based on the facts that conditioning does not increase the entropy and $(M_1,M_2)\rightarrow (X_{1},X_{2})\rightarrow Y_4$ forms a Markov chain.

\begin{figure*}[!t]
\normalsize
\setcounter{tempequationcounter}{\value{equation}}
\begin{itemize}
\item For the first block, $b=1$:{\small
\begin{IEEEeqnarray}{l}
\setcounter{equation}{138}
E_{enc2,1,m'',n''}=\left\{ \nexists \:(m'',n''):\hspace{-5pt}
\begin{array}{c}
\big(u_{2c}^n([1,m''],0),u_{2p}^n([1,n''],0),t_p^n(0,0),t_c^n(0),q^n\big)\in A_\epsilon^n(U_{2c},U_{2p},T_p,T_c,Q)
\end{array}\hspace{-5pt}\right\}\label{eqn:error_event_befI}\\
E_{dec2,1,i'',j''}=\left\{\hspace{-5pt}
\begin{array}{c}
\big(y_2^n(1),u_{1p}^n(j'',i'',0,0),u_{1c}^n(i'',0),u_{2c}^n([1,l_{2c,b}],0),u_{2p}^n([1,l_{2p,b}],0),t_p^n(0,0),t_c^n(0),q^n\big)\\
\in A_\epsilon^n(Y_2,U_{1p},U_{1c},U_{2c},U_{2p},T_p,T_c,Q)
\end{array}\hspace{-5pt}\right\}\\
E_{dec3,1,k,l,m,n}=\left\{\hspace{-5pt}
\begin{array}{c}
\big(y_3^n(1),u_{2c}^n([n,m],0),v_{1p}^n(l,k,0,0),v_{1c}^n(k,i),u_{1p}^n(1,1,0,0),u_{1c}^n(1,0),t_p^n(0,0),t_c^n(0), q^n\big)\\
\in A_\epsilon^n(Y_3,U_{2c},V_{1p},V_{1c},U_{1p},U_{1c},T_p,T_c,Q)
\end{array}\hspace{-5pt}\right\}\\
E_{dec4,1,i',j',l',m',n'}=\left\{\hspace{-5pt}
\begin{array}{c}
\big(y_4^n(1),u_{2c}^n([i',m'],0),u_{2p}^n([j',n'],0),v_{1c}^n(l',0),u_{1c}^n(1,0),t_c^n(0),q^n\big)\\
\in A_\epsilon^n(Y_4,U_{2c},U_{2p},V_{1c},U_{1c},T_c,Q)
\end{array}\hspace{-5pt}\right\}
\end{IEEEeqnarray}}

\item For block $b=2,\ldots,B-1$:{\small
\begin{IEEEeqnarray}{l}
E_{enc2,b,m'',n''}=\left\{ \nexists \:(m'',n''):\hspace{-5pt}
\begin{array}{c}
\big(u_{2c}^n([1,m''],1),u_{2p}^n([1,n''],1),t_p^n(1,1),t_c^n(1),q^n\big)\in A_\epsilon^n(U_{2c},U_{2p},T_p,T_c,Q)
\end{array}\hspace{-5pt}\right\}\label{eqn:error_event_I}\\
E_{dec2,b,i'',j''}=\left\{\hspace{-5pt}
\begin{array}{c}
\big(y_2^n(b),u_{1p}^n(j'',i'',1,1),u_{1c}^n(i'',1),u_{2c}^n([1,l_{2c,b}],1),u_{2p}^n([1,l_{2p,b}],1),t_p^n(1,1),t_c^n(1),q^n\big)\\
\in A_\epsilon^n(Y_2,U_{1p},U_{1c},U_{2c},U_{2p},T_p,T_c,Q)
\end{array}\hspace{-5pt}\right\}\label{eqn:error_event_II}\\
E_{dec3,b,i,j,k,l,m,n}=\left\{\hspace{-5pt}
\begin{array}{c}
\big(y_3^n(b),u_{2c}^n([n,m],i),v_{1p}^n(l,k,j,i),v_{1c}^n(k,i),u_{1p}^n(1,1,j,i),u_{1c}^n(1,i),t_p^n(j,i),t_c^n(i), q^n\big)\\
\in A_\epsilon^n(Y_3,U_{2c},V_{1p},V_{1c},U_{1p},U_{1c},T_p,T_c,Q)
\end{array}\hspace{-5pt}\right\}\label{eqn:error_event_III}\\
E_{dec4,b,i',j',k',l',m',n'}=\left\{\hspace{-5pt}
\begin{array}{c}
\big(y_4^n(b),u_{2c}^n([i',m'],k'),u_{2p}^n([j',n'],k'),v_{1c}^n(l',k'),u_{1c}^n(1,k'),t_c^n(k'),q^n\big)\\
\in A_\epsilon^n(Y_4,U_{2c},U_{2p},V_{1c},U_{1c},T_c,Q)
\end{array}\hspace{-5pt}\right\}\label{eqn:error_event_IV}
\end{IEEEeqnarray}}

\item For the last block, $b=B$: $E_{enc2,B,m'',n''}$, is the same as (\ref{eqn:error_event_I}), $E_{dec3,B,i,j,k,l,m,n}$ and $E_{dec4,B,i',j',k',l',m',n'}$ are given by (\ref{eqn:error_event_III}) and (\ref{eqn:error_event_IV}) with $b=B$, respectively.
\end{itemize}
\setcounter{equation}{\value{tempequationcounter}}
\hrulefill
\vspace*{4pt}
\end{figure*}
Now, the second bound in (\ref{eqn:outer_str2III}) can be obtained as
\begin{IEEEeqnarray*}{ll}
&n(R_1+R_2)- n(\delta_{1n}+\delta_{2n})\leq I(M_1;Y_3^n)+I(M_2;Y_4^n|M_1)\\
&\stackrel{(a)}{\leq} I(X_1^n;Y_3^n)+I(X_2^n;Y_4^n|X_1^n)\yesnumber\label{eqn:outer_fanoV}\\
&\stackrel{(b)}{\leq} I(X_1^n;Y_4^n)+I(X_2^n;Y_4^n|X_1^n)=I(X_1^n,X_2^n;Y_4^n)\\
&\leq\sum\limits_{i=1}^{n}I(X_{1,i},X_{2,i};Y_{4,i})\leq\sum\limits_{i=1}^{n}I(X_{1,i},X_{2,i},U_i,V_i,T_i;Y_{4,i})\\
&\stackrel{(c)}{=}\sum\limits_{i=1}^{n}I(X_{1,i},U_i,V_i,T_i;Y_{4,i})=n I(X_{1Q},U_Q,V_{Q},T_Q;Y_{4Q}|Q)\\
&\leq n I(X_{1Q},U_Q,V_{Q},T_Q,Q;Y_{4Q})=n I(X_{1},U,V,T;Y_{4})
\end{IEEEeqnarray*}
where (a) follows from (\ref{eqn:fanoIV_bef1}) and (\ref{eqn:fanoIV_bef2}), (b) from condition (\ref{eqn:cond_str_rec2^n}), and (c) from Lemma~\ref{lemma:MarkovII} for $u=4$. This completes the proof.
\end{IEEEproof}
\section{Analysis of the Probability of Error for Theorem~\ref{thm:zerodelay}}\label{app:error}
\addtocounter{equation}{8}%
Due to the symmetry of the random codebook generation, the probability of error is independent of the specific messages. Hence, without loss of generality, we assume that the message tuples $m_{1,b}=(m_{1cd,b},m_{1cn,b},m_{1pd,b},m_{1pn,b})=(1,1,1,1)$ and $m_{2,b}=(m_{2c,b},m_{2p,b})=(1,1)$ are encoded and transmitted in each block $b, b=1,\ldots,B$. Recall that, in the first block the cooperative information is defined as: $(m_{1pd,b-1},m_{1cd,b-1})=(m_{1pd,0},m_{1cd,0})=(0,0)$ and in the last block, a previously known message $(m_{1pd,b},m_{1cd,b})=(m_{1pd,B},m_{1cd,B})=(1,1)$ is transmitted. Furthermore, backward decoding is utilized at Rx1 and Rx2. Consider the events \eqref{eqn:error_event_befI}-\eqref{eqn:error_event_IV} at the top of this page.

Moreover, we define $\Fc_{b-1}$ to be the event in which no errors have occurred up to block $b$. Note that, in Rx1 and Rx2, up to block $b$ means blocks $b+1,\ldots,B$, due to backward decoding. We can write the overall probability of error as
\begin{IEEEeqnarray*}{l}
P_{e}=Pr\Bigg[\:\bigcup\limits_{b=1}^B\Big(\bigcup_{\left(m''\in[1,2^{nL_{2c}}],n''\in[1,2^{nL_{2p}}]\right)}E_{enc2,b,m'',n''}\Big)\\
\quad\qquad\cup\bigcup\limits_{b=1}^{B-1}\Big(E_{dec2,b,1,1}^c\cup\bigcup\limits_{(i'',j'')\neq(1,1)}E_{dec2,b,i'',j''}\Big)\\
\cup\bigcup\limits_{b=1}^{B}\Big(E_{dec3,b,1,1,1,1,l_{2c,b},1}^c\cup\bigcup\limits_{(i,j,k,l)\neq(1,1,1,1)}E_{dec3,b,i,j,k,l,m,n}\Big)\\
\cup\bigcup\limits_{b=1}^{B}\Big(E_{dec4,b,1,1,1,1,l_{2c,b},l_{2p,b}}^c\cup\bigcup\limits_{(i',j')\neq
(1,1)}E_{dec4,b,i',j',k',l',m',n'}\Big)\Bigg]\\
\leq\sum\limits_{b=1}^BPr\left(E_{enc2,b}|\Fc_{b-1}\right)+\sum\limits_{b=1}^{B-1}\left(E_{dec2,b}|E_{enc2,b}^c,\Fc_{b-1}\right)\\
\quad+\sum\limits_{b=1}^{B}\left(E_{dec3,b}|E_{enc2,b}^c,\Fc_{b-1}\right)+\sum\limits_{b=1}^{B}\left(E_{dec4,b}|E_{enc2,b}^c,\Fc_{b-1}\right)
\end{IEEEeqnarray*}
where we define
\begin{IEEEeqnarray}{l}
E_{enc2,b}=\bigcup_{\left(m''\in[1,2^{nL_{2c}}],n''\in[1,2^{nL_{2p}}]\right)}E_{enc2,b,m'',n''}\qquad\\
E_{dec2,b}=E_{dec2,b,1,1}^c\cup \bigcup\limits_{(i'',j'')\neq (1,1)}E_{dec2,b,i'',j''}\\
E_{dec3,b}=E_{dec3,b,1,1,1,1,l_{2c,b},1}^c\\
\qquad\qquad\cup\quad\bigcup\limits_{(i,j,k,l)\neq (1,1,1,1)}E_{dec3,b,i,j,k,l,m,n}\\
E_{dec4,b}=E_{dec4,b,1,1,1,1,l_{2c,b},l_{2p,b}}^c\\
\qquad\qquad\cup\quad\bigcup\limits_{(i',j')\neq (1,1)}E_{dec4,b,i',j',k',l',m',n'}
\end{IEEEeqnarray}
and $E^c$ denotes the complement of the event $E$.

Hence, assuming that no errors have occurred up to block $b$, bounding the probability of encoding or decoding error in block $b$ for each user is sufficient for bounding the overall probability of error.

First, we bound the probability of encoding error for the cognitive user (Tx2) at the beginning of block $b$, defined as $P_{e,enc2,b}$:
\begin{IEEEeqnarray*}{rl}
P_{e,enc2,b}&=Pr\left(E_{enc2,b}\right)\\
&=Pr\Big(\bigcup_{\left(m''\in[1,2^{nL_{2c}}],n''\in[1,2^{nL_{2p}}]\right)}E_{enc2,b,m'',n''}\Big)
\end{IEEEeqnarray*}

Using mutual covering lemma \cite{ElgKim10, ElgVan81}, $P_{e,enc2,b}\rightarrow 0$ if $n\rightarrow\infty$ and (\ref{eqn:zerodelay_Tx2_enc_begin})-(\ref{eqn:zerodelay_Tx2_enc_end}) hold.

Next, we bound the probability of decoding error for the cognitive user (Tx2) at the end of block $b$, defined as $P_{e,dec2,b}$:{\small
\begin{IEEEeqnarray}{l}
P_{e,dec2,b}{=}Pr\left(E_{dec2,b}|E_{enc2,b}^c\right)\nonumber\\
=Pr\Big(E_{dec2,b,1,1}^c\cup\bigcup\limits_{(i'',j'')\neq(1,1)}E_{dec2,b,i'',j''}|E_{enc2,b}^c\Big)\nonumber\\
{\leq}Pr(E_{dec2,b,1,1}^c|E_{enc2,b}^c){+}\sum\limits_{i''\neq1}Pr(E_{dec2,b,i'',1}|E_{enc2,b}^c)\label{eqn:error_Tx2_dec_first}\\
{+}\sum\limits_{j''\neq1}(E_{dec2,b,1,j''}|E_{enc2,b}^c){+}\sum\limits_{i''\neq1,j''\neq1}(E_{dec2,b,i'',j''}|E_{enc2,b}^c)\IEEEeqnarraynumspace\label{eqn:error_Tx2_dec_first2}
\end{IEEEeqnarray}}
Due to the asymptotic equipartition property (AEP) \cite{CovTho06} and considering the codebook generation of Theorem \ref{thm:zerodelay}, $Pr\left(E_{dec2,b,1,1}^c|E_{enc2,b}^c\right)\rightarrow 0$ as $n\rightarrow\infty$. Utilizing \cite[Theorem 15.2.3]{CovTho06} for the other terms in (\ref{eqn:error_Tx2_dec_first}) and (\ref{eqn:error_Tx2_dec_first2}), we have
\begin{IEEEeqnarray}{rcl}
P_{e,dec2,b}&{\leq}&\epsilon +
2^{nR_{1cd}}2^{-n(I(U_{1c}U_{1p};Y_2|U_{2c}U_{2p}T_pT_cQ)-6\epsilon)}\label{eqn:error_Tx2_dec_second}\\
&&+2^{R_{1pd}}2^{-n(I(U_{1p};Y_2|U_{2c}U_{2p}U_{1c}T_pT_cQ)-6\epsilon)}\\
&&+2^{(R_{1cd}+R_{1pd})}2^{-n(I(U_{1c}U_{1p};Y_2|U_{2c}U_{2p}T_pT_cQ)-6\epsilon)}\qquad
\end{IEEEeqnarray}

Now, it can easily be shown that when (\ref{eqn:zerodelay_Tx2_dec_begin}) and (\ref{eqn:zerodelay_Tx2_dec_end}) hold, $P_{e,dec2,b}$ tends to zero as $n\rightarrow\infty$. Note that the second term in the right side of (\ref{eqn:error_Tx2_dec_second}) imposes no constraint on $R_{1cd}$, because the events of the second terms in the right side of (\ref{eqn:error_Tx2_dec_first}) and (\ref{eqn:error_Tx2_dec_first2}) share the same p.m.f.

In a similar manner, the probability of the decoding error for Rx1 at the end of block $b$ (defined as $P_{e,dec3,b}$) can be bounded as
\begin{IEEEeqnarray}{l}
P_{e,dec3,b}{=}Pr\left(E_{dec3,b}|E_{enc2,b}^c\right)\nonumber\\
=Pr\Big(E_{dec3,b,1,1,1,1,l_{2c,b},1}^c\cup\nonumber\\
\qquad\qquad\qquad\bigcup\limits_{(i,j,k,l)\neq(1,1,1,1)}E_{dec3,b,i,j,k,l,m,n}|E_{enc2,b}^c\Big)\nonumber\\
{\leq}Pr\left(E_{dec3,b,1,1,1,1,l_{2c,b},1}^c|E_{enc2,b}^c\right)\nonumber\\
\qquad{+}\sum\limits_{(i,j,k,l)\neq(1,1,1,1)}Pr\left(E_{dec3,b,i,j,k,l,m,n}|E_{enc2,b}^c\right)\nonumber\\
{\leq}\epsilon + \sum\limits_{(i,j,k,l)\neq
(1,1,1,1)}Pr\left(E_{dec3,b,i,j,k,l,m,n}|E_{enc2,b}^c\right)\label{eqn:error_Rx1_dec_first}
\end{IEEEeqnarray}

For the second term in the right side of (\ref{eqn:error_Rx1_dec_first}), there are sixty cases that cause an error. However, some of these cases share the same p.m.f and so there are only nine distinct cases. Now, using the packing lemma \cite{ElgKim10} (or \cite[Theorem 15.2.3]{CovTho06}), we bound the probability of these events (conditioning on $E_{enc2,b}^c$ suppressed). Note that in the following, when the value of an index is unspecified, e.g., $i$, that index can take any value from its set, e.g., $i=1$ or $i\neq1$. First, consider
\begin{IEEEeqnarray*}{l}
Pr(E_{dec3,b,1,1,1,l\neq1,l_{2c,b},1})\\
=\sum\limits_{(y_3^n,u_{2c}^n,v_{1p}^n,v_{1c}^n,u_{1p}^n,u_{1c}^n,t_p^n,t_c^n,q^n)\in A_\epsilon^n}
p(v_{1p}^n|v_{1c}^n,t_p^n,t_c^n,q^n)\\
\qquad\qquad\qquad\qquad\times p(y_3^n,u_{2c}^n,v_{1c}^n,u_{1p}^n,u_{1c}^n,t_p^n,t_c^n,q^n)\\
{\leq}2^{-n(I(V_{1p};Y_3|U_{2c}V_{1c}U_{1p}U_{1c}T_pT_cQ)-6\epsilon)}\yesnumber\label{eqn:error_Rx1_dec_secondI}
\end{IEEEeqnarray*}

Similarly, $E_{dec3,b,i\neq 1,j,k,l,m,n}$ obtains
\begin{IEEEeqnarray*}{rl}
Pr&(E_{dec3,b,i\neq 1,j,k,l,m,n})\\
&=\sum\limits_{(y_3^n,u_{2c}^n,v_{1p}^n,v_{1c}^n,u_{1p}^n,u_{1c}^n,t_p^n,t_c^n,q^n)\in A_\epsilon^n}p(t_p^n,t_c^n,q^n)\\
&\qquad\times p(v_{1p}^n,v_{1c}^nu_{1p}^n,u_{1c}^n|t_p^n,t_c^n,q^n)p(u_{2c}^n|t_c^n,q^n)p(y_3^n|q^n)\\
&{\leq}2^{-n(I(U_{2c}V_{1p}V_{1c}U_{1p}U_{1c}T_pT_c;Y_3|Q)+I(U_{2c};T_p|T_cQ)-6\epsilon)}\quad\yesnumber\label{eqn:error_Rx1_dec_secondII}
\end{IEEEeqnarray*}
\addtocounter{equation}{2}
The probabilities of the other error events in (\ref{eqn:error_Rx1_dec_first}) can be bounded in a similar manner to obtain the bounds in \eqref{eqn:error_Rx1_dec_secondIII}.
\begin{figure*}[!t]
\normalsize
\setcounter{tempequationcounter}{\value{equation}}
\begin{IEEEeqnarray*}{rcl}
\setcounter{equation}{160}
Pr\left(E_{dec3,b,1,j\neq1,k\neq1,l,l_{2c,b},1}\right)&{\leq}&2^{-n(I(V_{1p}V_{1c}U_{1p}T_p;Y_3U_{2c}|U_{1c}T_cQ)-6\epsilon)}\\
Pr\left(E_{dec3,b,1,j\neq1,1,l,l_{2c,b},1}\right)&{\leq}&2^{-n(I(V_{1p}U_{1p}T_p;Y_3U_{2c}|V_{1c}U_{1c}T_cQ)-6\epsilon)}\\
Pr\left(E_{dec3,b,1,j\neq1,1,l,(m,n)\neq(l_{2c,b},1)}\right)&{\leq}&2^{-n(I(U_{2c}V_{1p}U_{1p}T_p;Y_3|V_{1c}U_{1c}T_cQ)+I(U_{2c};T_p|T_cQ)-6\epsilon)}\\
Pr\left(E_{dec3,b,1,1,k\neq1,l,l_{2c,b},1}\right)&{\leq}&2^{-n(I(V_{1c}V_{1p};Y_3|U_{2c}U_{1p}U_{1c}T_pT_cQ)-6\epsilon)}\yesnumber\label{eqn:error_Rx1_dec_secondIII}\\
Pr\left(E_{dec3,b,1,1,1,l\neq1,(m,n)\neq(l_{2c,b},1)}\right)&{\leq}&2^{-n(I(V_{1p}U_{2c};Y_3|V_{1c}U_{1p}U_{1c}T_pT_cQ)+I(U_{2c};T_p|T_cQ)-6\epsilon)}\\
Pr\left(E_{dec3,b,1,j\neq1,k\neq1,l,(m,n)\neq(l_{2c,b},1)}\right)&{\leq}&2^{-n(I(U_{2c}V_{1p}V_{1c}U_{1p}T_p;Y_3|U_{1c}T_cQ)+I(U_{2c};T_p|T_cQ)-6\epsilon)}\\
Pr\left(E_{dec3,b,1,j,k\neq1,l,(m,n)\neq(l_{2c,b},1)}\right)&{\leq}&2^{-n(I(U_{2c}V_{1p}V_{1c};Y_3|U_{1p}U_{1c}T_pT_cQ)+I(U_{2c};T_p|T_cQ)-6\epsilon)}
\end{IEEEeqnarray*}
\begin{IEEEeqnarray*}{rcl}
Pr\left(E_{dec4,b,i',1,1,1,m',l_{2p,b}}|(i',m')\neq(1,l_{2c,b})\right)&{\leq}&2^{-n(I(U_{2c};Y_4U_{2p}|V_{1c}U_{1c}T_cQ)-6\epsilon)}\\
Pr\left(E_{dec4,b,1,j',1,1,l_{2c,b},n'}|(j',n')\neq(1,l_{2p,b})\right)&{\leq}&2^{-n(I(U_{2p};Y_4U_{2c}|V_{1c}U_{1c}T_cQ)-6\epsilon)}\\
Pr\left(E_{dec4,b,i',j',k',l',m',n'}\right|k'\neq1)&{\leq}&2^{-n(I(U_{2c}U_{2p}V_{1c}U_{1c}T_c;Y_4|Q)-6\epsilon)}\yesnumber\label{eqn:error_Rx2_dec_second}\\
Pr\left(E_{dec4,b,i',1,1,l'\neq1,m',l_{2p,b}}|(i',m')\neq(1,l_{2c,b})\right)&{\leq}&2^{-n(I(U_{2c}V_{1c};Y_4U_{2p}|U_{1c}T_cQ)-6\epsilon)}\\
Pr\left(E_{dec4,b,1,j',1,l'\neq1,l_{2c,b},n'}|(j',n')\neq(1,l_{2p,b})\right)&{\leq}&2^{-n(I(U_{2p}V_{1c};Y_4U_{2c}|U_{1c}T_cQ)-6\epsilon)}\\
Pr\left(E_{dec4,b,i',j',1,l'\neq1,m',n'}|(i',m')\neq(1,l_{2c,b}),(j',n')\neq(1,l_{2p,b})\right)&{\leq}&2^{-n(I(U_{2c}U_{2p}V_{1c};Y_4|U_{1c}T_cQ)-6\epsilon)}\\
Pr\left(E_{dec4,b,i',j',1,1,m',n'}|(i',m')\neq(1,l_{2c,b}),(j',n')\neq(1,l_{2p,b})\right)&{\leq}&2^{-n(I(U_{2c}U_{2p};Y_4|V_{1c}U_{1c}T_cQ)-6\epsilon)}
\end{IEEEeqnarray*}
\setcounter{equation}{\value{tempequationcounter}}
\hrulefill
\vspace*{4pt}
\end{figure*}

Considering (\ref{eqn:error_Rx1_dec_first})-(\ref{eqn:error_Rx1_dec_secondIII}), it can easily be shown that $P_{e,dec3,b}\rightarrow 0$ as $n\rightarrow\infty$ if (\ref{eqn:zerodelay_Rx1_begin})-(\ref{eqn:zerodelay_Rx1_end}) hold.

Finally, employing an approach similar to that utilized for Rx1, we bound the probability of decoding error for Rx2 at the end of block $b$ (defined as $P_{e,dec4,b}$):
\begin{IEEEeqnarray}{l}
P_{e,dec4,b}{=}Pr(E_{dec4,b}|E_{enc2,b}^c)\nonumber\\
=Pr\Big(E_{dec4,b,1,1,1,1,l_{2c,b},l_{2p,b}}^c\nonumber\\
\qquad\cup\bigcup\limits_{(i',j',k')\neq (1,1,1)}E_{dec4,b,i',j',k',l',m',n'}|E_{enc2,b}^c\Big)\nonumber\\
{\leq}Pr\left(E_{dec4,b,1,1,1,1,l_{2c,b},l_{2p,b}}^c|E_{enc2,b}^c\right)\nonumber\\
\;\:{+}\sum\limits_{(i',j',k')\neq(1,1,1)}Pr\left(E_{dec4,b,i',j',k',l',m',n'}|E_{enc2,b}^c\right)\nonumber\\
{\leq}\epsilon + \sum\limits_{(i',j',k')\neq(1,1,1)}Pr\left(E_{dec4,b,i',j',k',l',m',n'}|E_{enc2,b}^c\right)\qquad\label{eqn:error_Rx2_dec_first}
\end{IEEEeqnarray}

There are fifty six cases that cause an error for the second terms in the right side of (\ref{eqn:error_Rx2_dec_first}) with only seven distinct p.m.fs. Applying the packing lemma \cite{ElgKim10, CovTho06}, the probabilities of these events (conditioning on $E_{enc2,b}^c$ suppressed) can be bounded as described in \eqref{eqn:error_Rx2_dec_second}. Combining (\ref{eqn:error_Rx2_dec_second}) and (\ref{eqn:error_Rx2_dec_first}), we can see that when (\ref{eqn:zerodelay_Rx2_begin})-(\ref{eqn:zerodelay_Rx2_end}) hold, $P_{e,dec4,b}\rightarrow 0$ as $n\rightarrow\infty$.

\section*{Acknowledgment}
The authors would like to thank all the anonymous reviewers and the associate editor for their constructive comments and suggestions on the paper. The authors also wish to thank M. H. Yassaee and R. Bayat for their helpful comments.

\begin{IEEEbiographynophoto}{Mahtab Mirmohseni}
received the  B.Sc. and M.Sc. degrees in Electrical Engineering from Sharif University of Technology (SUT), Tehran, Iran, in 2004 and 2007, respectively. She is currently a Ph.D. candidate at Sharif University of Technology, Tehran, Iran, under the supervision of Prof. M. R. Aref. Her research interests include areas of communication theory and multiuser information theory with emphasis on cognitive networks.
\end{IEEEbiographynophoto}

\begin{IEEEbiographynophoto}{Bahareh AKhbari}
received the B.Sc. degree in 2003, the M.Sc. degree in 2005 and PhD degree in 2011 all in Electrical Engineering from Sharif University of Technology (SUT), Tehran, Iran. She was also a visiting PhD student at the University of Minnesota, MN from October 2010 to  September 2011. Her research interests include network information theory and communication theory.
\end{IEEEbiographynophoto}

\begin{IEEEbiographynophoto}{Mohammad Reza Aref}
was born in city of Yazd in Iran in 1951. He received his B.Sc. in 1975 from University of Tehran, his M.Sc. and Ph.D. in 1976 and 1980, respectively, from Stanford University, all in Electrical Engineering. He returned to Iran in 1980 and was actively engaged in academic and political affairs. He was a Faculty member of Isfahan University of Technology from 1982 to 1995. He has been a Professor of Electrical Engineering at Sharif University of Technology since 1995 and has published more than 230 technical papers in communication and information theory and cryptography in international journals and conferences proceedings. His current research interests include areas of communication theory, information theory and cryptography with special emphasis on network information theory and security for multiuser wireless communications. At the same time, during his academic activities, he has been involved in different political positions. First Vice President of I. R. Iran, Vice President of I. R. Iran and Head of Management and Planning Organization, Minister of ICT of I. R. Iran and Chancellor of University of Tehran, are the most recent ones.
\end{IEEEbiographynophoto}




\end{document}